# Toward a Circular Nanotechnology for Biofuels: Integrating Sustainable Synthesis, Recovery, and Performance Optimization

Lydia Lonardi[1], Caitlyn Lew-Tong[1], Brian Dunsten Miranda[2], Chandraleka V.S.[3], Evelyn (Wenxuan) Li[4], Bhaavyaa Sahukari[5], Harshit Poddar[6], Keerthana Satheesh[7], **Utkarsh Chadha[4, *]**

[1]*Department of Mechanical and Industrial Engineering,* ***University of Toronto****, Toronto, Ontario, Canada M5S 3G8*

[2]***Monash University****, Wellington Rd, Clayton VIC 3800, Australia*

[3]***University College Dublin****, Ireland*

[4]*Department of Materials Science and Engineering,* ***University of Toronto****, Toronto, Ontario, Canada M5S 3E4*

[5]***Vellore Institute of Technology****, Bhopal, Madhya Pradesh, India*

[6]***KTH Royal Institute of Technology****, Stockholm, Sweden*

[7]***Vellore Institute of Technology****, Vellore, Tamil Nadu, India*

*Corresponding: utkarshchadha1302@gmail.com;

## Abstract

This review exhaustively evaluates the role of nanomaterials across the synthesis, characterization and application stages of biofuel systems. Common types of nanomaterials that are used for biofuel applications include metal oxides, carbon-based structures, and hybrids, which are evaluated for their effectiveness in efficient biofuel production. The properties of such nanomaterials are being utilized as an aid to produce biofuels through improved catalysis, enzyme immobilization and thermal stability. Common synthesis methods, such as sol-gel, coprecipitation, and green synthesis, are compared, alongside characterization tools, such as TEM, SEM, FTIR, and BET. This study focuses on transesterification, biomass pretreatment, and fermentation processes, where nanomaterials significantly improve yield and reusability. There are several challenges, despite the merits of using nanomaterials, and the trade-offs include cost, scalability, and environmental impact, which further expand into evaluating the life cycle of such materials. This review outlines the practical potential of nanomaterials in enabling efficient and sustainable biofuel production.

**Keywords:** Nanostructure; Green synthesis; Life cycle assessment, Catalytic, Sustainability

***This is an accepted manuscript with modified figures and post-acceptance corrections. Please consult the published version for the final record.***

## Graphical Abstract

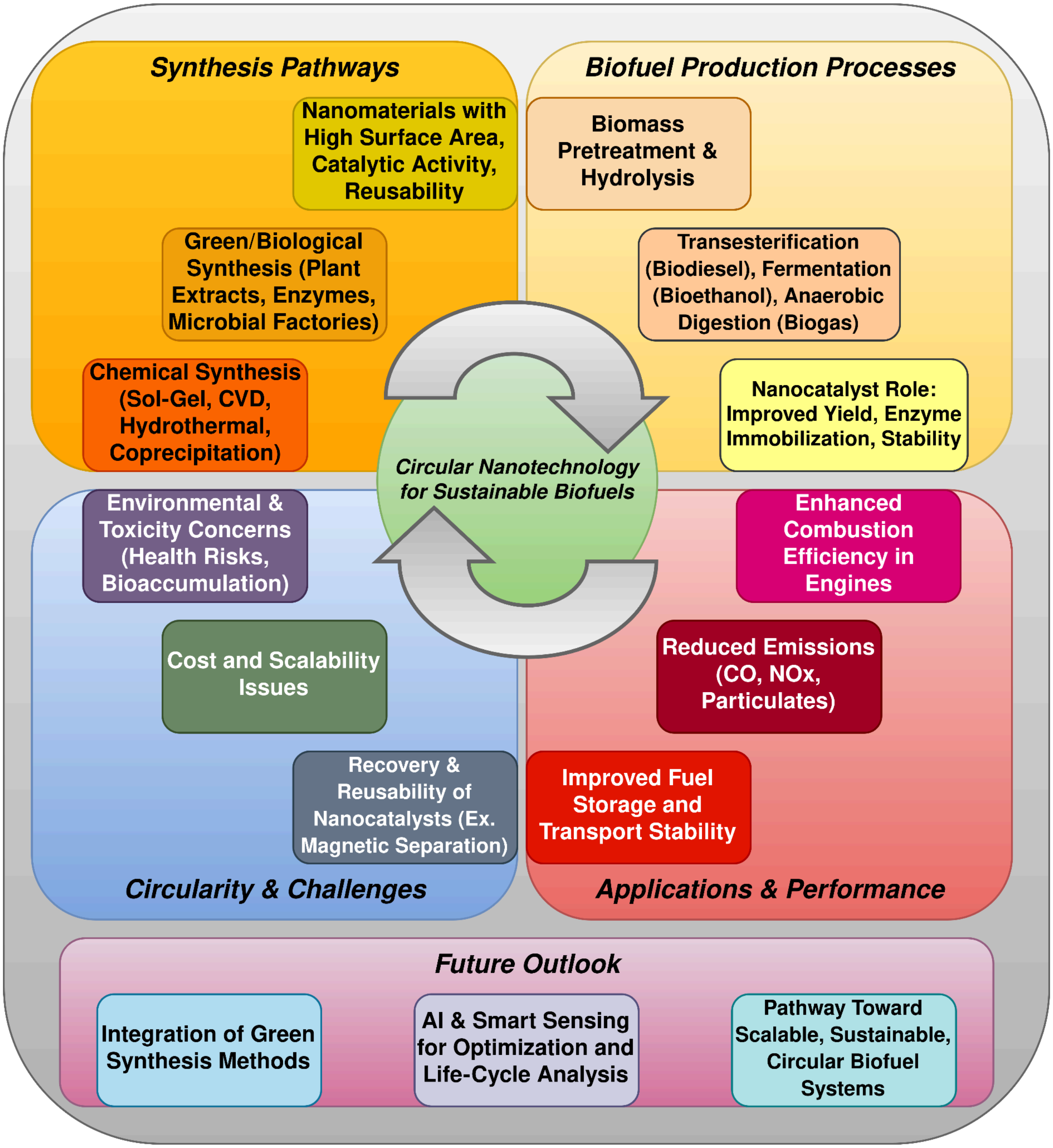

## 1. Introduction

Climate change is increasingly recognized as one of the most urgent global issues, largely resulting from the extensive use of fossil fuels, which are responsible for almost 90% of carbon dioxide emissions and more than 75% of greenhouse gas emissions, according to the United Nations [1]. As a result, there is a pressing need to transition to cleaner, renewable energy sources. Biofuels serve as a particularly promising solution, as they are produced from organic matter (often matter that would not otherwise be utilized), reduce emissions, and provide more favorable long-term environmental impacts [2, 3]. Owing to their unique properties, the integration of nanomaterials has shown potential for further enhancing the use of biofuels.

Nanomaterials are more frequently employed in biofuel applications to improve the efficiency of several processes, particularly the reactions involved in the pretreatment and production of biofuels such as biodiesel, bioethanol, biogas, and the conversion of fuel cells. Their functions typically involve catalyzing reactions, enabling enzyme immobilization, and aiding in the separation of reaction products [4]. These roles are supported by their unique properties, including a high surface-to-volume ratio [5], reusability [6], thermal and chemical stability [7] and capacity for surface functionalization [8]. In addition, nanoparticle structures can be zero-dimensional (0D), such as quantum dots and fullerenes; 1D, such as nanotubes; 2D, such as graphene; and 3D, such as graphite [9]. The properties of nanomaterials vary depending on their composition and structure, which forms the basis for their classification. Commonly used nanoparticles in biofuel applications include metal/metal oxide-based and carbon-based materials, as well as numerous composites and hybrids [10].

NPs can be synthesized for biofuel applications via chemical, green, biological, and enzymatic methods [11]. These synthesis methods play important roles in determining the characteristics of a nanoparticle, which makes it suitable for different roles as a catalyst in biofuel production [12]. The most sought-after qualities include having a large surface area, acidic properties, and high porosity, among others [12, 13]. Owing to the expensive and sometimes unsustainable nature of nanoparticle synthesis, more companies are starting to move toward green synthesis methods, as they have significantly lower environmental impacts and can be made from free waste products such as waste cooking oil [14, 15]. They are categorized into several classes of nanomaterial microstructures, with even slight changes resulting in significantly different characteristics [16]. One such characterization of nanomaterials consists of bulk solids that have building blocks of a singular length scale, which can be classified into two classes. Nanoparticle characterization techniques include transmission electron microscopy (TEM) [12], scanning electron microscopy (SEM) [17], Fourier transform infrared spectroscopy (FTIR) [18], and the Brunauer–Emmett–Teller (BET) method [19]. These methods are all used to distinguish different properties of the nanoparticle and may be used in conjunction with other characterization techniques for more specific results [12].

Nanomaterials have become increasingly important in improving biofuel production, offering advancements in efficiency, yield, and environmental impact. Their unique properties, such as high surface area and stability, allow for more effective processes for biomass treatment as well as biodiesel, bioethanol, and biogas production. For example, these materials help optimize reaction conditions [2], increase enzyme performance through attachment (immobilization) [20], and enable catalyst recovery and reuse [21]. Nanomaterials offer multiple advantages for biofuel performance. They serve as components in both microbial and enzymatic biofuel cells, enhancing conductivity, electron transfer, and biocompatibility, which leads to improved output and efficiency [22-24]. Additionally, nanomaterials increase combustion efficiency by facilitating more complete fuel oxidation and reducing emissions such as nitrogen oxides and carbon monoxide [25-27]. Furthermore, they improve biofuel storage and transportation by increasing stability and minimizing contamination [22, 28].

The rapid expansion of engineered nanoparticles (ENPs) has introduced several potential challenges and limitations. Environmental and toxicological concerns have been identified due to their durability, which leads to potential bioaccumulation, and health risks associated with the emission of ultrafine particles [29-31]. Economically, industrial biofuel production encounters major cost challenges, particularly due to expensive biomass pretreatment and the complex nanomaterial synthesis process [32, 33]. Current life cycle analyses of nanoparticles are often insufficient and lack thorough consideration of their toxicity and potential concerns with their disposal [34]. Future nanotechnology research should focus on enhancing sustainability and efficiency to ultimately increase industrial feasibility. Green synthesis methods are emerging as an eco-friendly approach for producing bionanoparticles using renewable biological resources, reducing the ecological impact while enhancing the efficiency of biofuel production [35]. Similarly, the integration of artificial intelligence (AI) and smart sensing technologies for monitoring, optimization, and control of biofuel processes will also increase sustainability by increasing efficiency and minimizing emissions [36-38]. There is also potential for nanomaterials to model biological systems through biomimicry or function as nanohybrids to enhance efficiency [39, 40].

At this point in time, nanomaterials must be engineered not only for performance in their own biofuel systems but also for retrofitting with existing infrastructure. This minimizes cost and accelerates potential use across biofuel production environments. Petroleum refineries are a major source of global energy but are also among the leading contributors to air pollution [41]. While renewable energies offer a more sustainable future, they still require certain advancements; therefore, the oil and gas industry remains predominant [42]. Owing to their central role in energy production and the issues they present, they would greatly benefit from the integration of nanomaterials.

### 1.1. Research objectives and methodology

#### 1.1.1. Objectives

The objective of this comprehensive review is to provide a system-level analysis of the role of nanomaterials in the development of biofuel production. This study specifically aims to be a dominant source on developments in the field of biofuels and the role of nanomaterials in research on sustainable fuels and current developments in nanomaterial-based catalysis, pretreatment, and conversion processes for biofuels are synthesized. It also aims to focus on sustainable synthesis routes for nanomaterials, strategies for the recovery, recyclability, and environmental compatibility of nanocatalysts, and a circular framework connecting nanomaterial design to sustainable biofuel.

#### 1.1.2. Research Questions

This study is guided by the following key questions:

- What types of nanomaterials are most effective in different stages of biofuel production?

- How do synthesis methods influence the performance, reusability, and environmental impact of nanomaterials?
- What are the critical challenges in integrating nanotechnology into scalable and circular biofuel systems?
- How can material recovery, lifecycle performance, and sustainability metrics be improved through nanomaterial design?

### 1.1.3. Methodology

This review utilizes literature from databases such as Scopus, Web of Science, Springer Nature, ScienceDirect, and the Wiley Online Library, with a focus on keywords such as nanotechnology, biofuels, green chemistry, and sustainability. This is a structured literature analysis of peer-reviewed articles, and industrial reports from 2000--2025. Studies were classified and analyzed according to material type, synthesis route, application domain (biodiesel, bioethanol, biogas, biohydrogen), recovery methods, and sustainability metrics. Emphasis has been placed on recent advancements that demonstrate circularity principles, environmental impact mitigation, and technoeconomic viability.

# 2. Nanomaterials for biofuel applications

## 2.1. Types of nanomaterials used

### 2.1.1. Metal and Metal Oxide Nanoparticles (NPs)

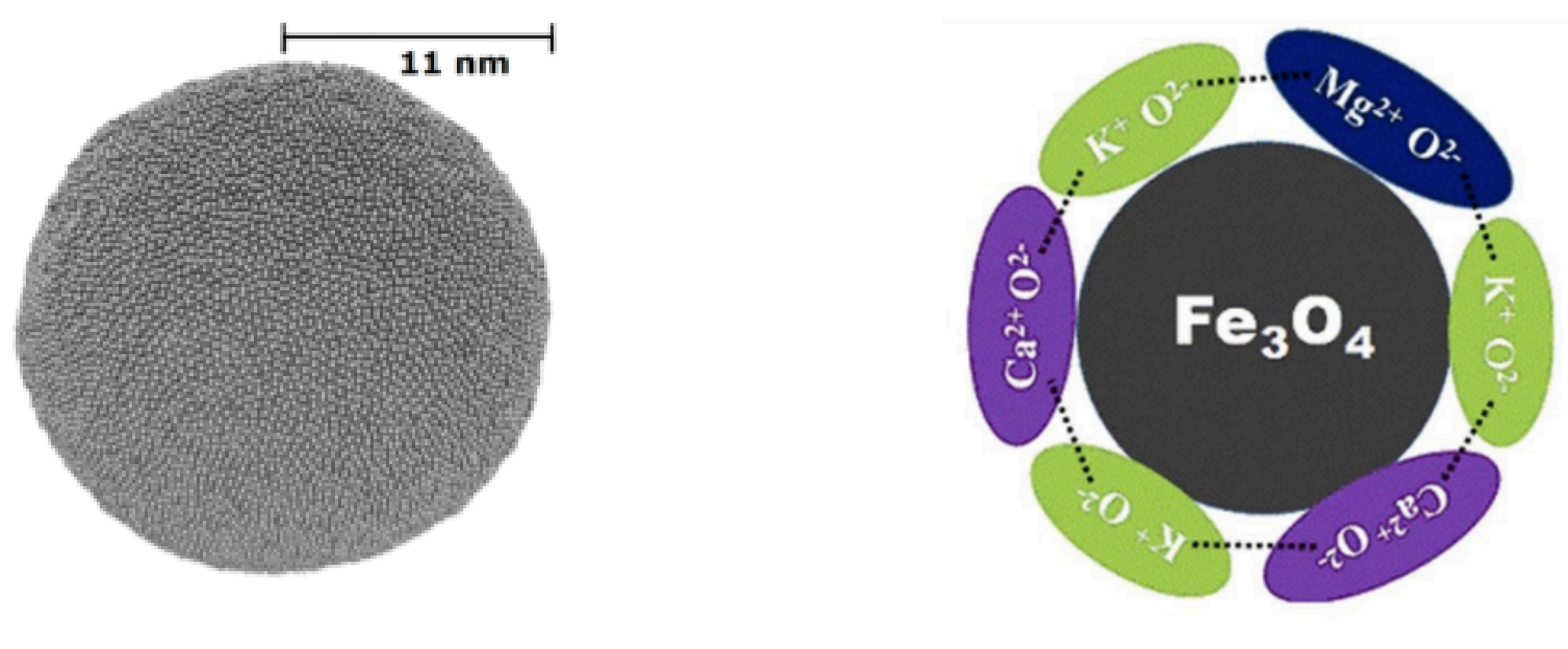

**(a)** **(b)**

**Figure 1.** (a) Amorphous Iron Metal Nanoparticle [43].; (b) $Fe_3O_4$ Magnetic Nanocatalyst Coated with Citrus Sinensis Peel Ash [51].

Metal nanoparticles, ranging from 1 to 100 nm in diameter, are monomers composed of a single element, either as isolated atoms or groups of atoms [44]. Common examples include Iron (as displayed in **Figure 1(a)**), Gold, Copper, Silver, or Palladium [45]. Metal oxide nanoparticles, which fall within the same size range as metal nanoparticles, involve metal atoms chemically bonded to oxygen atoms, such as $Fe_2O_3$, MgO, $SiO_2$, and $Al_2O_3$ [46]. The morphologies of metal and metal oxide nanoparticles are diverse and include 0D nanoparticles and 1D, 2D, and 3D nanomaterials [47]. Metal nanomaterials serve as key additives and often act as catalysts to improve yield and stability in the production of biofuels such as biodiesel, bioethanol, biogas, and their pretreatment or hydrolysis. For example, metal nanoparticles are widely used for diesel fuel to increase combustion efficiency and reduce emissions. Their high thermal conductivity, catalytic activity, and high surface area help accelerate the beginning of combustion, enhance fuel–air mixing, and reduce soot and carbon monoxide emissions. Metal oxides can also reduce hazardous nitrogen oxide emissions when they catalyze oxidation and reduction reactions by absorbing oxygen [12, 48-50]. Furthermore, metal oxide particles prevent fungal growth, are actively resistant to microbes, are semiconducting, are optically absorbent, possess chemical sensing properties and have a large surface-to-volume ratio. It is common practice to use metal oxides in sensor and biosensor technologies because they are desirable because of their high sensitivity, fast response and recovery time, and high stability and reproducibility [47]. They are also less costly but more difficult to produce.

In addition, certain metal-based nanoparticles have magnetic properties, meaning that they can be controlled via magnetic fields [52]. However, they are typically superparamagnetic, meaning that their magnetic properties are only displayed in the presence of an external magnetic field [53]. These nanoparticles can be made from pure magnetic metals such as iron and nickel or from oxides such as nickel oxide [54]. Magnetic nanoparticles are utilized in several biofuel production processes, including transesterification for biodiesel or hydrolysis and fermentation for bioethanol. They typically serve as catalysts to increase reaction rates and improve yields [55]. Their large surface area provides numerous active sites for catalysis, and their chemical and thermal stability allows them to withstand higher temperatures typical of fuel production [56]. Furthermore, their magnetic nature enables easy separation from mixtures via magnetic fields once the reactions are complete, allowing for efficient recovery and reuse for multiple reaction cycles [55]. As shown in **Figure 1(b),** magnetic nanocatalysts are often coated with additional material, forming a composite (as described in section 2.1.3), to tailor certain behaviors, such as preventing corrosion and increasing hydrophobicity. For example, a study coated Citrus sinensis peel ash (CSPA) on a magnetic $Fe_3O_4$ nanocatalyst to produce biodiesel from cooking oil. The CSPA increased the surface area and provided basic components, including calcium and potassium, which enhanced the catalysis of the transesterification reaction. The nanocatalyst helped synthesize a maximum biodiesel yield of 98% [51].

### 2.1.2. Carbon-based nanomaterials

Owing to their unique properties, carbon-based nanomaterials (CBNs) are widely used for biofuel applications. They are typically thermally and electrically conductive, chemically stable, and have a large surface-to-volume ratio to increase the number of reaction sites [60]. For example, their conductivity, large surface area, and biocompatibility (based on the presence of carbon in many biological systems) make them effective catalysts for enhancing fuel cell activity [61]. CBNs can also be easily modified for diverse applications because of their versatile bonding enabled by sp, $sp^2$, and $sp^3$ hybridization, which allows the formation of various allotropes, such as graphene, carbon nanotubes (CNTs), and fullerenes [60]. CBNs can be modified via both covalent and noncovalent methods, tailoring specific electronic, optical, hydrophilic, and mechanical properties. Noncovalent functionalization involves electrostatic and van der Waals forces; in contrast, covalent functionalization often employs oxidation, creating oxygen-containing groups that enable further reactions [62].

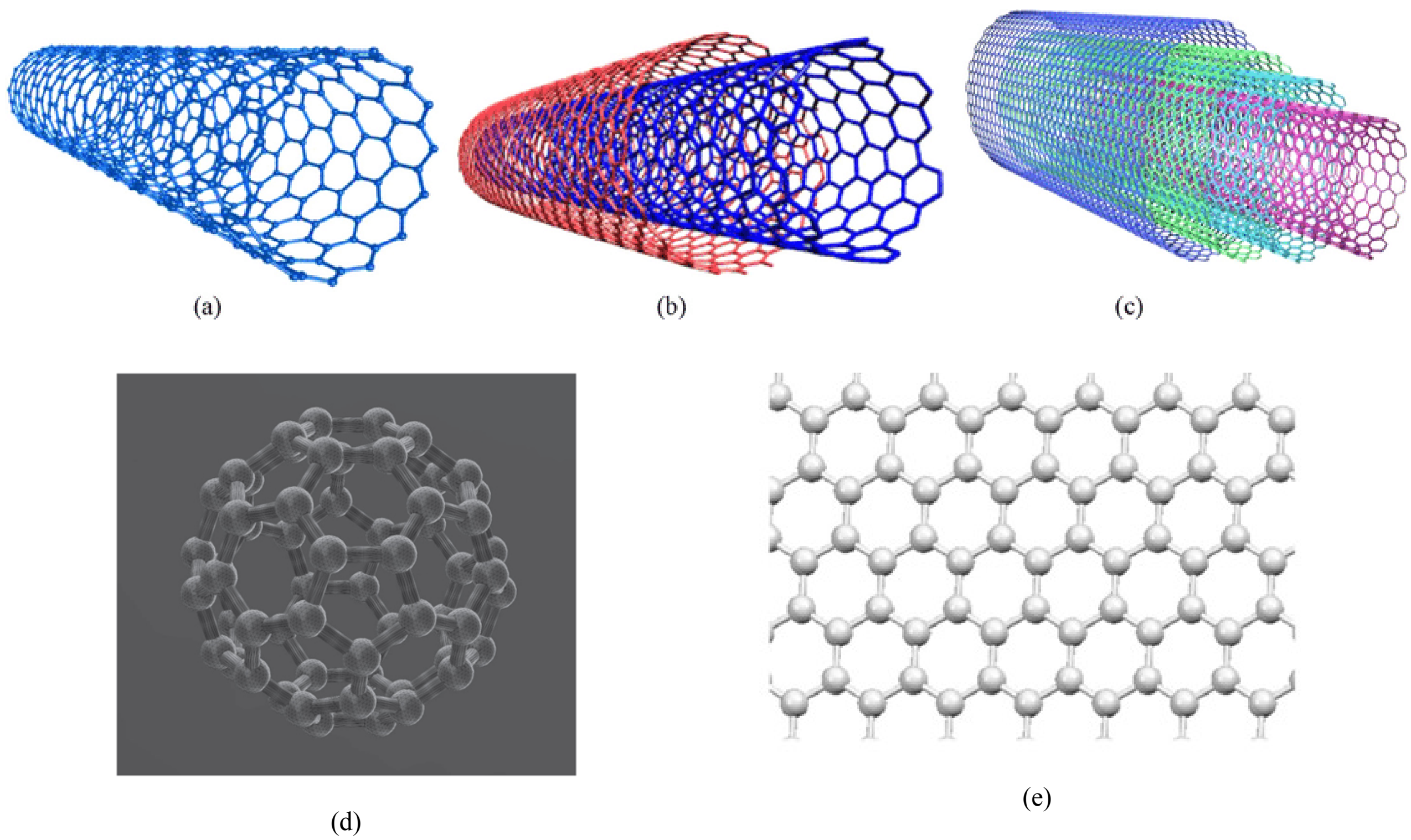

**Figure 2.** Carbon-based nanomaterials: (a) SWCNTs, (c) DWCNTs, (b) MWCNTs, (d) fullerene, and (e) graphene. (a), (b), (c) Adapted from [57], (d) Adapted from [58], (e) Adapted from [59].

CNTs are 1D cylindrical nanostructures formed from graphene sheets that can be rolled into various lengths, diameters, and layer counts. They are commonly used as catalysts to increase reaction rates because of their large surface area (with a length–to–diameter ratio of up to 132 million [63]), electrical conductivity, and chemical stability, making them useful for biofuel use and production [61]. The three forms of CNTs are single-walled CNTs (SWCNTs) **(Figure 2** (a)), multiwalled CNTs (MWCNTs) (**Figure 2** (b)), and double-walled CNTs (DWCNTs) (**Figure 2** (c)), each with distinct properties [64]. SWCNTs consist of a single graphene layer with a diameter of 1.5–15 nm and high electrical conductivity ($10^6$ $m^{-1}$) [65]. MWCNTs contain multiple concentric graphene layers and are preferred for biodiesel production because of their reduced cost [66]; they have a diameter of 20–50 nm, lower conductivity ($10^5$ $m^{-1}$), and greater mechanical strength [65]. DWCNTs contain two nested SWCNTs, combining the properties of SWCNTs and MWCNTs [64]. Fullerenes are spherical, hollow-structured carbon allotropes composed of at least 60 atoms, with a combination of single and double bonds, as shown in **Figure 2** (d). They are highly stable and chemically inert; however, they are soluble in organic solvents, which increases the potential for functionalization through addition and redox reactions, allowing their properties to be tailored for various applications [61]. Graphene (**Figure 2** (e)) is a lightweight 2D sheet of $sp^2$-hybridized carbon atoms organized in a hexagonal lattice structure. It is the base unit for fullerenes and CNTs, offering properties such as electrical conductivity, a large surface area, and mechanical strength. It can be used for catalysis of reactions and energy storage. For example, sulfonated graphene has been used as a catalyst in the transesterification of palm oil with methanol to produce biodiesel [67].

### 2.1.3. Hybrid and Composite Nanostructures

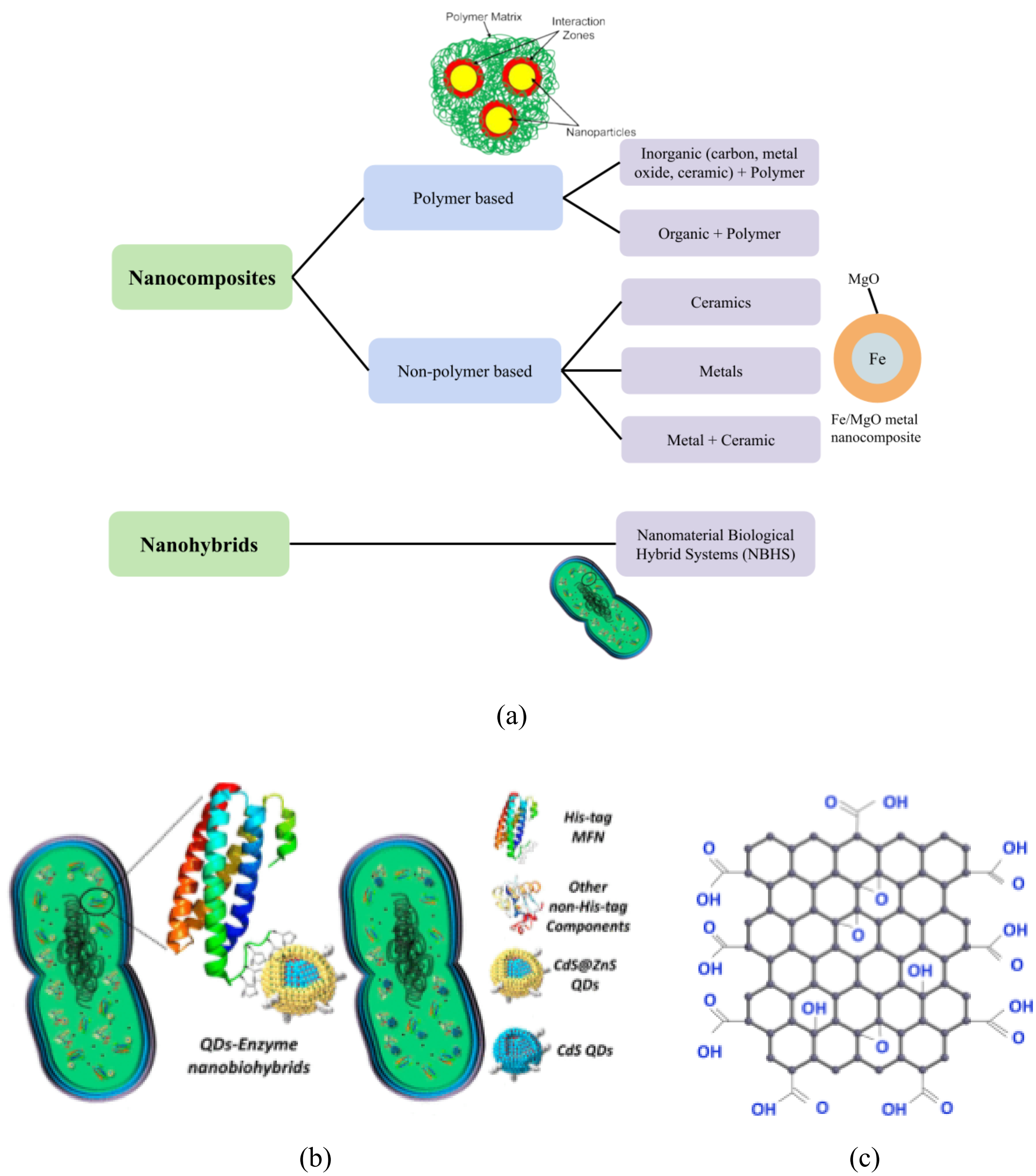

**Figure 3.** (a) Nanocomposite and nanohybrid classification. Adapted and recreated from [72], the NBHS molecule was adapted from [73], and the polymer matrix was adapted from [74].

Examples of nanocomposites and nanohybrids for biofuel applications (b) "Nanorg" NBHS, composed of carbon quantum dots and microbes. Adapted from [73], (c) graphene oxide. Adapted from [75].

Nanohybrids combine at least two nanomaterials, whereas nanocomposites combine nanoparticles and other materials, such as polymers [68]; a general classification of each is detailed in **Figure 3(a)**. Both create unique structures with particular characteristics, such as modified surface areas, catalytic properties, and biocompatibility properties [69, 70]. Nanocomposites are typically categorized as either polymer-based or nonpolymer-based. Polymer nanocomposites, usually less than 100 nm in length, feature either inorganic materials such as carbon-based nanoparticles, metal oxides, and ceramics or organic materials such as cellulose and lignin dispersed throughout a polymer, increasing strength and conductivity [71]. Nonpolymer nanocomposites combine metals and/or ceramics with more than one solid phase, offering enhanced properties such as greater mechanical strength [71]. Various hybrid and composite nanoparticles, such as nanomaterial biological hybrid systems (NBHSs), metal-based composites, and carbon-based composites, are currently being developed for biofuel production and use.

Composite nanomaterials can enhance biofuel synthesis. For example, photovoltaic thermal systems (PVSs) generate heat and electricity to support processes in biofuel production. Hybrid nanofluids, such as $TiO_2/Al_2O_3$ composites, enhance thermal regulation in PVSs, leading to improved stability and increased electrical output [76, 77]. NBHS combines biological systems and photocatalytic nanomaterials (typically metal oxides [78]) to speed up photosynthetic processes in traditional biofuel-producing organisms; for example, electron transfer across membranes in bacteria or algae. NBHS acts as an environmentally friendly catalyst that improves the efficiency of $CO_2$ fixation into fuels and chemicals [79]. For example, **Figure 3(b)** depicts a detailed example of "Nanorg" nanobiohybrids, which are microbes enhanced with carbon quantum dots (CQDs). They use sunlight to produce ethanol and hydrogen directly from $CO_2$, nitrogen, and water, eliminating the need for sugar or fossil fuel inputs [73]. Furthermore, **Figure 3(c)** displays the structure of graphene oxide (GO), i.e., graphene functionalized with various oxygen-containing groups, which serves as a catalyst for the transesterification of fats and oils into biodiesel. Its structural, electrical and mechanical properties allow it to be useful in many areas of the oil and gas industry, such as drilling, lubrication, oil–water separation, and prevention of corrosion [42, 80].

## 2.2. Functional Requirements of Nanomaterial Properties in Biofuel Systems

Nanomaterials must exhibit various properties to effectively support the production and function of nanomaterial-based biofuels, including high surface-to-volume ratios, reusability, thermal and chemical stability and the capacity for surface functionalization [6, 7]. These characteristics not only contribute to the efficient and high-quality generation of biofuels but also increase the overall feasibility and practicality of the use of biofuels as alternatives to fossil fuels. For biofuels to serve as an optimal replacement, they must also be cost-effective and environmentally sustainable [81].

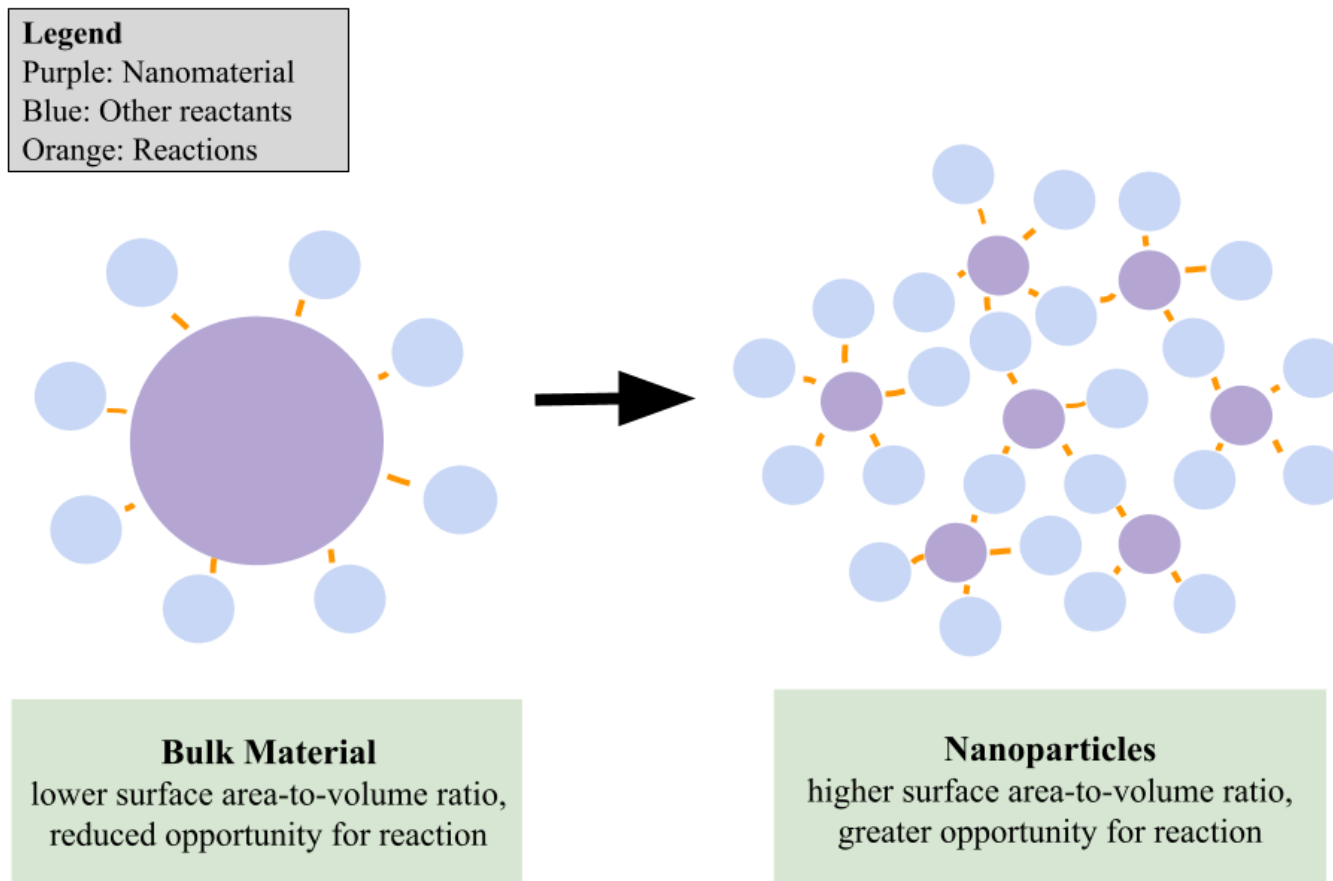

(a)

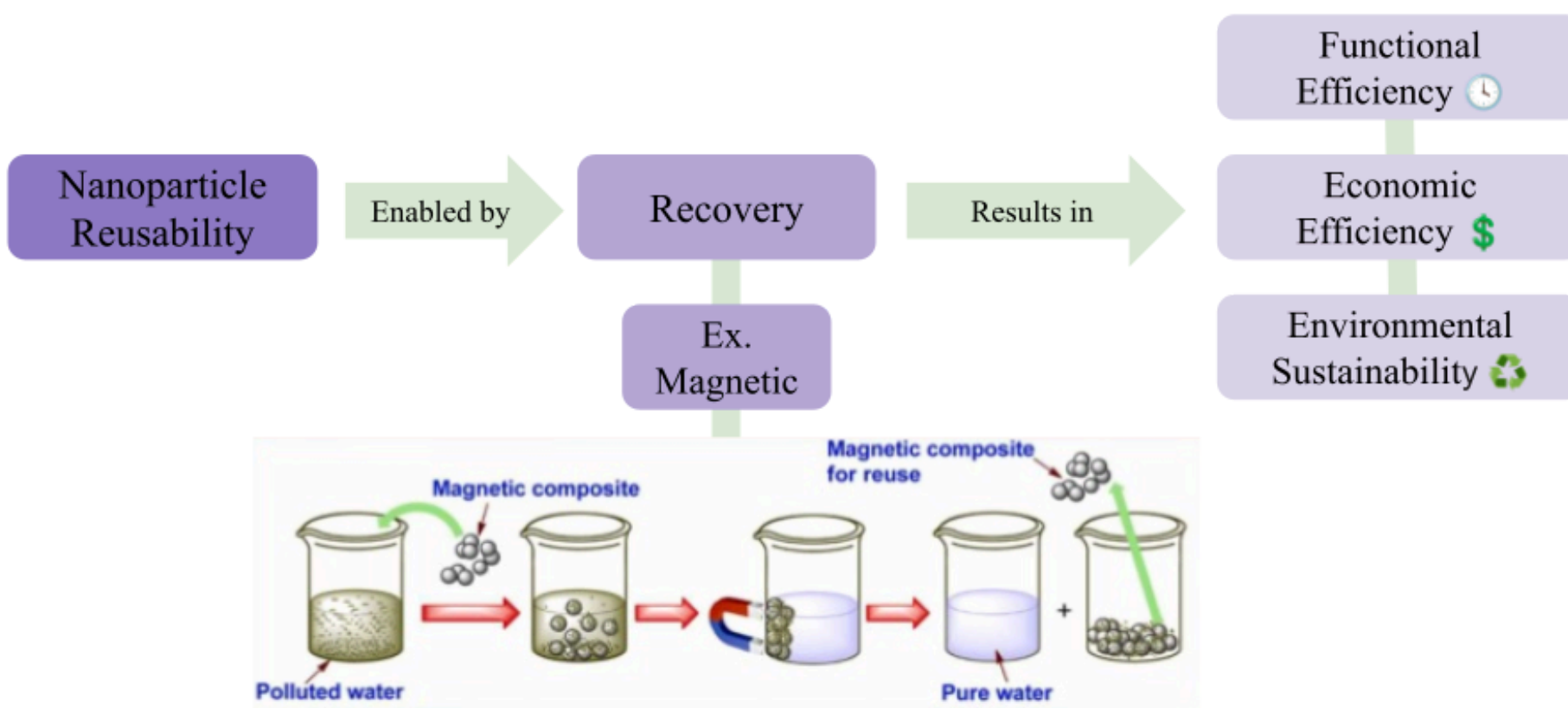

(b)

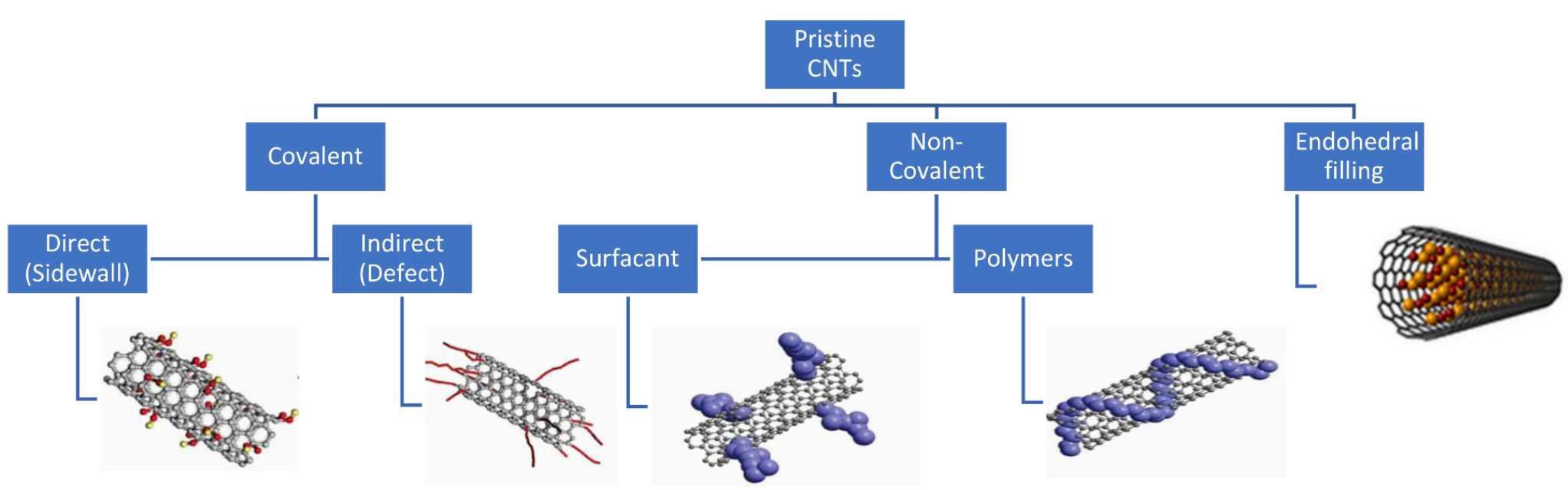

π–π

(c)

**Figure 4.** (a) Bulk material versus nanoparticle reactions**.** Recreated from [84]; (b) Advantages of Nanoparticle Reusability Enabled by Recovery, Exemplified by Magnetic Nanoparticle Retrieval. Derived from [88]; (c) Several ways to Functionalize Carbon Nanotubes [95].

### 2.2.1 High surface area-to-volume ratio

As displayed in **Figure 4 (a)**, a large surface-area-to-volume ratio of nanomaterials contributes to many of their unique properties, as it enables more reactive sites to be exposed to the environment, enabling faster and more efficient reactions. Specifically, atoms located at the edges and vertices of nanoparticles have fewer bonds with neighboring atoms (low coordination numbers) and are therefore more reactive, as they have a greater need to form additional bonds to stabilize themselves [82]. As a result, their large surface area makes nanomaterials highly useful in catalysis, thereby increasing yield. For example, they have more sites to bind to and immobilize large quantities of enzymes involved in biofuel production, such as those used in bioethanol fermentation [83].

### 2.2.2 Reusability

Reusability of nanomaterials is essential for economic and functional efficiency, as well as environmental sustainability, since it reduces the need for frequent material replacement and minimizes waste production [6]. **Figure 4 (b)** generalizes the importance of nanoparticle reusability. Nanocatalysts are often preferred over other materials because of their reusability. For example, homogeneous catalysts, such as potassium hydroxide and sodium hydroxide, which exist in the same phase as reactants, have been commonly used in biodiesel production because of their low initial cost. However, heterogeneous nanocatalysts, which exist in different phases than reactants do, are becoming more favored because of their ability to be reused [50, 85]. Nanomaterials that retain their catalytic efficiency and structural integrity over several reaction cycles demonstrate consistent performance and are valuable for applications that require long-term operations (without interruptions to replace catalysts).

Consequently, it is also important for nanoparticles and their bound substrates, such as enzymes, to be easily removed from mixtures to be ready for reuse. Various methods are employed to separate and reuse catalysts from reaction mixtures, such as centrifugation, filtration, and manipulation of magnetic nanoparticles with magnetic fields [86]. Reusability and recovery are constantly being monitored and optimized in development processes. For example, a lignocellulosic biomass pretreatment strategy uses cerium-doped iron oxide nanoparticles to achieve 44% biomass breakdown, also known as delignification, in the first run. Fifty percent of the nanocatalysts were then recovered for reuse via magnetic fields; 37.1% achieved delignification in the second run, followed by 30.8% in the third run [87].

### 2.2.3 Stability: Chemical and Thermal

Nanoparticle stability refers to the ability to preserve properties such as size, morphology, and surface chemistry over time and under various conditions, enabling the nanoparticle to perform its intended functions effectively. Stability therefore implies durability, efficiency, and longevity, which increases the potential for reusability, ultimately reducing costs and materials [7]. Stability plays a critical role in biofuel applications, as it directly impacts fuel performance, safety, storage life, and engine compatibility [89]. Instability may prevent nanoparticles from fulfilling their intended role. Particular consideration is required when nanomaterials are involved, as they tend to agglomerate in biofuel blends and must often function under more extreme conditions [90]. For example, a study reported that a metal–oxide nanocatalyst used for biodiesel production became unstable at high temperatures, leading to deformation. This deformation prevents the catalyst from binding to its target, resulting in deactivation, which hinders biodiesel yield [91, 92].

Stability in nanomaterials is a major criterion in biofuel development to ensure reliable performance in these applications. For example, a study aimed to improve the poor stability of cerium oxide nanoparticles in a diesel-biodiesel blend (80% diesel, 20% biodiesel from waste cooking oil), which agglomerates over time, thereby reducing their effectiveness. Consequently, cerium oxide was combined with alumina, a particle known for its high stability, to produce a nanocomposite. This functionalized nanoparticle showed 44% greater stability within one month. [90].

Recent research has focused on improving nanocatalyst stability. For example, to address certain nanoparticles' tendency to cluster together, i.e., aggregation, highly active core/yolk shell nanocatalysts have emerged to improve the stability of processes that occur at high temperatures, such as the operation of fuel cells [93]. They achieve this by confining the active material to the interior core, which physically restricts movement, prevents aggregation, and shields sites from chemical or thermal degradation [94].

### 2.2.4 Capacity for Surface Functionalization

The surface functionalization of nanoparticles refers to the addition of chemical groups to alter their surface or the attachment of various inorganic or organic materials. As exemplified in **Figure 4 (c),** with CNTs, this process involves different types of bonds, such as covalent, noncovalent, hydrogen bonding, or van der Waals forces, and can even fill nanoparticles in certain cases [95]. As a result, many nanoparticle variations can be produced, each with unique properties. This process is useful for modifying nanoparticle characteristics, such as stability, catalytic activity, and compatibility, to make them well suited for a wide range of applications [96].

Various studies have demonstrated that functionalizing nanoparticles can improve their capacity for enzyme immobilization, meaning that enzymes can attach more effectively to the nanomaterial and work with greater stability and efficiency. This is particularly valuable in biofuel production, where enzymes play key roles in catalyzing reactions, such as assisting in the decomposition of biomass during pretreatment or accelerating reactions during biofuel production. For example, a study aimed to immobilize lipase enzymes onto mesoporous silica nanoparticles as catalysts to improve biodiesel production. Hydrophobic groups were functionalized onto the silica surface to mimic the natural substrate of lipases and enable them to better conform to nanomaterials, thereby increasing their enzymatic activity and stability during transesterification reactions [97].

Additionally, functionalizing magnetic nanoparticles coated with mesoporous silicon with amine ($–NH_2$) and aldehyde (-CHO) groups allowed for improved immobilization with Rhizopus oryzae lipase, followed by uncomplicated magnetic recovery of the enzyme–nanoparticle complex from the reaction mixture for reuse in various cycles [98]. Similarly, functionalizing nanocarbons is beneficial for biofuel cells, as enzyme immobilization is a crucial step in creating enzymatic electrodes for these cells. Functionalized carbon nanomaterials, which feature reactive groups such as hydroxyl and amino groups, possess properties such as high strength, adhesion, and film-forming capabilities, which make them ideal for processes such as enzyme immobilization [99].

## 3. Methods for the Synthesis of Nanomaterials for Biofuel Applications

Nanomaterial synthesis can employ chemical, physical, enzymatic, or biological methods [11]. This section will focus primarily on chemical and green synthesis methods, with an emphasis on how certain techniques are advantageous in biofuel applications.

### 3.1. Chemical

The efficacy of nanocatalysts in biofuel production heavily relies on the synthesis method, as they control the characteristics, size, porosity, surface area, and morphology of the nanoparticles [12]. When combined with other procedures, including hydrogenation, pyrolysis, gasification, and anaerobic digestion, nanotechnology has proven to be especially useful in the production of biofuels [13]. Nanocatalysts exhibit improved catalytic ability in biofuel conversion and synthesis when they are highly porous and have a large surface area with acidic properties, all of which are characteristic of chemical synthesis methods [12, 13].

Chemically synthesized nanocatalysts are essential for increasing the reaction and yield rates in biodiesel production. For metal nanoparticles specifically, chemical synthesis methods are most commonly used; however, there is now a larger transition toward green synthesis methods because of the significantly lower environmental impact [14]. Chemical synthesis is more costly in terms of time and money and requires hard labor, making it less preferable to other synthesis methods. One of the most common nanoparticles synthesized for use in biofuel applications is ZnO [100]. Various chemical synthesis methods exist for producing nanoparticles, each offering advantages and disadvantages that influence the resulting material's characteristics. **Table 1** provides a brief outline of certain chemical synthesis methods.

**Table 1.** Chemical synthesis methods

| Names of Methods | Brief explanation | Remarks | Example Nanomaterial Types | Suitability for Biofuel Synthesis | Advantages | Disadvantages | Ref. |
|---|---|---|---|---|---|---|---|
| Electrodeposition | Creates bimetallic nanostructures, and occasionally combines plasmonic and catalytic characteristics. | Allows for selection and modification of the current and the total charge being transferred - overall controlling the amount of metal ions being reduced. | Pd, Ni–Pd, Ni–$Al_2O_3$, Ni–SiC, Co, Co–W, Co–P, Ni, Ni–P, Ni–Mo, Ni–Zn, Ni–Fe, Ni–Fe–Cr | Moderate | Large specific surface area, low cost, less agglomeration, high safety, environmentally friendly, simple processing. | Requires harsh environments, high temperatures, and additionally capping agents or surfactants. Nanomaterial microstructure and morphology cannot be finely or easily controlled. Highly toxic waste products, reduced long term catalytic activity. | [101-103] |
| Sol-Gel Processes | It goes through two steps driven by a catalyst which result in a 3D network of polycondensed alkoxy and hydroxyl groups. | Enables controlled resultant nanoparticle properties. | $TiO_2$, ZnO, $SiO_2$, $TiO_2$ doped with Fe or N, Mesoporous silica, hybrids | High | Immobilizes catalysts for continuous reuse in biodiesel production applications. Can reach yields of 100% at ambient conditions. Very stable. | Catalysts may transform during synthesis which may impact performance. Different nanocatalysts obtain various performances and yields. | [11, 104, 105] |
| Chemical Vapor Deposition | This method covers all processes that coat ceramic and metallic compounds, where a vapor releases a solid material through a chemical reaction. | This includes a wide range of methods, with the most commonly used being thermal and laser chemical vapor deposition. | CNTs, graphene, $MoS_2$, $WS_2$, Si, GaN, $TiO_2$, $Al_2O_3$, $Si_3N_4$ | High | Produces high quality biofuel and CNTs, Economically feasible, lower defect density, high graphitization, predicted profitability over a ten year period. | High energy demands due to high temperature conditions needed, increased process complexity, requires specialized additional equipment. | [11, 106, 107] |
| Langmuir-Blodgett Method | Used to microfabricate incredibly thin films applied as separation membranes, biological and chemical sensors, | Films produced from this method are ideal for molecular assembly study as | Au, Ag, Graphene oxide, $MoS_2$, Polymer-metal NP films, Phospholipid monolayers, | High | Enables for precise control over material properties, low energy and water usage needed, may use available precursors at low cost, | Posttreatment processes add complexity, must undergo chemical treatment and particle treatment, long-term | [108-110] |
| | optical filters, and for photoelectric applications. | well as for studying aggregation in 2D | amphiphiles | | potential for enhanced reproducibility, offers alternatives to toxic organic solvents, enables automation and large scale production. | material stability has not yet been studied, no demonstration of reproducibility yet. | |
| Hydrolysis | Procedure uses water and an acid or base catalyst to break down units. | The final composition of the hydrolyzed substance after hydrolysis is greatly dependent on the conditions of the procedure and the source of the substance. | ZnO, $Fe_3O_4$, Magnetic $Fe_3O_4$ NPs, Mesoporous silica (MSNs), Graphene oxide (GO), CNTs, CNCs from biomass | High | Can be applied under mild reaction conditions, environmentally friendly, a necessary step in second generation bioethanol production. | May result in toxic waste products and degradation of sugars which will interfere with fermentation. | [111-113] |
| Co precipitation | Nanoparticles are formed when a stabilizing agent is mixed with the metal oxide. | Among the most widely used methods of synthesis. | $Fe_3O_4$, γ-$Fe_2O_3$, CaO–NiO, Mn–$Fe_3O_4$, Zn–$Fe_3O_4$, $Fe_3O_4$@$SiO_2$, $Fe_3O_4$@Au, $Fe_3O_4$–PEG, $Fe_3O_4$–chitosan, $Al_2O_3$, $TiO_2$, ZnO | High | Operation conditions include low temperatures and ambient pressure, low energy consumption, more cost effective, widely used. | Often leads to agglomeration of particles when unstabilized, which can result in particles of nonuniform size. | [11, 114-116] |
| Wet chemical methods | Synthesizes nanoparticles through a much simpler process using low temperatures. | Umbrella term for methods like coprecipitation, colloidal synthesis, and solvothermal methods. | Au, Ag, Pt, Pd, Nanocubes, nanorods, nanoplates, Hollow spheres/tubes, Au–Pd, fimPt–Ag, | High/Moderate | Relatively simple, cost-effective, morphology, composition, and size can be controlled, high surface area nanostructures are created → improved electrochemical performances, can achieve mass activity over 6 times commercial catalysts, exhibits overpotentials of approximately 10 mV at 10 | Control of morphology and size is imprecise, often uses toxic organic chemicals, health concerns, and exhibits sensitivity to moisture. | [117] |
| | | | | | mA $cm^{-2}$ in 1 M KOH, highly scalable. | | |
| Spray Pyrolysis | Particles are formed using atomization, combustion, evaporation, and nucleation procedures. | Flame conditions and burner design alter the properties of the nanoparticles. | Au, Ag, ZnO, $Fe_2O_3$, $Fe_3O_4$, $TiO_2$, Au/$TiO_2$, Ag/$TiO_2$, Au/$Fe_2O_3$, Au/$Fe_3O_4$, Ag/$(Y_{0.95}Eu_{0.05})_2O_3$, Fe@Au (core–shell), yolk-shell Au or $TiO_2$ types, Fe-Au, Ag-Au (via cospray or precursor mixing) | High | Cost-effective, does not require a vacuum, easily adaptable for large area deposition and industrial use. General nanoparticle size and morphology can be controlled, synthesis parameters can be modified to produce different nanoparticle properties. | Very dependent on initial synthesis conditions, incomplete vaporization at low temperatures may occur, require high temperatures, toxic waste may result, and non uniform catalysts may be observed. | [11, 118, 119] |

### 3.1.1. Sol-gel

The sol-gel synthesis method is distinct because it has the ability to control various properties of the produced nanoparticles due to its dependence on a wide variety of factors. This low-cost, highly effective method allows for the selection of the precursor and can adjust the form, size, and particle distribution. The optimal conditions produce tiny, homogeneous nanoparticles of uniform size with equal degrees of polymerization. The sol-gel method is split into 4 primary steps. Even a minor adjustment to the initial conditions can have a drastic effect on the finished product, which enables many different types of catalysts to be produced via the same method [11]. An overview of several potential drying processes is provided in **Table 2.**

**Table 2.** Drying processes for the sol–gel method

| Drying Processes | Brief Overview | Nanomaterial Type | Application in Biofuels | Advantages | Disadvantages | Ref. |
|---|---|---|---|---|---|---|
| Freeze Drying | Enables materials to maintain their form when used to create composite materials. Based on ice expansion. | Carbon cryogel microspheres | Esterification of oleic acid to biodiesel (FAME) | Able to reduce dehydration damage and sometimes able to maintain molecular structure when being dried in a colloidal state. | Occasionally causes microcracks.<br>Never before used with an organic solvent. | [125-128] |
| Supercritical Drying | Splits a gel substance that is not subject to capillary stress into its solid and liquid components by removing pore liquid without collapsing its structure. | Metal oxide aerogel catalyst | Methanolysis of sunflower oil to produce FAME (biodiesel) | Better at preventing cracking, irreversible shrinking, and pore collapse. | Usually, expensive and time consuming, therefore limited in practical applications. | [125, 129-131] |
| Thermal Drying | One example of this is oven drying: samples are dried using the hot air of the oven at approximately 70 degrees celsius, turning the precipitate into a dry powdery substance. | Metal oxide nanocatalyst | Transesterification of waste cooking oil to produce biodiesel (FAME) | Low complexity, cost effective, and commonplace. | Evaporating liquid from inside pores causes capillary stresses to develop, gel network often shrinks and cracks as a result. | [132-134] |
| Spray Drying | Involves spraying incredibly fine droplets onto a solid surface and subsequently turning them from a liquid to small dried particles only nanometers in diameter. Particle size is influenced by a number of factors, including | Core–shell hybrid nanocomposite microspheres, Metal–oxide-supported nanoparticle catalyst | Catalytic cracking of bio-oil vapors to produce hydrocarbons (biofuels) | May enable the design of diverse materials, dries fast, decides internal structure, and has superior tunability. | One of the most expensive techniques, highly energy demanding, and environmentally challenging. Can lead to fractured or hollow particles due to crust formation. May lead to low yields, | [135-139] |
| | solvent viscosity, rate of flow, surface tension, and atomizer type. | | | | agglomerates, and clogging. The low exhaust temperature is difficult to integrate into other process units. | |

Agglomeration occurs when nanoparticles collide and cluster during the drying process, accelerating aging through polycondensation reactions and leading to more complex agglomeration. The hydrodynamic effect and Brownian motion also contribute to agglomeration, as do condensation and capillary drag [120-122]. In silica nanoparticles, these effects can be reduced with alcohol dehydration [11]. In the context of biofuels, heterogeneous catalysts synthesized via the sol–gel technique are effective for biodiesel production on the basis of criteria related to kinematic viscosity, acidity index, copper corrosivity, and other factors [123]. Additionally, nanomaterials such as $BF_3$ and $AlCl_3$ can be incorporated through sol–gel synthesis routes into silica matrices to create a continuous biodiesel production process [105]. These catalysts immobilized through the sol–gel route inside the silica matrices are able to induce oleic acid esterification, which leads to the creation of biodiesel [105]. Continuous biodiesel production has been made possible via ultrasonic activation by catalysts immobilized via the sol–gel method; when Lewis acid catalysts are incorporated into the sol–gel matrix, they can be reused to continuously produce biodiesel [105]. Sol-gel synthesized nanocatalysts, specifically $SiO_2$ and $ZrO_2$, have been used to increase biodiesel yields [124].

Catalysts immobilized by sol-gel methods are not yet commonly used in the production of biodiesel through ultrasonic activation; however, there is a firm possibility of achieving this in the near future [105]. Adding Lewis acid catalysts into a sol-gel matrix enables continuous biodiesel production, as the catalysts can be reapplied and reused, which ultimately decreases the cost of production [105, 123].

#### 3.1.2. Co-Precipitation

This method is one of the most common methods for synthesizing nanomaterials [15]. In addition, this method is the most widely used method to prepare ZnO materials [140, 141]. Synthesis via this method is used to produce mixed rare earth solid catalysts for use in biofuel production. The catalysts made via this method exhibit enhanced catalytic activity in terms of both accelerating the rate of the initial reaction and converting FFAs (free fatty acids) and triglycerides into FAME (fatty acid methyl ester). The coprecipitation-synthesized catalysts are used in both the transesterification and esterification reactions of the biofuel production process. Additionally, catalysts synthesized through coprecipitation convey notable adaptability in producing biodiesel from waste and unrefined oils [143].

Coprecipitation produces simple or high-acidity catalysts to create complex two-component, solid salt-saturated carriers that will eventually form mesoporous basic nanocatalysts used in biofuel production. The metal oxide nanomaterial CaO–NiO is an example of a superior nanocatalyst formed through the coprecipitation method, which has more primary active sites and less area due to mixing a substantial amount of the active component with oxides. It then receives further modifications by potassium fluoride through impregnation to create more active sites, improving its performance in biofuel applications [115]. $Al_2O_3$, $TiO_2$, $Fe_3O_4$, and ZnO are examples of nanomaterials synthesized via coprecipitation, as the process is simpler and more cost effective than many others, with $Fe_3O_4$ and ZnO already being used to generate biofuels in industry. However, structural properties are lacking, and thermal treatment must be implemented to improve them. Coprecipitated $Fe_3O_4$ nanoparticles are also used as catalysts to create biodiesel from waste cooking oil [15, 142].

#### 3.1.3. Hydrothermal Method

The hydrothermal method is another synthesis method for producing nanoparticles that can be used in biofuel applications. Hydrothermal liquefaction is used during biofuel production through the thermochemically conversion of several wet biomass feedstocks into four primary products: biocrude oil, solid residue, aqueous phase, and gaseous phase. These wet biomass feedstocks include residues from forestry, the food processing industry, and agriculture. For thermochemical conversion, water (or another type of solvent, although mostly water) is used in an inert atmosphere. As a result, this method is exceptionally sustainable [144].

Liquefaction method converts most of the macromolecules contained in the biomass feedstock into biocrude oil components of differing yields, enabling it to lower the cost of evolved biofuels when combined with the biorefinery concept. It is also the most versatile synthesis method in terms of producing biofuels, as it can produce advanced biofuels such as alcohol, jet, diesel, and gasoline. The next steps that involve this synthesis method involve applications in upgrading urban residues that are low cost to reduce the quality of sustainable diesel and gasoline fuels. Hydrothermal liquefaction is one of the most flexible methods used to convert renewable wet biomass feedstocks into liquid energy-dense fuels (ex. biocrude) [144, 145]. Carbonization method is also used to synthesize solid hydrochar, which can be used directly as a biofuel, unlike biocrude, which is a precursor to biofuel. Hydrocars are favorable for use in energy generation because they have a calorific value similar to that of coal, high chemical stability, and low cost. The hydrothermal gasification method has high reaction rates, produces high-quality syngas, and does not require wet biomass to dry. However, hydrothermal methods generally consume large amounts of freshwater, which is a disadvantage of the synthesis process [145].

$SnO_2$ is an example of nanoparticles synthesized via hydrothermal methods. The hydrothermal method is favored for its scalability, low environmental impact, and versatility, with a simplicity that allows for less room for error [146]. Solid acid catalysts are also formed through this method. Hydrothermal carbonization allows carbonaceous materials to be produced at low temperatures, and the maximum esterification efficiency of oleic acid to biodiesel (FAME) is 96.4%. This method is also cost effective with reduced energy consumption and is relatively environmentally friendly [147].

#### 3.1.4. Chemical vapor deposition (CVD)

Chemical vapor deposition (CVD) is an umbrella term for processes that coat elements, metals and their alloys, and virtually any other ceramic or metallic compounds, including intermetallic compounds. CVD has been used to produce nitride, oxide, and carbide nanopowders [11]. It is one of the more reliable chemical processes specifically used to produce 2D nanomaterials and thin films [148, 149], involving depositing the vapor form of precursor materials on the substrate or exposing them to reactions. This can be done with or without catalysts in an empty chamber at high temperatures. When this technique is implemented, a solid substance is deposited from a vapor when a chemical reaction occurs near or on a substrate surface that is usually heated. The most common types of CVD include low-pressure, hot-wall, laser-assisted, atmospheric pressure, cold-wall, photoassisted, and plasma-enhanced CVD [150]. Plasma-enhanced chemical vapor decomposition (PECVD) is a lower-cost alternative to conventional CVD that does not require a catalyst and allows for scaling and affordable production of 2D materials that are high in quality. This would potentially significantly increase the number of applications in sensors, electronics, etc. [151].

CVD is a sustainable and green method that effectively transforms biomass derived from agriculture into high-value carbon materials while also creating an efficient potential synthesis pathway for carbon-based electrodes in future energy storage systems. Converting biomass into electrode materials is a sustainable route to clear energy and helps reduce the carbon footprint. Combining PECVD (plasma-enhanced chemical vapor deposition) with steam activation enables the fabrication of high-performance carbon materials, which simplifies the creation process, improves the performance of the material, and raises the question of large-scale production [152]. Additionally, biochar (a byproduct of the biomass conversion process) has been utilized as a catalyst in multiple fields, including biofuel production and CVD processes. Biochar is also produced from waste and biomass, which makes it a sustainable, green catalyst. The use of carbon nanomaterials is also an effective plastic conversion method when plastic pollution is addressed. It is also an effective catalyst for CVD, as it includes the production of carbon nanomaterials (where it excels in that area) and syngas (a biofuel) [152, 153].

### 3.1.5. Spray Pyrolysis

Spray pyrolysis is an uncomplicated, efficient, lower-cost technique intended to synthesize thin films of a higher quality than other processes [154, 155]. Spray pyrolysis is known for its ability to scale up and yield high-purity homogeneous samples. To produce metal oxide nanoparticles, the technique used is referred to as flame pyrolysis, which alters the nanoparticle depending on the design of the burner and flame conditions. Further modification can be accomplished by altering the synthesis conditions. For example, increased preparation temperature results in increased crystallite size [119]. Spray pyrolysis is also used as a combinatorial method to evaluate photocatalysts and various metal oxide materials for the production of solar energy. This allows for renewable solar energy to be created via photocatalysts produced via the spray pyrolysis method, where renewable solar energy is a broader area of biofuel research. However, spray pyrolysis is not the only synthesis method for creating photocatalysts that can be used in the production of solar energy [156].

Additionally, biomass is transformed into liquid biofuels via catalysts synthesized via spray pyrolysis and flame spray pyrolysis (FSP) methods (among other synthesis methods), which also serve to upgrade biomass vapors via CFP (catalytic fast pyrolysis). Furthermore, catalysts synthesized via FSP demonstrated higher selectivity in favor of yielding aromatics, alkenes, and compounds that were moderately deoxygenated, whereas traditional materials predominantly yield alkanes. FSP-synthesized catalysts also exhibited similar or superior total carbon product yields, demonstrating more efficient use of the expensive noble metals used in the synthesis process [157]. FSP has proven to be a readily scalable continuous process with a significantly smaller number of processing steps required than conventional pathways do, making it a valuable and promising synthesis method in this field. This result also indicated an intensification of the synthesis procedure while changing the catalyst morphology in a way that has the potential to be advantageous for tuning activity. FSP also has the potential to be much more affordable than conventional catalysts because of its scaling ability, as it results in a decrease in the amount of solvent required and more efficient gas usage [157].

This method (among other methods) aims to improve biodiesel production efficiency by creating nanocatalysts that have a large surface area and a greater number of active sites [158]. For example, spray-pyrolysis-synthesized ZnO nanoparticles enhance catalytic activity in biodiesel production for transesterification reactions [13].

#### 3.1.6. Other Processes

This subsection covers the other chemical synthesis methods outlined in Table 1 in subsection 3.1.

##### 3.1.6.1. Other Wet Chemical Methods

***Colloidal synthesis***

Compared with other synthesis methods, colloidal synthesis is relatively simple and is comparatively cost-effective [117]. It is well known for its efficiency in producing monodisperse nanoparticles (specifically metallic alloy nanomaterials) while accurately controlling their size, morphology, and composition [12, 117]. This leads to enhanced catalytic performance in biofuel production during the transesterification phase and reforming reactions in bio-oil and biodiesel production [159].

***Solvothermal Synthesis***

Solvothermal methods are straightforward and allow for a multitude of crystal lattice formations, with a high dependency on the solvent composition. Even at lower temperatures, it is effective in making uniform crystalline nanoparticles. This method is able to synthesize a wide range of nanoparticles because of the numerous synthesis conditions on which the process is dependent [117]. High activity has been found in phosphide nanoparticles of transition metals synthesized via the solvothermal method in bio-oil-upgrading hydrodeoxygenation reactions. Solvothermal environments have a relatively high pressure, which facilitates the growth of mesoporous structures, especially those with large surface areas that allow for catalytic biomass conversion [160].

***One-Pot Method (Heat-Up Procedure)***

The one-pot method can integrate multiple sequential steps performed under the same reaction conditions, increasing interest in the field of modern organic synthesis. When applied to biofuel synthesis, it reduces the complexity of the production process for this reason, as free fatty acid esterification and triglyceride transesterification can be performed simultaneously [161, 162]. Pd@meso-ZSM-5 is an example of a successful one-pot synthesized nanocatalyst that has achieved superior conversion catalytic efficiency in high-density biofuel [162]. Catalysts synthesized via this method can reach biodiesel conversion rates of 99.4% and a biodiesel yield of 71.3% [163]. The incorporation of the catalyst 1,8-diazabicyclo[5.4.0]undec-7-ene (DBU) in this method has the potential to make the biodiesel production process more environmentally friendly and even reduce the reaction time [161]. With DBU incorporated, 97.1% was the maximum total biodiesel yield for the one-pot process [161].

##### 3.1.6.2. Electrodeposition

Electrodeposition is a cost-effective, easily scalable method to produce electrocatalysts with varying compositions and morphologies. The method can produce catalysts with enhanced catalytic performance at a lower cost and includes several electrodeposition approaches, each with unique advantages and disadvantages that allow control over different nanocatalyst properties [164]. Therefore, electrodeposition is a more cost-effective method for fuel cell catalyst synthesis than many other synthesis methods [165]. In a previous study, MES (microbial electrosynthesis systems) operated with an electrodeposited carbon felt hybrid cobalt electrode had the highest yield of alcohol and acetic acid, both of which are biofuel precursors. This was followed by carbon felt/stainless steel, plain carbon felt, and plain stainless steel. Additionally, the electrodeposited hybrid biocathode, referred to as MES-4:CF/SS/Co-O, was shown to exhibit better performance than the other electrodes tested in terms of lower electronic losses, improved reduction capabilities, and superior acetic acid synthesis. This leads to greater biofuel synthesis efficiency [166].

### 3.2. Green and Biological Synthesis Approaches

Green and biological synthesis methods are relatively new in nanotechnology and use natural materials as stabilizing/reducing agents. This method is economically and environmentally preferable to chemical and physical approaches and typically involves producing nanoparticles from plant and natural polymer extracts mixed with a metal oxide precursor before heating [167].

Biological synthesis is efficient, as it results in a higher concentration of nanoparticles in the reaction mixture. The time required to synthesize the same nanoparticles varies greatly with the biological material chosen for the synthesis process. In an experiment by Plesnicute et al., the biological method for synthesizing the same type of nanoparticle resulted in a significantly higher concentration of nanoparticles being produced than when chemical synthesis methods were used. After carrying out the proposed synthesis protocol (as listed above), the researchers of the aforementioned experiment reported that allowing for resuspension and aging (for approximately 72 hours) led to even higher concentrations of nanoparticles, which could prove to be an added step to the proposed biological synthesis method in the future [167].

Biologically synthesized catalysts are often used in the process of biofuel production, specifically for the production of biodiesel. Biological catalysts have advantages in fat conversion and do not result in high reaction temperatures, soap formation, or the additional complex process of purifying glycerol, which are common chemical synthesis methods. As the name suggests, green/biological methods also have a much greater overall impact on the environment. Additionally, the use of enzymes produced through green synthesis methods in the process of biodiesel formation also prevents separation from occurring during transesterification and simplifies the overall production process [168].

Using green/biological processes lowers the required temperature of enzymatic processes and the catalyst separation cost, as well as ensuring that final products emit less greenhouse gas emissions than finished products that are chemically catalyzed. This contributes to the biological synthesis method requiring less heat to be provided in the process. Enzyme-catalyzed biodiesel also performed much better in all categories pertaining to environmental impact while also reducing production costs from easier separation, minimal side reactions, and reducing the amount of wastewater that must be treated. In addition, when lipase is used as a catalyst, low-quality oils may be used, including substrates that contain high amounts of FFAs. In terms of the next steps, research is currently being conducted to replace food-based feedstocks with waste-based feedstocks, including cooking oils, municipal sludge, and spent coffee grounds. Owing to their fast biomass generation and high oil content, algae are promising sources of biodiesel. Immobilization has also been explored because of its potential to increase catalytic activity and enable the enzyme to be reused in subsequent reaction cycles [168]. This phenomenon is associated with enzymatic synthesis approaches.

Microorganisms, including yeast, algae, marine algae, plant extracts, bacteria, seaweeds, and fungi, can synthesize nanoparticles for use in biofuel applications by behaving as biological factories, as they can reduce metal ions [35, 169]. NPs can also be used to synthesize biogas, biodiesel, and bioethanol, among other biofuels [170]. NPs are released into the reaction media after being synthesized either extracellularly or intracellularly before physical methods are used to separate them. Compounds that are already present within the media are used to prevent aggregation and stimulate their formation. These compounds have hydroxyl, terpenoid, carbonyl, amine, phenolic, and alkaloid functional groups, as well as the pigments chitosan, laureate, and starch. Algae is commonly used as a nanobiofactory for biofuel production because of its ability to reduce metal ions from the metals it collects and is especially useful where metallic nanoparticles are synthesized from dried live or dead biomass. When synthesized in this way, a simple aqueous medium can be used under conditions of normal pH and ambient temperature and pressure. Furthermore, when sugarcane leaves are converted to ethanol (a biofuel precursor) with improved conversion due to the immobilization of cellulase on $MnO_2$ nanoparticles, after 5 cycles, the enzyme retains 60% of its catalytic activity and 75% of its binding efficiency. [169].

### 3.3. Enzymatic Synthesis Approaches

Enzyme-mediated nanoparticle synthesis is relatively cost-effective, environmentally friendly, and highly scalable; it is also one of the most promising synthesis methods in the field of nanobiotechnology [171]. The immobilization of enzymes on nanomaterials is preferable, as it preserves or improves enzyme activity while improving enzyme loading, stability, and reusability [172, 173]. In this synthesis method, the enzyme itself is immobilized and used to form part of the nanomaterial structure [174]. The enzyme, which is usually a lipase or cellulase in large-scale enzymatic biofuel production, is encapsulated to form a nanoparticle before producing biofuel through transesterification reactions [174]. Biodiesel production catalyzed by lipase is subject to the high cost of the enzyme, but the use of immobilized lipases reduces this cost by increasing the reusability of the enzyme [175]. Enzymes that are bound by nanoparticles are more stable with improved biocatalytic efficiency in biofuel production and industrial applications, paving the way for their commercial use in the production of biofuels and ethanol, a biofuel precursor [174].

Nanoimmobilized enzymes are one way to lower the cost of biodiesel production, and nanomaterials are integrated with enzymes to immobilize carriers [175]. Nanomaterials are increasingly being used as enzyme carriers, and when bound to enzymes, nanomaterials display higher enzymatic activity than free enzymes because they undergo Brownian motion [175]. An experiment carried out by Wang et al. used immobilized lipase on $Fe_3O_4$ nanoparticles with a biodiesel conversion yield of over 88% for 192 hours, with a 100% yield over the first 3 cycles [176]. Immobilized cellulase has also been proven to increase yields of ethanol production, and immobilized enzymes have been shown to enhance complex lignocellulosic substrate hydrolysis when used with free (not immobilized) cellulase [174, 175].

# 4. Material characterization

## 4.1. Classification of Nanomaterial Microstructures

Nanomaterials are divided into different microstructure classifications, where even a minor variation in structure results in vastly different properties. A well-known illustration of this difference can be found in the wide variation in the hardness of various forms of carbon, such as graphite and diamond [177]. One such category consists of devices and materials that are small in size and possess structures such as nanoparticles, thin films, or nanowires that are either supported and embedded by a substrate or isolated. Catalysts and semiconductor devices that use multilayer or single quantum well structures are the most commonly known technological applications where material properties rely on these microstructures. Another category consists of devices and materials that have nanometer-scale particles where the microstructure is confined solely to a thin, nanometer-scale surface layer of the bulk material. To modify the chemical composition or atomic structure of these microstructures, CVD, PVD, laser beam, and ion implantation treatments are the most frequently used. A principal subgroup of this category consists of materials with a lateral nanometer-scale structured surface region created by 'writing' this nanometer-scale structural design onto the free surface [177]. The third category consists of bulk solids (also with nanometer-scale microstructures) whose length scale of building blocks (for example, their atomic arrangement of chemical composition) varies by only a few nanometers throughout the entire bulk. This category can be further broken down into two groups referred to as classes [177].

The first class has an atomic structure and/or chemical composition that continuously varies in space on an atomic scale throughout the bulk solid. Nanomaterials of this class are most commonly produced by quenching a structure at both equilibrium and high temperature, for example, by melting a solution of solids until the structure is far from being at equilibrium. Examples of this class include supersaturated solid solutions, gels, glasses, and implanted materials. Furthermore, the second class of nanomaterials is composed of nanometer-scale building blocks, although primary crystals. It is unimportant for the chemical composition, atomic structure, and crystallographic orientation to remain the same. Depending on the three factors stated above, crystallite building blocks result in either coherent or incoherent interface formation, and materials that are built in this way are said to be microstructurally heterogeneous. This inherent property is the cause of many of the material characteristics and distinguishes them from other microstructurally homogeneous structures. This class has only recently been synthesized and studied over the past two decades [177].

### 4.1.1. Influence of Microstructural Properties on Biofuel Performance

In an experiment by N. Upadhyay et al. (2024), fuel experienced a relatively high peak pressure and fast combustion, seemingly due to a few factors, i.e., the surface-to-volume ratio of the nanoparticles, proper fuel mixing, oxidation provided by an oxygen buffer, and an increased evaporation rate of the fuel droplets [178]. The high surface area is possibly a result of morphology and nanoscale size, which contributed to the increase in peak pressure and combustion rate. More specifically, larger surface area to volume ratios cause an increase in catalytic activity in the nanobiofuel, which leads to elevated peak pressure in the cylinder and an increased rate of net heat release (improving heat transfer rates). This directly enhances the combustion efficiency in diesel engines that use biodiesel–diesel blends, although the ratio of biofuel to diesel in these fuels remains low. In addition, nanoparticles with higher surface-to-volume ratios enable better fuel atomization and fast evaporation, which increases the thermal efficiency of the brake system [178]. Furthermore, incorporating nanoparticles into fuel blends notably lowers hydrocarbon emissions, and the nanoparticles themselves behave as oxidation catalysts. This accelerates flame propagation inside the cylinder, increases the surface-to-volume ratio, lowers the temperature required for carbon activation, and fosters rapid burning [178].

In another experiment, when nanomaterials are constructed, the metal oxides undergo electrochemical, surface, optical, and structural reconstructions that affect the parameters of their cell, symmetry of their lattice, configurations of their surface, and extreme band gap enhancements [179]. Additionally, tempering with respect to the surface area of a catalyst, porosity, configuration of the surface, particle size, and other factors directly affects the conversion efficiency. Other characteristics that are affected include the rate of reaction and consumption of energy during the reaction. Furthermore, in regard to metal oxide catalysts, the surface area available for reactions can be increased as a result of decreasing the particle size. This is accomplished by ensuring a reduced activation energy and higher rates of conversion of reactants to products. Lower activation energies allow for the use of fewer catalysts, shorter reaction times, and reactions under moderate pressure and temperature conditions. Nanoscaling metal oxides has also been proven to provide more space for the reagents and reactants to react while increasing the number of active sites of the material [179].

### 4.2. Characterization Techniques

Imaging and spectroscopic techniques, including scanning electron microscopy (SEM), transmission electron microscopy (TEM), Fourier transform infrared spectroscopy (FTIR), and Brunauer–Emmett–Teller (BET) methods, are employed to analyze the microstructures of the nanomaterials in greater detail. These methods help reveal specific properties that influence the behavior and performance of nanomaterials in various applications, including biofuel processes. For example, nanomaterials often immobilize enzymes to increase their stability, increase their catalytic activity, and allow for reusability. Understanding the interactions between nanocarriers and enzymes at the structural level is necessary to optimize nanomaterial applications in biofuel systems. As a result, FTIR, SEM, and TEM techniques are used to analyze the biophysical and conformational changes in enzymes postimmobilization. These methods reveal how immobilization affects enzyme stability and activity by identifying the structural integrity and intermolecular interactions between the enzyme and the nanomaterial [180, 181].

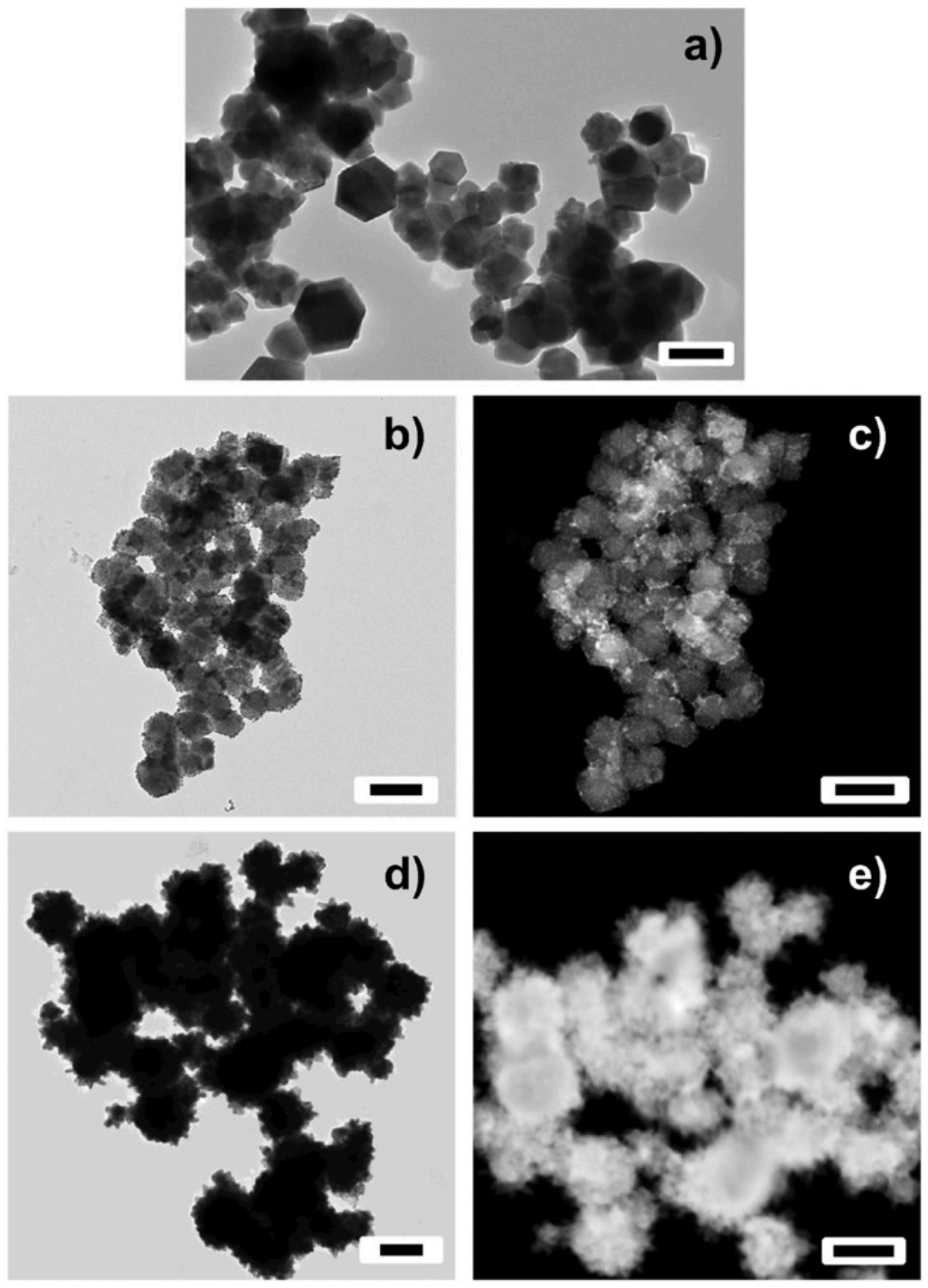

**(i)**

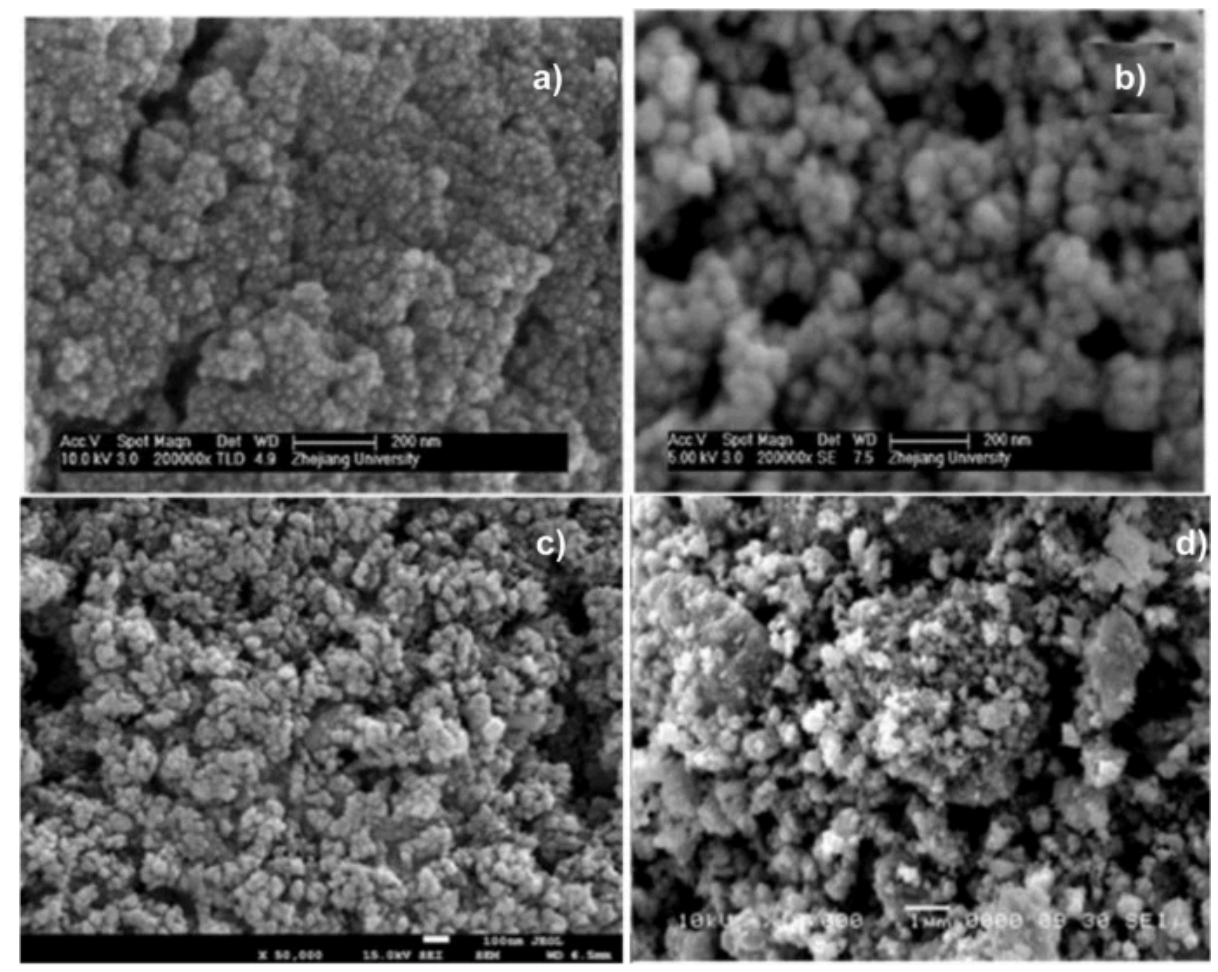

**(ii)**

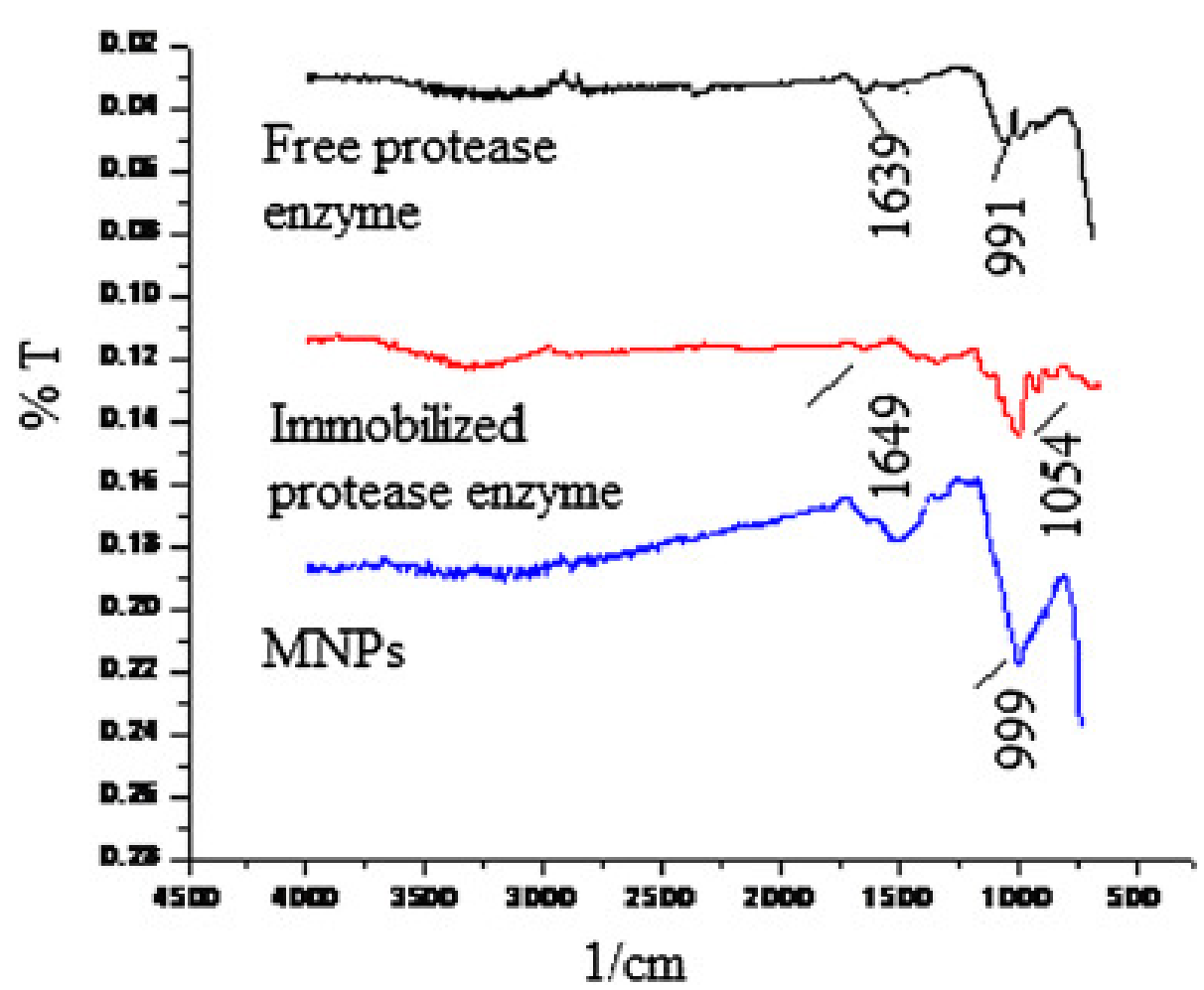

**(iii)**

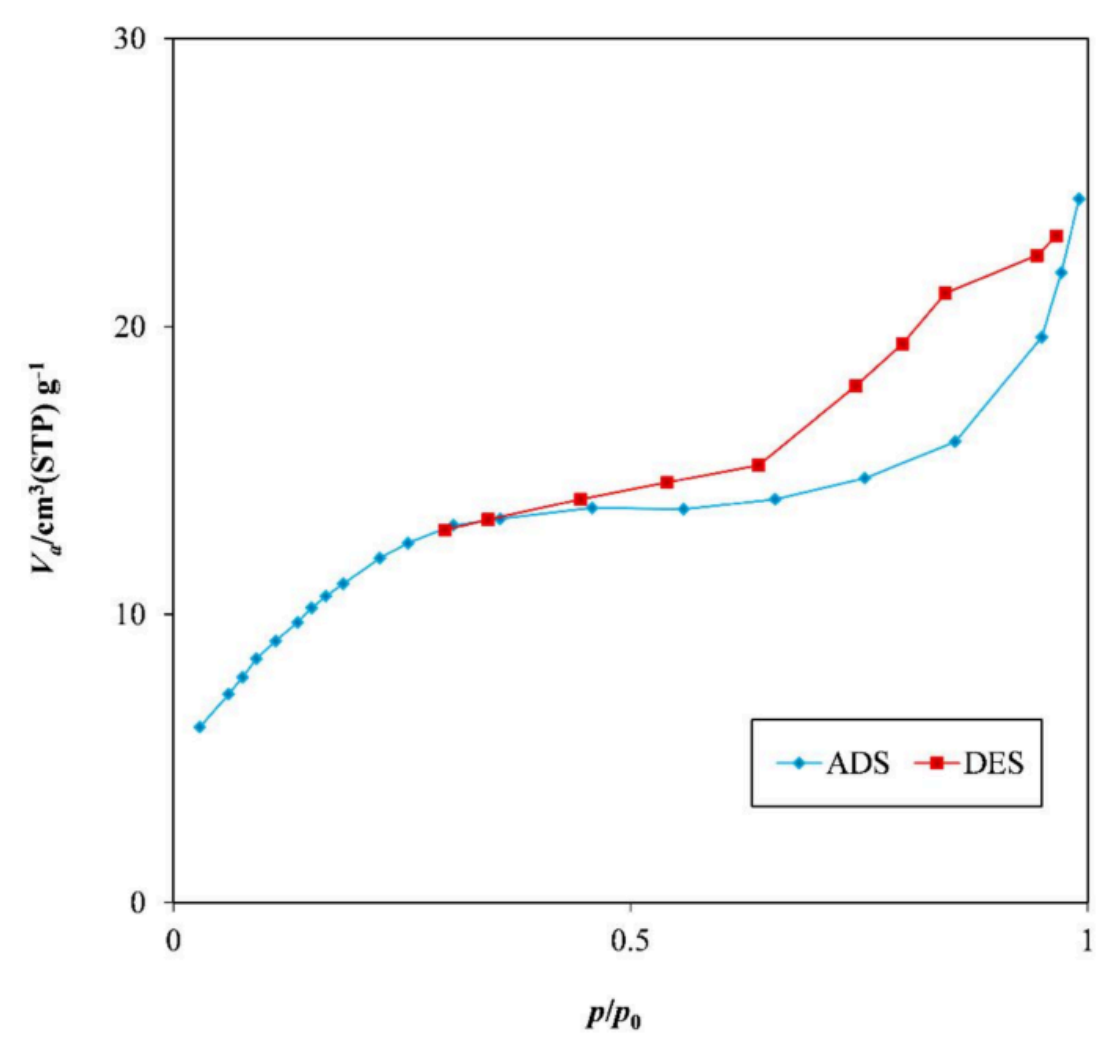

**(iv)**

**Figure 5. (i)** TEM imaging of the nanoparticles under various conditions: (a) PEI-coated $Fe_3O_4$ nanoparticles, 100 nm; (b) Au seed-decorated $Fe_3O_4$–PEI nanoparticles, bright field, 200 nm; (c) Au seed-decorated $Fe_3O_4$–PEI nanoparticles, dark field, 200 nm; (d) bright field, 100 nm; (e)

Au-coated $Fe_3O_4$ nanoparticles, dark field, 100 nm. Adapted from [186]; **(ii)** SEM images of spherical nanoparticles with sizes of (a) approximately 12 nm at 200°C, (b) approximately 53 nm at 250°C, (c) less than 100 nm in size, and (d) 24--65 nm in size. Adapted from [191]; **(iii)** FTIR Analysis of Free and Immobilized Protease Enzymes on MNPs and MNPs. Adapted from [193]; **(iv).** BET isotherms of the $\alpha Fe_2O_3/CuO$ nanocomposite. Adapted from [206].

### 4.2.1. Transmission electron microscopy (TEM)

TEM is widely used as an analysis tool in biofuel research, allowing characterization at the nanoscale. Its main usage is in identifying the size of the nanoparticles used in the production of biodiesel, along with SEM [12]. TEM can also be used to sample and analyze soot particles from biodiesel and allows for easy differentiation of the properties of various biodiesel types, and is able to compare the properties of soot particles between conventional diesel fuel and biodiesel [182]. TEM passes electrons through an ultrathin sample to provide two-dimensional imaging of internal structures, such as shape, crystallinity, heterogeneity, morphology, defects, and dispersion, with higher resolution and magnification than SEM [183]. For example, researchers have reported that while magnetite nanoparticles have inert surfaces, applying a silica coating increases their reactivity by creating more active sites and allows functionalization with specific groups that can enhance certain characteristics [184].

TEM can be used to verify the qualities of materials during nanomaterial synthesis. This study aimed to develop a green method for synthesizing cadmium sulfide (CdS) nanoparticles using extracts from the green alga *Chlamydomonas reinhardtii.* TEM imaging revealed various nanoparticle features, such as shape and size, and revealed well-defined spherical nanoparticles ranging from 2 to 7 nm in size. The results also revealed lattice fringes, which confirmed that the nanoparticles were crystalline [187].

Moreover, a study reported the fabrication of $Fe_3O_4$@Au core–shell nanoparticles, approximately 100 nm in size, through a multistep assembly method aimed at applications involving catalysis, which is relevant to biofuels. As illustrated in **Figure 5 (i)**, TEM imaging allowed researchers to visualize the nanostructure's core–shell morphology, which confirmed the formation of the nanoparticles as described previously [186]. Similarly, a study focusing on the production of gold nanoparticles via the green microalga *Chlorella vulgaris* used TEM to verify the presence, location, and size of the nanoparticles within the algal cells. The nanoparticles were located in the cytoplasm and measured 40–60 nm in diameter [188].

### 4.2.2. Scanning electron microscopy (SEM)

Scanning electron microscopy is an imaging technique used to analyze the surface features of dry materials by bombarding samples with a high-energy electron beam and detecting scattered electrons to obtain three-dimensional images that reveal the surface features of the particles [189]. In the context of nanomaterials, SEM can analyze various characteristics, such as size, the thickness of films or coatings, chemical composition, the form and distribution of nanoparticles within composites, elemental makeup, surface texture and structure. This information is essential for comparing samples, evaluating their properties, and guiding the modification or development of nanomaterials for specific applications [183]. **Figure 5 (b)** shows SEM images of various-sized nanoparticles [191].

SEM provides information for catalyst research at different scales, ranging from the nanometer scale to the millimeter scale, making it more useful than TEM for attaining averages and learning the inhomogeneity of a sample. In addition, SEM images are more intuitive and easily interpreted, and the samples can be in the form of an extrudate, a powder, or a monolith [17]. Like TEM, SEM is also widely used as an analysis tool in biofuel research [12]. It is utilized mainly for identifying the size of the nanoparticles used in the production of biodiesel. SEM can also be used to analyze the absorption characteristics of char, a solid residue produced during biomass gasification, which is used to produce syngas [192]. For example, a study aimed to immobilize the protease enzyme from *Bacillus subtilis* on magnetic nanoparticles for the purpose of synthesizing glycinamides while also demonstrating the enzyme's stability and reusability. SEM and FTIR techniques were employed to verify these qualities. SEM analysis of the surface morphology of the magnetic nanoparticles before and after protease immobilization revealed that they maintained their size and shape throughout the process, which confirmed that the enzymes attached solely to the surface without causing particle clumping or size changes [193].

For example, researchers have studied the synthesis, characterization, stability, and engine performance of biodiesel and diesel fuel blends, with a focus on the use of magnetite nanoparticles as additives. SEM was used to confirm the shape consistency and particle growth trends. Various elements related to nanomaterial form and distribution have been identified, and these nanoparticles, which are generally spherical with a size of 100 nm, increased in size with increasing laser energy during synthesis [194].

Furthermore, a study aimed to improve pyrolysis-based biofuel production from biomass by increasing fuel yield and quality while simultaneously producing silver nanomaterials from the waste char of the biomass via a process called biotemplating. SEM was used to confirm the validity of nanostructuring via the biotemplating process by visualizing the 20–200 nm silver nanoparticles, revealing the formation of a "mesh" of semispherical nanoparticles from the samples. This finding also indicated that nanoparticle properties could vary with feedstock; for example, nanoparticles derived from corn stover feed are typically larger and more distributed in size [195].

### 4.2.3. Fourier Transform Infrared Spectroscopy (FTIR)

Fourier transform infrared (FTIR) spectroscopy is a technique that measures the vibrational properties of molecules, such as amino acids and cofactors, to detect subtle structural changes. It works by passing infrared light through a sample to determine the quantity of light that is absorbed at various wavelengths. It is a fast and cost-effective analysis method and is compatible with gases, liquids, solids, powders, and thin film samples. Each chemical bond vibrates at a specific frequency, so FTIR can analyze properties such as functional group attachments and chemical bonding [181, 196].

FTIR is particularly useful in analyzing biofuels. It can accurately detect fuel properties and impurities and can measure biodiesel concentrations in diesel. Additionally, FTIR distinguishes biofuel origins from different feedstocks, helping assess fuel quality [197]. For example, one study aimed to understand how the use of several lignocellulosic biomasses (lignin, cellulose, etc.) affects the chemical structure and mechanical strength of a solid biofuel called a biocoke. FTIR was used to analyze the chemical structure of the biocoke, identifying functional group changes that were linked to the presence of these biomass materials. For example, the 1800–1500 $cm^{-1}$ region of the FTIR spectrum reflected structural changes caused by lignin breakdown, with stronger spectral signals appearing as the lignin content increased, which was associated with greater strength. In contrast, the 800–400 $cm^{-1}$ area presented spectral changes linked to the cellulose content, where increased cellulose was associated with weaker biocoke [198].

FTIR offers several advantages, as follows [199]:

- Possesses high reproducibility
- This method requires significantly lower sample quantities (usually a few microliter samples).
- Intricate sample preparation procedures are not necessary
- Greater preservation of sample integrity

Spectral accuracy faces two main challenges, i.e., water molecule-induced interference and complex data analysis. Interference caused by water molecules results from the samples all containing water molecules within them, which have different vibrational modes than the sample itself. Because signals from the water overshadow the unit's other signals, infrared radiation is strongly absorbed and interferes. Minimizing this effect often involves reducing the water content, utilizing desiccation methods, and enhancing the accuracy and clarity of characterizing the other parts of the sample. However, the complexity of data analysis is a declining issue that is on track to become obsolete, as the track record of technological developments in this field strongly suggests that data analysis will become increasingly less complex [199].

FTIR can also be employed to confirm and track the progress of chemical reactions. It is used primarily to analyze the chemical composition and structure of a multitude of materials, including biological substances [200]. In transesterification reactions, the portion of biodiesel in the reaction can be detected via FTIR via infrared mid-range analysis and absorption bands. This enabled the FTIR method to distinguish the biodiesel from the sample of soybean oil used in an experiment by Rosset et al., using differences found in the spectra of both samples [200]. In this case, FTIR was used to monitor and evaluate the transesterification of soybean oil into biodiesel. By detecting specific absorption bands corresponding to specific chemical bonds, FTIR allows researchers to track the transformation of raw oil into methyl esters. Key spectral regions, such as 1425–1447 $cm^{-1}$ and 1188–1200 $cm^{-1}$, appear in biodiesel but not unreacted oil, whereas other regions, such as 1370–1400 $cm^{-1}$, are present in oil but not biodiesel [200]. FTIR is also a reliable and relatively fast technique for analyzing the content of fatty acid methyl esters in biodiesel [201]. The ability of FTIR to detect functional groups allows it to distinguish between refined oil and biodiesel [200, 201].

Furthermore, FTIR is employed to study how enzymes interact with nanomaterials when immobilized for biofuel production. This technique helps characterize the structural and functional changes in enzymes when bound to nanoparticles, enabling researchers to draw conclusions regarding their stability and performance [202]. For example, building on the glycinamide synthesis study in section 3.3.1, FTIR confirmed the successful binding of protease to magnetic nanoparticles (MNPs) by identifying key chemical interactions. As shown in **Figure 5 (c)**, in the spectra of the immobilized enzyme, strong peaks at 1054 $cm^{-1}$ and 1649 $cm^{-1}$ indicate the presence of Fe-O-Si bonds, whereas a band at 1639 $cm^{-1}$ confirms the presence of an amide bond [193].

Additionally, a study employed FTIR when synthesizing a core–shell nanocatalyst, where palladium nanoparticles were deposited on poly-methyldopa (PMDP)-coated $Fe_3O_4$ magnetic nanoparticles. FTIR analysis revealed the surface composition and interactions of the nanocomposite. This confirms the successful coating of $Fe_3O_4$ nanoparticles with the PMDP polymer by identifying functional groups such as hydroxyl, amine, and carboxyl vibrations. By observing that the number of carboxyl and amine groups shifted to lower wavenumbers, FTIR confirmed the presence of strong bonds and the effective immobilization of Pd on the magnetic nanoparticle surface [203].

#### 4.2.4. Brunauer–Emmett–Teller (BET)

The Brunauer–Emmett–Teller (BET) method is a widely used calculation-based technique for measuring the specific surface area (SSA) of porous materials. It fits gas adsorption isotherms, typically nitrogen, to a semiempirical model that estimates how much gas forms a monolayer on the material's surface [204, 205]. BET can be applied to various nanomaterial and nanomaterial-enhanced biofuel applications by providing insight into SSA, which can help researchers draw conclusions regarding properties, such as catalytic efficiency, adsorption, and porosity [206]. For example, part of a study that characterized graphene nanosheets employed BET analysis to determine that the material was mesoporous (the diameter of the material ranged from 2–50 nm [207]), with an average pore diameter of 6.28 nm. The SSA was also determined by calculating the amount of nitrogen gas absorbed onto its surface at the liquid nitrogen temperature (-196°C) [208]. In the context of graphene materials, a study investigated the relationships among physical properties, such as the SSA, to understand how they influence the electrical conductivity of graphene. BET analysis was employed to measure the SSA of various graphene materials, which was then compared to their electrical conductivity. Materials with higher SSAs generally demonstrate better electrical conductivity, which is logical, as this implies a greater number of reaction sites [209].

BET is effective in optimizing biofuel efficiency given that surface area is directly related to catalytic efficiency. A study employed BET to characterize a nanocatalyst's surface properties when developing calcium oxide nanocatalysts for biodiesel production from soybean. BET analysis confirmed the mesoporous nature and high surface ability of the material by revealing an SSA of 67.781 $m^2/g$ and an average pore diameter of 3.302 nm. These features are critical to catalytic activity, providing greater reactant accessibility. As a result, the biodiesel reaction obtained a high conversion of 97.61% under optimal conditions [210].

BET (along with FTIR and SEM) is used to analyze the absorption characteristics of char, which is a solid residue produced during biomass gasification and is used to produce syngas. Char actually has properties similar to those of coal, and the BET surface area is used to determine the sorption performance [192]. A study employed a wide range of characterization techniques, including FTIR, SEM, TEM, and BET analysis, when developing a copper oxide (CuO) on an alpha iron oxide ($\alpha Fe_2O_3$) nanocatalyst to catalyze the transesterification biodiesel production process from waste cooking oil.

As detailed in **Figure 5 (d)**, BET analysis was critical in quantifying the necessary surface properties for catalytic performance. The adsorption curve (ADS) shows how much nitrogen the sample absorbs, whereas the desorption curve (DES) shows how much gas is released as the pressure is reduced. Analyzing this curve allows the BET method to calculate the SSA. The integration of CuO nanoparticles notably increased the SSA of the catalyst, resulting in a surface area of 74.45 $m^2/g$ for the nanocatalyst and reaching a total surface area of 334.45 $m^2/g$. For the nanocatalyst, this indicates a porous structure, large adsorption ability, and a high number of accessible active sites for the transesterification reaction. SEM was used to analyze the morphology and size of the catalysts, whereas TEM was used to assess the size and shape of the individual nanocatalyst particles [206].

## 5. Nanomaterials in Biofuel Production

### 5.1. Enhancement of Biomass Pretreatment and Hydrolysis

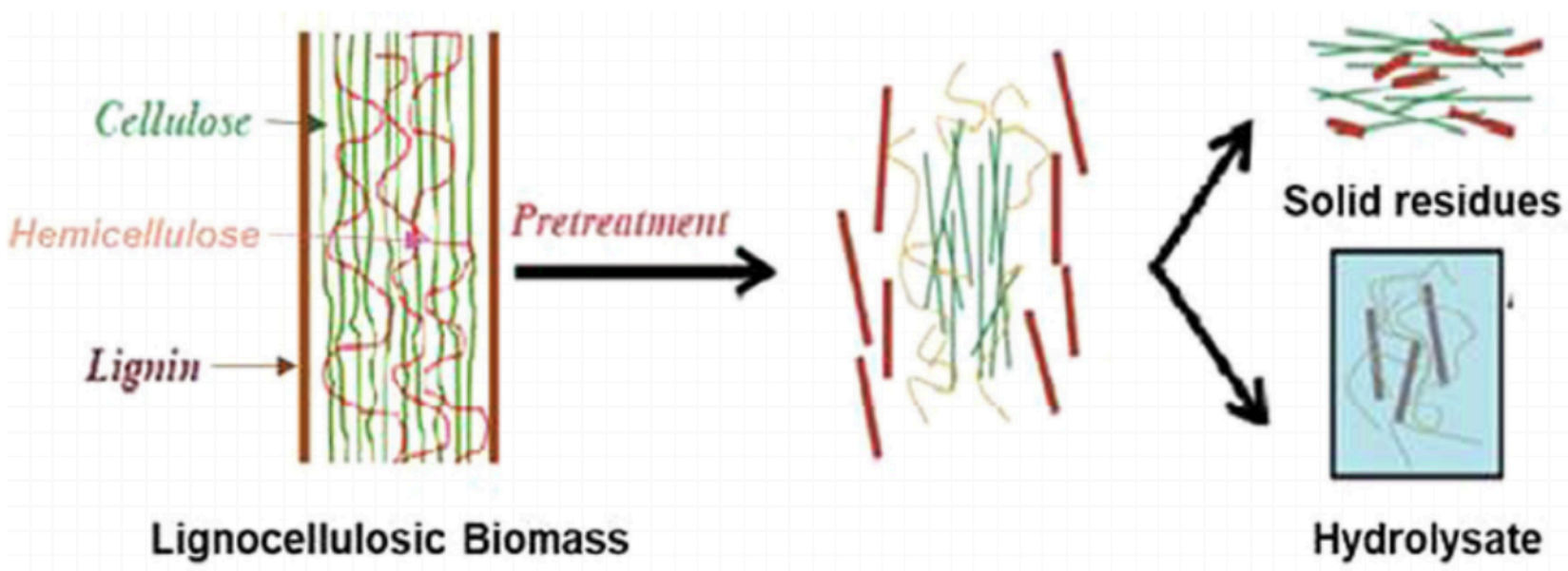

(a)

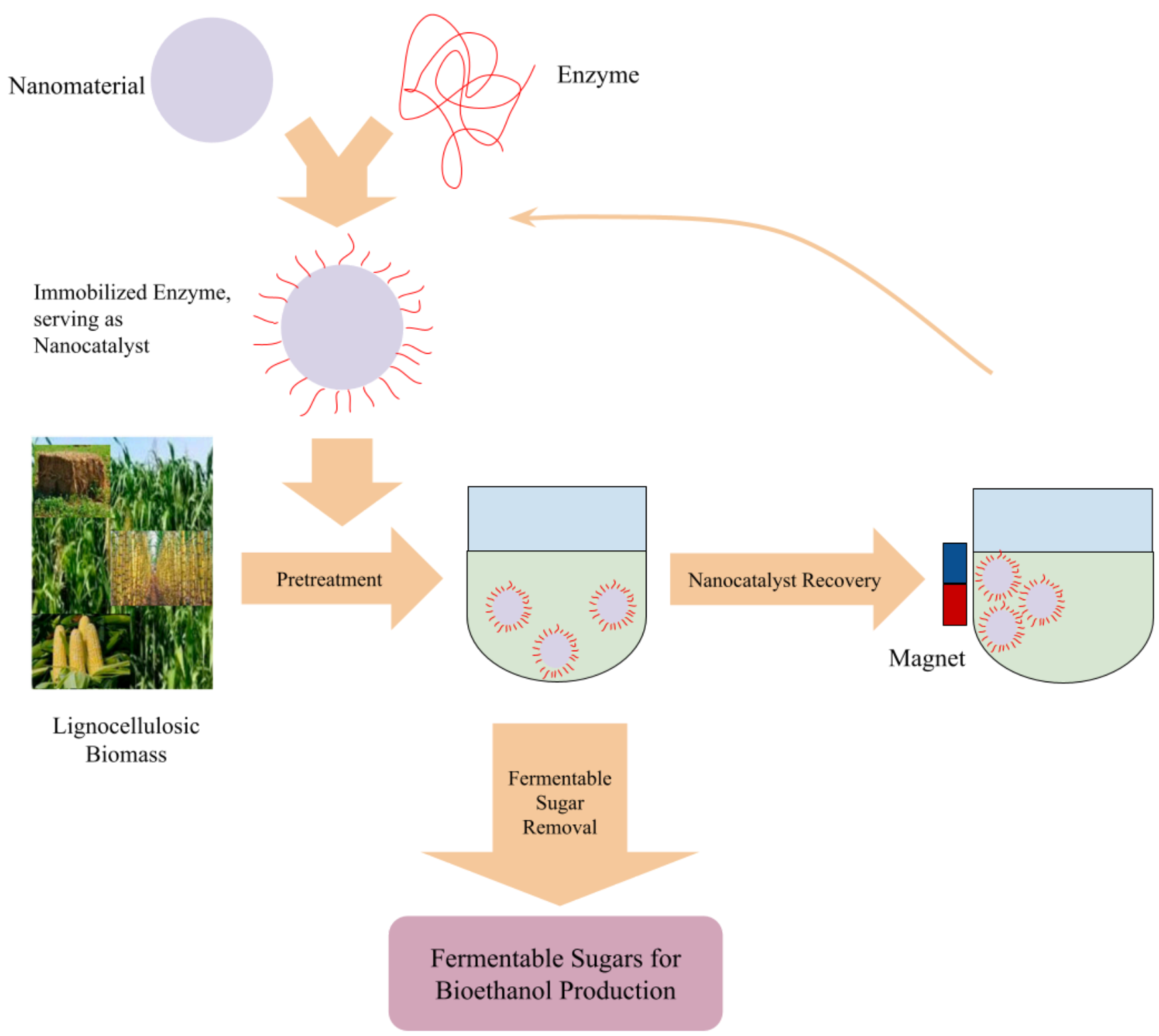

(b)

**Figure 6.** (a) Lignocellulosic Biomass Pretreatment results in Simpler Cellulose, Hemicellulose, and Lignin Units. Adapted from [212]. (b) Enzyme-Nanomaterial Immobilization for Hydrolysis of Lignocellulosic Biomass. Lignocellulosic Biomass Adapted from [220]. Information adapted from [221].

Biomass, a renewable energy source derived from organic material, is produced either directly (from plants) or indirectly (from living organisms that consume glucose) through the process of photosynthesis [213]. It serves as an alternative to traditional fossil fuels, enabling the generation of bioenergy in the form of bioethanol, biodiesel, and biogas through various processing techniques. The initial stage of these processes is pretreatment, which typically involves breaking down the biomass into smaller components, making it more accessible to additional enzymes for conversion into biofuels [2]. **Figure 6 (a)** shows an example in which lignocellulosic biomass is pretreated to fragment into simpler base units. Similarly, to increase biodiesel yield, oils with high free fatty acid (FFA) contents must be pretreated under acidic conditions to convert FFAs into fatty acid esters that have a lower, acceptable pH for oils before proceeding to the biodiesel production process [214]. Enzymatic hydrolysis is another process in which enzymes degrade polysaccharides, with the addition of water, into simpler monosaccharides that can be fermented by microorganisms and enzymes [215].

#### 5.1.1. Role of Nanoparticles in Improving Biomass Pretreatment and Hydrolysis

Nanomaterials play a valuable role in both the pretreatment and enzymatic hydrolysis processes of biomass conversion [216]. They serve to bind with and immobilize enzymes involved in these processes [217], enhancing their stability and performance [218]. Owing to their catalytic properties and large surface area, nanoparticles can improve the speed and efficiency of biomass breakdown, which can accommodate greater quantities of enzymes and acids [219]. For example, the distinct structure of lignocellulosic biomass is composed of various polymers, such as linear cellulose chains, branched hemicellulose, and lignin. The biomass must undergo pretreatment to break down its structure and improve the accessibility of cellulose and hemicellulose to enzymes that convert them into fermentable sugars. Fermentable sugars can then undergo fermentation to become liquid biodiesel. **Figure 6 (b)** details these steps in a situation in which nanocatalysts have magnetic properties, allowing them to be recovered via magnets and reused. Pretreatment methods aim to disrupt or remove lignin and loosen cellulose and hemicellulose [219-222]. NPs are increasingly being applied to enhance these processes by increasing the efficiency of key enzymes, such as hemicellulase, cellulase, and ligninolytic enzymes. Owing to their small size, nanoparticles can increase reactions and help penetrate the lignocellulosic cell wall, facilitating the release of a greater yield of fermentable sugars [216].

Similarly, in terms of improved yield with nanomaterials, the use of acid-functionalized magnetic nanoparticles as catalysts in the hydrolysis of a disaccharide called cellobiose achieved 78% conversion to simpler sugars within one hour at 175°C, significantly surpassing the control experimental rate of 52% without nanoparticles [223]. As described in section 2.1.1, magnetic nanomaterials are particularly useful because they can be recovered with magnetic fields for various pretreatment cycles.

### 5.2. Enhancement of Biodiesel Production (Transesterification)

Biodiesel is produced from vegetable oils or animal fats that undergo a transesterification reaction to reduce the oil velocity to a state comparable to that of typical diesel [227]. The multistep reaction first breaks triglycerides into fatty acids, producing glycerol as a byproduct.

The fatty acids subsequently react with methanol to yield fatty methyl esters, which constitute biodiesel [228].

Homogeneous catalysts that dissolve in the reaction medium [229] are typically employed because of their speed of reaction [230] but cannot be reused and often generate soap as a byproduct. This further complicates the process since large amounts of hot water, associated with high costs, must be employed to neutralize or remove the catalyst and soap from the mixture after the reaction [231]. Homogeneous nanocatalysts are also not resistant to fatty acids, water, or other impurities commonly found in waste and low-quality feedstocks, which limits their effectiveness [232].

### 5.2.1 Role of Nanomaterials in Improving Transesterification

The use of nanoparticles as heterogeneous catalysts, which are typically recoverable and reusable since they do not dissolve in the medium, enhances the sustainability and efficiency of biodiesel production. The large surface area of these materials provides more reactive sites, increasing their catalytic activity and increasing their conversion rates. Additionally, their small size allows easy separation and recovery from the reaction, enabling reuse across multiple cycles and thereby reducing costs and material requirements [233, 234]. Nanocatalysts improve biodiesel production through four primary stages [235]:

- Adsorption, where reactants attach to the nanocatalyst's large surface area;
- Activation, where reactant molecules are activated at specific catalytic sites;
- Reactions, where triglycerides are broken down; and
- Desorption, where newly formed products are separated from the nanocatalyst.

The biodiesel production rate is influenced by the concentration of the nanocatalyst, as an excessive amount can obstruct interactions between reactants and active sites or cause other transfer issues. Furthermore, the type of nanomaterial significantly impacts production efficiency and speed since different nanoparticles possess distinct properties (as discussed in section 2) [236]. Nanocatalysts can be derived from various nanoparticles, including carbon-based materials such as graphene oxide and carbon nanotubes, metal oxide nanocatalysts, and magnetic catalysts [234]. Additionally, nanobiocatalysts also permit the immobilization of enzymes, such as lipases, to improve both stability and reusability [230]. For example, magnetic $Fe_3O_4$ nanoparticles were encapsulated within graphene oxide sheets and used as carriers to immobilize *Candida rugosa* lipase for biodiesel production. The immobilization efficiency was 85.5%, a biodiesel yield of 92.8% was achieved at 40°C, and the immobilized enzyme sustained its activity over five cycles of reuse [237].

The catalysts must maintain an initial catalytic activity of a minimum of 90% over 10 reaction cycles to be regarded as practical for industrial usage, which must be considered when employing catalysts in biofuels. Heterogeneous catalysts (including nanocatalysts) are reusable and can be specifically adjusted and easily recovered, making them usable for multiple catalytic reaction cycles and thereby decreasing the cost and labor of biodiesel [232, 238]. Solid transesterification nanocatalysts (a form of heterogeneous catalyst) are advantageous over traditional nanocatalysts in terms of thermal and chemical stability and serve as alternatives for future biodiesel production routes. Their solid-state enables heterogeneous catalysts to easily separate and recover from reaction mixtures, ultimately enabling reuse and minimizing chemical waste. Unlike homogeneous catalysts, they are also more resistant to waste and low-grade feedstocks and can process them without compromising the biofuel yield. For example, a heterogeneous solid-base $C_4H_4O_6HK$-loaded ZrO nanocatalyst was employed for the transesterification process to produce biodiesel from soybean and methanol. Employing the nanocatalyst resulted in an efficiency of 98.03% under optimal conditions and allowed for longer use by maintaining the activity for five cycles [239]. To extend the lifespan of these catalysts, protective layers and regeneration techniques can be applied. Nevertheless, the reusability of a catalyst significantly decreases over time as deactivation mechanisms are implemented, including thermal sintering (high temperatures) and surface fouling (which blocks the catalyst's surface) [232].

Numerous studies have demonstrated the efficiency of nanocatalysts in achieving high biodiesel yields. For example, zinc-doped calcium oxide nanomaterials significantly enhance biodiesel production, achieving yields of up to 96.5% from cost-effective feedstocks such as waste cooking oil, thereby improving both efficiency and economic viability [240]. Additional examples are presented in **Table 3**, which provides an exhaustive comparison of the examples of nanocatalysts that have been used in enhancing the biodiesel yields from various biomasses and nanocatalysts. This further evaluates their yield, sources, effectiveness, pollution tendency, reaction time, temperature during the reaction process, etc.

**Table 3.** Biodiesel yields from various biomasses and nanocatalysts. Structure adapted from [230].

| Nanocatalyst | Fat/Oil Source | Highest Yield (%) | Cost Effectiveness | Pollution Tendency | Additional details | | Ref |
|---|---|---|---|---|---|---|---|
| | | | | | Reaction Time (minutes) | Temperature (°C) during reaction process | |
| PCDFe nanocubes, with carboxyl and amino-functionalized carbon dots (CDs) and Fe nanoparticles | Waste cooking oil | 99.79 | High | Low | 90 | 70 | [241] |
| $Na_2Ti_6O_{13}$ (1.5% wt) | Pure and cooked oils | 95.9 | Moderate | Low | 120 | 80 | [242] |
| Zn-doped CaO (5% wt) | Waste cooking oil | 96.5 | High | Low | 120 | 57.5 | [240] |
| $MgO/MgFe_2O_4$ (4% wt) | Sunflower oil | 91.2 | High | Low | 240 | 110 | [243] |
| CaO (1.4% wt) | Jatropha oil | 96.73 | Moderate | Low | 60 | 55 | [244] |
| $TiO_2$–ZnO | Palm Oil | 92 | Moderate | Low | 300 | 60 | [245] |
| MGO@$TiO_2$Ag (4.15% wt) | Waste cooking oil | 96.54 | High | Low | 2.82 | 65 | [246] |
| CaTiO3 (1.8% wt) | Ochrocarpus longifolius leaves extracted as a novel fuel | 97.7 | High | Low | 45 | 65 | [247] |
| ZrO loaded with $C_4H_4O_6HK$ (6.0% wt) | Soybean oil with methanol | 98.03 | High | Low | 120 | 60 | [239] |

### 5.3. Enhancing bioethanol production

Bioethanol can either substitute for or be blended with conventional gasoline, serving as a more sustainable option. [248, 249]. Bioethanol is divided into four generations (Gs), each distinguished by slightly different production methods and biomass sources [250, 251]:

- First-generation (1G) bioethanol produced from food crops
- Second-generation (2G) bioethanol derived from lignocellulosic biomass
- Third-generation (3G) bioethanol produced from algae-based sources
- Fourth generation (4G) bioethanol derived from genetically engineered algae 1G has traditionally been the primary source for biofuel; however, the use of 2G has recently increased because of its greater sustainability, as it does not rely on edible food sources [252].

Currently, 3G and 4G have only been employed in laboratory settings, as they remain underdeveloped [250]. **Figure 7** provides a general summary of the steps involved in 1G, 2G, 3G, and 4G bioethanol production.

1G bioethanol production involves four main steps: hydrolysis to break down starch or sugar into fermentable sugars, fermentation by microorganisms to produce ethanol, distillation to separate ethanol, and dehydration to reach fuel-grade purity [252-254]. Second-generation (2G) bioethanol includes additional pretreatment steps to deconstruct the complex structure of lignocellulosic biomass [256]. It can proceed via biochemical conversion and thermochemical conversion pathways. Biochemical conversion involves pretreatment, detoxification of biomass from unwanted byproducts, and then 1G steps. Thermochemical conversion involves gasifying biomass to produce a carbon monoxide, hydrogen, and carbon dioxide mixture (syngas). This syngas is either cooled and fermented by microbes or passed over metal catalysts at high temperatures to transform into ethanol and other alcohols [255]. 3G utilizes either microalgae or macroalgae as biomass. The algae are dried to prevent gel formation, crushed to increase the surface area, hydrolyzed into fermentable sugars by acids or enzymes, fermented into bioethanol and carbon dioxide by microorganisms, and then purified for fuel use. Owing to their high water content, algae produce less ethanol per gram than the 1G and 2G feedstocks do. The widespread adoption of 3G is constrained by unresolved economic, technological, and policy-related barriers requiring further research [249].

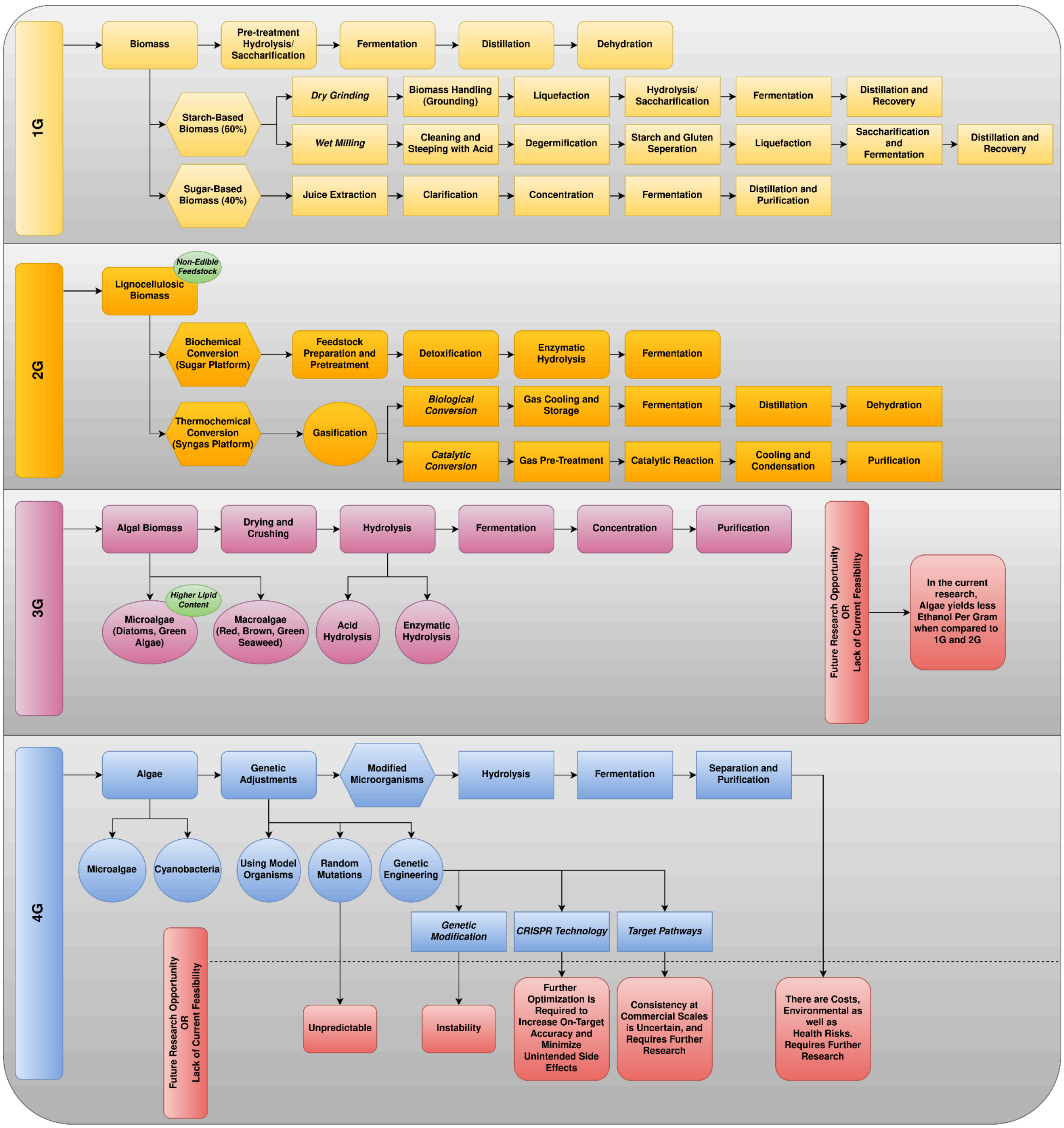

**Figure 7.** 1G, 2G, 3G, 4G, bioethanol production steps with consideration of feasibility.

4G bioethanol production relies on the use of genetically modified microorganisms, mainly algae, as feedstock. Traditional genetic engineering manipulates enzymes and metabolic pathways and inserts genes into genomes to improve quality and yield. However, the metabolic stability of organisms remains a challenge. Furthermore, targeting pathways redirects cellular activity to increase biofuel-relevant traits, and random mutagenesis (through exposure to UV light or chemicals) can induce genetic variation but these methods are currently inconsistent at a commercial scale. CRISPR genome editing allows for cost-effective and simultaneous gene targeting via synthetic guide RNA or gene repression for particular purposes, such as increasing lipid synthesis or reducing lignin [257, 258]. However, further refinement is required to increase target accuracy and minimize unintended side effects. 4G approaches currently require advanced tools, and further research on risk mitigation, stability, and effectiveness is needed. Health and environmental concerns regarding genetically modified algae must also be addressed before their industrial use [249, 259].

### 5.3.1. Role of Nanomaterials in Improving Bioethanol Production

Nanomaterials, particularly nanocomposites, have the potential to increase the efficiency and sustainability of key ethanol production stages, including hydrolysis and saccharification, fermentation, and distillation. Various types can be employed on the basis of their unique properties, such as metals and carbon-based and hybrid composite nanoparticles [254]. Owing to their high surface area and catalytic efficiency, nanomaterials can enhance the conversion of biomass to fermentable sugars by immobilizing enzymes and improving their stability [254].

One study immobilized cellulase onto functionalized multiwalled carbon nanotubes (MWCNTs), as shown in **Figure 8 (a),** to improve stability and reuse. The system achieved an optimal enzyme loading of 97% onto the nanoparticle, and the MWCNT-cellulase composite retained 52% of its activity after six uses, demonstrating that nanoparticle-enzyme performance and yield can be sustained across multiple uses [262]. Metal oxide nanoparticles, such as $Fe_3O_4$ and $Fe_2O_3$, are also commonly used as catalysts and supports to immobilize enzymes, increasing yield and reaction time [263]. For example, researchers have developed magnetic nanoparticles using a hydrophobic monomer (MAPA), which exhibit strong stability and a high capacity for loading an α-amylase enzyme due to their large surface area, in turn facilitating a greater number of reactions. The enzyme–nanoparticle composite also maintained 83% of its activity after eight uses [264].

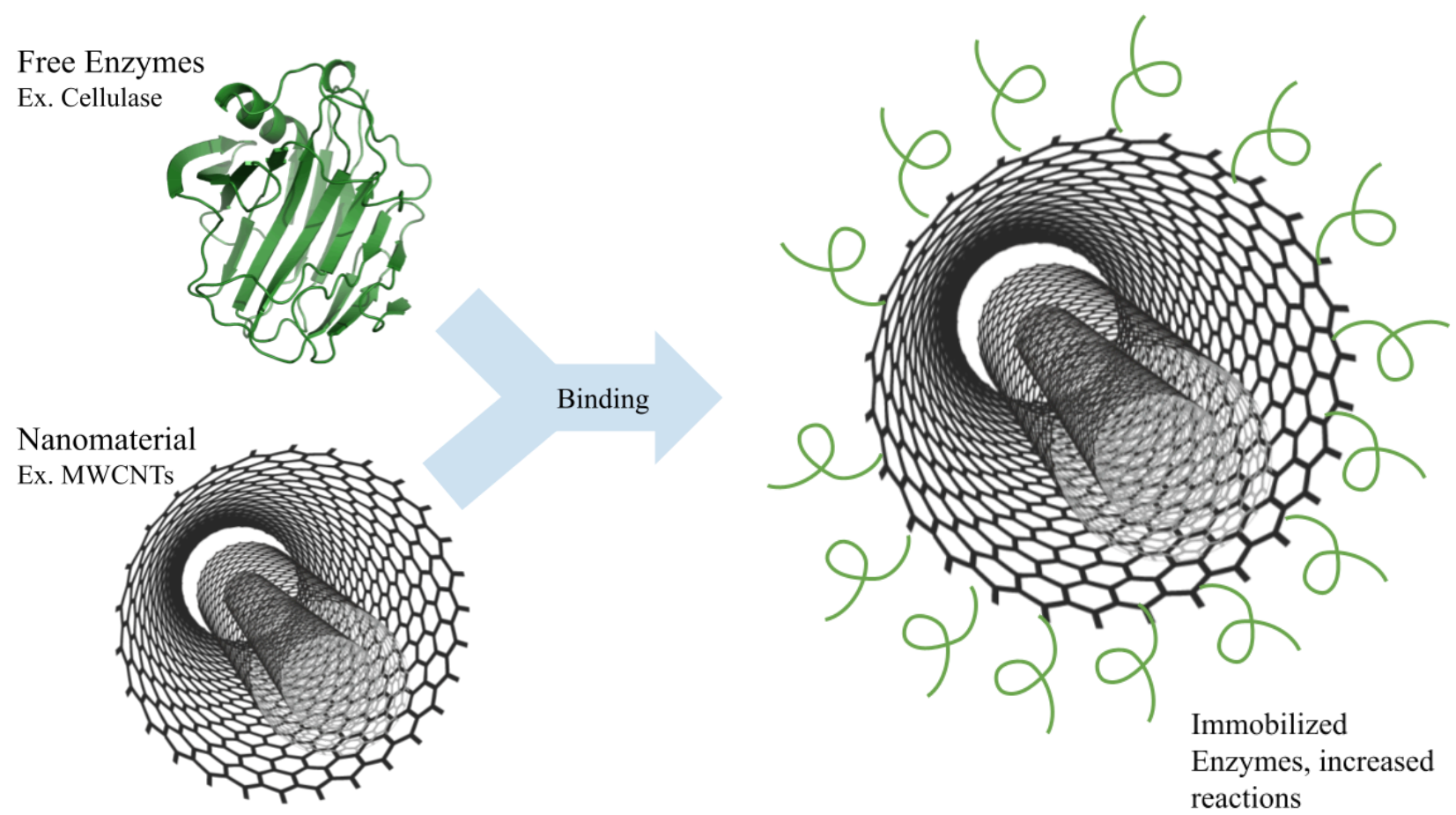

(a)

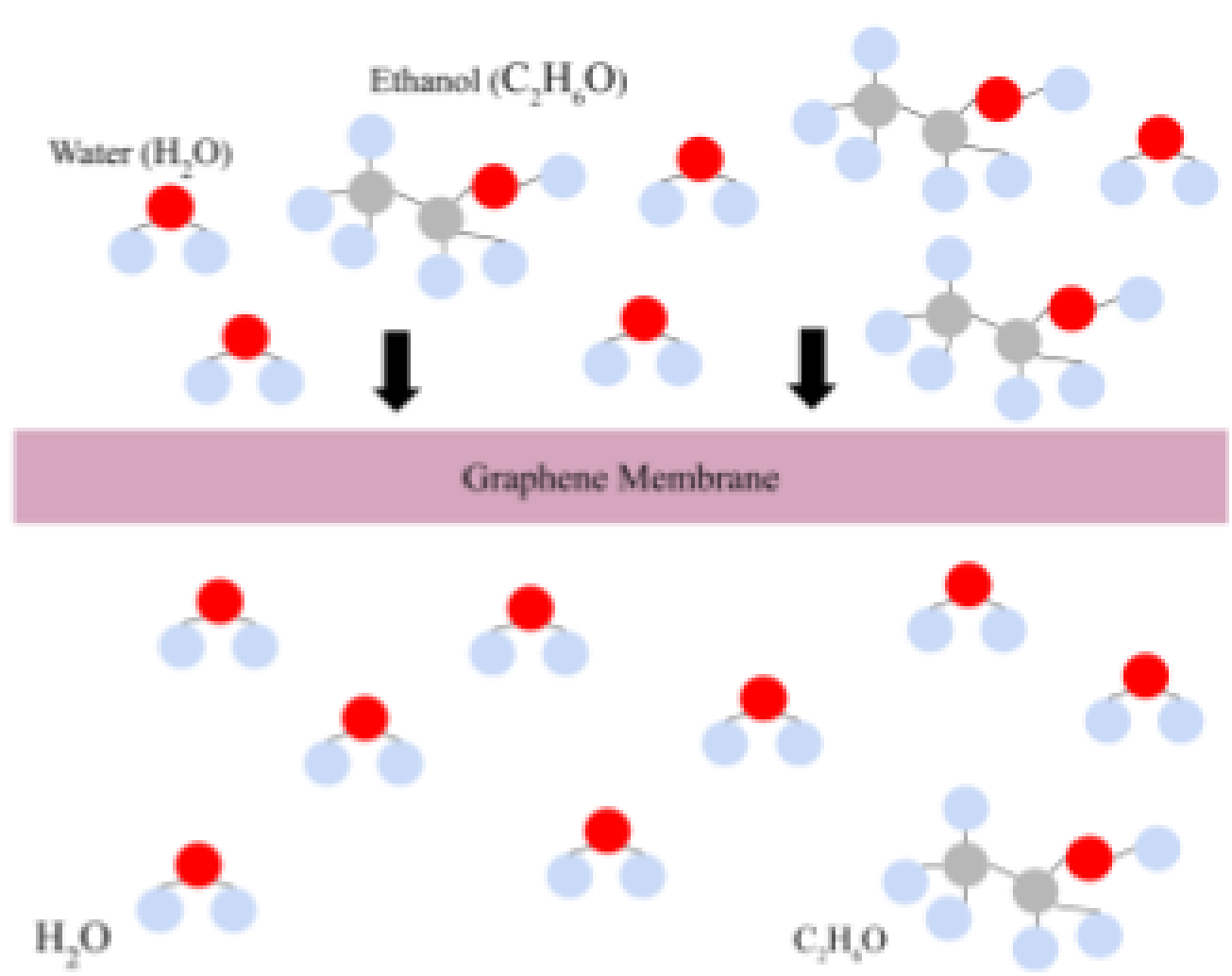

(b)

**Figure 8.** (a) Immobilization of Cellulase on Multiwalled Carbon Nanotubes. MWCNTs adapted from [260]. Cellulase adapted from [261]; (b) Graphene Microfiltration Membrane Controlling Water and Ethanol Flow [265].

Additionally, nanoparticles can be used to create microfiltration membranes that aid in the distillation process by separating and purifying ethanol from impurities and other unwanted substances. Membrane filtration has several benefits, such as low energy use and compatibility with other separation methods [266]. The incorporation of nanomaterials into these membranes can significantly increase their performance by improving their selectivity, stability, and permeability due to properties such as high surface area, thermal stability, and electrical conductivity [267]. Graphene is particularly effective because of its two-dimensional honeycomb structure. When integrated into membranes, graphene forms narrow, well-ordered nanochannels between polymer chains that act as selective filters. As visualized in **Figure 8 (b)**, these channels block ethanol molecules of 0.44 nm while allowing smaller water molecules of 0.28 nm to pass, thereby improving the separation efficiency [267, 268].

### 5.4. Nanomaterials in biogas production

Biogas, a mixture of primarily carbon dioxide and methane [269], is produced via anaerobic digestion, a process that converts biomass into biogas in the absence of oxygen [270]. Anaerobic digestion is composed of four stages [271]:

- Hydrolysis, where complex polymers are broken down into simpler, soluble molecules with the support of enzymes;
- Acidogenesis, where the products of hydrolysis serve as substrates for acidogenic bacteria and are further broken down into short-chain organic acids;
- Acetogenesis, where fatty acids and alcohols are converted into acetate ($C_2H_3O_2$-), carbon dioxide ($CO_2$), and hydrogen ($H_2$) by acetic acid-forming bacteria; and
- Methanogenesis, where methanogenic archaea convert acetate, $H_2$, $CO_2$, and other single-carbon compounds into methane ($CH_4$) and some $CO_2$.

#### 5.4.1. Role of nanomaterials in biogas production

Nanomaterials are increasingly used to increase biogas production, often by stimulating bacterial activity. The impact of carbon-based nanomaterials on anaerobic digestion varies and is influenced by factors such as their type, concentration, and thus interaction with microbial communities. Certain nanomaterials can be toxic and hinder the functioning of anaerobic bacteria. As a result, bioactive nanomaterials, such as carbon nanotubes, are emerging as promising alternatives [273] that can alter cell actions and cause biological responses in living tissues [274]. A study revealed that SWCNTs increased methane production rates. The toxicity of SWCNTs was also minimized because their presence triggered the production of extracellular polymeric substances that prevented the nanoparticles from entering and damaging the cell [275]. Similarly, multiwalled carbon nanotubes (MWCNTs) enhance methane production in anaerobic digestion by increasing microbial activity. A maximum increase of 1.1 to 2.2 times was found relative to the control. The addition of iron to form a composite (MWCNT-Fe) further increases production by 4.5-- to 10.2-fold [276].

Trace metals, such as iron, nickel, and cobalt, are essential for the growth of methanogenic bacteria involved in anaerobic digestion. The nanoscale form of zero-valent iron (NZVI) is particularly effective because of its high surface-to-volume ratio, facilitating increased reactions. Hydrogen sulfide ($H_2S$), often a toxic byproduct of aerobic digestion, reacts with the oxide shell of NZVI, forming FeS and $FeS_2$, thereby reducing $H_2S$ levels and increasing methane production. NZVI doses ranging from 0.05–0.10 wt% can increase methane production by up to 9.8%, whereas higher 0.20 wt% doses may inhibit biogas production due to potential toxicity [273]. Another study investigated the influence of metal and metal–oxide particles on biogas production from microalgae. The yield was moderate for an initial sixty hours but subsequently significantly increased biogas production. A 10 mg/L concentration of $Fe_3O_4$ resulted in the highest yield, increasing the biogas output by 1.24 times. It stimulates microbial activity by releasing $Fe^{2+}$ ions, which support microbial metabolic functions. Additionally, $Fe_3O_4$ nanoparticles help disrupt and breakdown tough biomass structures, making it easier for microbes to access and digest the material [277].

### 5.5. Case Studies of Improved Yields with Nanomaterials

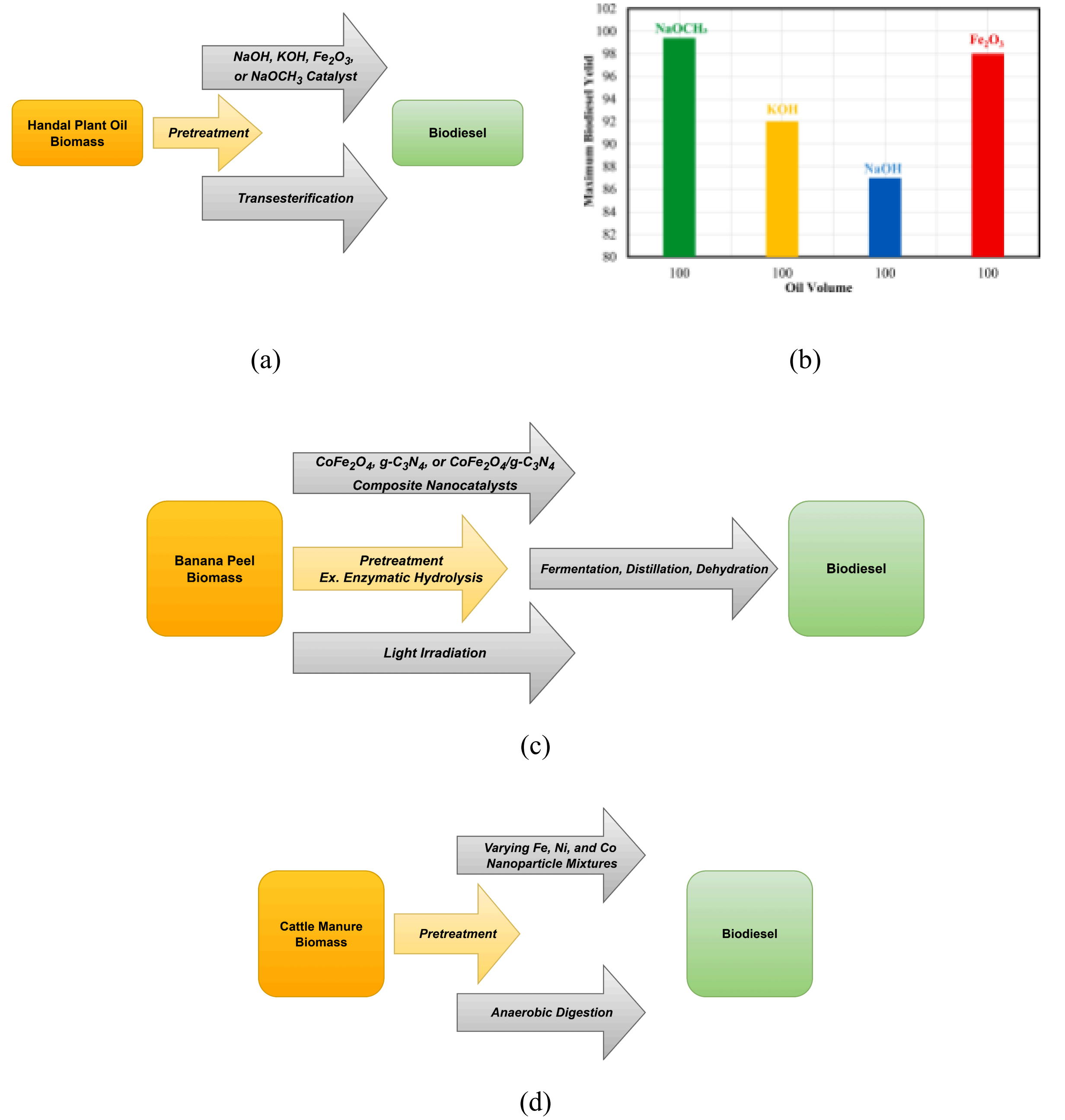

**Figure 9.** (a) Biodiesel production steps from Handal plant oil and (b) maximum biodiesel yield (%) with various nanocatalysts. Adapted from [278]. (c) Bioethanol production steps from banana peel biomass. (d) Biogas production steps from cattle manure.

Several case studies highlight the advantages of nanomaterial use for biofuel applications in terms of efficiency [278, 279] and emission reduction [280], with specific examples focusing on biodiesel, bioethanol, and biogas production. These are explained in the subsections 5.5.1 to 5.5.3.

### 5.5.1 Improving Biodiesel Production via Nanoparticles with Handal Plant Oil

In a case study from Jordan, researchers evaluated the catalytic effectiveness of different nanocatalysts (NaOH, KOH, $Fe_2O_3$, and $NaOCH_3$) in the transesterification process via oil from the handal plant. The general biodiesel production steps are represented in **Figure 9 (a)**. As exemplified in **Figure 9 (b)**, the study demonstrated an overall efficiency of yield when nanomaterials were used as catalysts. For example, $Fe_2O_3$ allowed for a one-step transesterification process that produced a 98% biodiesel yield in 50 minutes, whereas conventional methods require approximately 24 hours. A 12:1 methanol-to-oil ratio was used, with a 14% catalyst concentration. Additionally, $NaOCH_3$ delivered the largest biodiesel yield of 99.4% at a 12:1 methanol-to-oil ratio and 0.8% catalyst concentration [278].

### 5.5.2. Improving bioethanol production with banana peels via nanoparticles

A recent study demonstrated how the addition of cobalt ferrite ($CoFe_2O_4$), graphitic carbon nitride (g-$C_3N_4$), and their composite ($CoFe_2O_4$/g-$C_3N_4$)) during the bioethanol production process can significantly increase bioethanol yields via the use of a banana peel biomass. The general production process is outlined in **Figure 9 (c)**. Banana peels contain large amounts of fermentable sugars after being subjected to suitable pretreatment processes [281], such as enzymatic hydrolysis. A. The Ethanol production increased from 11.16% without nanoparticles to 52.16% when 100 ppm $CoFe_2O_4$ was employed [279].

### 5.5.3 Improving biogas production via nanoparticles with cattle manure

A recent study explored the use of nickel (Ni), iron (Fe), and cobalt (Co) nanoparticles at various concentrations to increase the yield and efficiency of biogas production from cattle manure. **Figure 9 (d)** generalizes the production process steps. Compared with the addition of cattle manure alone, the addition of an optimal concentration of 30 mg/L Fe, 2 mg/L Ni, or 1 mg/L Co resulted in a 14.61% increase in total biogas yield and a 19.3% increase in methane production. This nanoparticle-enhanced process also reduced the production of hydrogen sulfide ($H_2S$), a harmful byproduct, by 35.1%, thereby improving biogas quality. To be more specific about the role of these metal nanoparticles, iron primarily acts as an electron donor, facilitating the conversion of carbon dioxide to methane and facilitating the breakdown of volatile solids. They also mitigate $H_2S$ toxicity by forming iron sulfide. Nickel and cobalt support the growth and activity of bacteria that execute methanogenesis, a process that converts molecules to methane. Moreover, adding 2 mg/L nickel nanoparticles (20 nm) or 1 mg/L cobalt nanoparticles (20 nm) resulted in significant methane production increases of 101% and 86%, respectively [280].

## 5.6. Economic and Energy Input–Output Analysis

Cost-effective systems of bioenergy production require technoeconomic assessments of the cost of the nanoparticle production process, as this will impact the overall cost of bioenergy synthesis. In other words, the cost of nanoparticle synthesis is an important contributing factor to the cost of biofuel production, so cost-effective nanoparticles must be used in this process and developed further for the overall production cost to be commercially feasible [22]. The cost is also greatly dependent on the biomass that is being converted into biofuels, with maize ethanol having a cost of $17 US/GJ, whereas algal fuel systems have a significantly higher cost of $28--65 US/GJ, high-energy input cost, and negligible energy returns on investment, which has been shown in various studies [282].

The use of nanotechnology in biofuel synthesis comes at high operational costs, especially when it is used to convert lignocellulosic biomass as nanocatalysts. In addition to the high synthesis cost of nanoparticles, nanocatalysts must undergo complex pretreatment to improve the efficiency of conversion, which is often very costly. This method also has high capital and obstructs scalability, which is currently a setback from large-scale implementation [4]. However, high yields can be achieved via this method. In an experiment by Srivatava et al., a maximum yield of 93.56% in biodiesel was achieved under optimum conditions. This same experiment also showed that cellobiose conversion can reach up to 96.0%, whereas conventional conversion without nanoparticle catalysts is only 32.8% [283]. Reusability is also an important factor, as in an experiment by Chen et al., the $LaTiO_3$ nanocatalyst had a 74% production yield following an 8-round test, and the $Fe_3O_4$4/ZnMg(Al)O nanocatalyst maintained an 82% biodiesel yield over 7 operations [284]. After 5 rounds of immobilization, the $MnO_2$ nanoparticles were able to retain 60% of the catalytic capacity of cellulose [284]. These numbers must be improved to make biofuel synthesis using nanomaterials more feasible on an industrial scale. Carbon-based nanotubes, however, are a more affordable option, as they are widely accessible [4].

An improved combustion performance has been recorded when nano additives are incorporated with fuel, as the higher the concentration of nanoparticles is, the higher the cetane and calorific values are [22, 283, 285]. Metal oxide nanoparticles improve biodiesel fuel efficiencies by behaving as oxygen buffers, which reduce the emission of oxides from nitrogen, leading to simultaneous hydrocarbon processing. Additionally, aluminum- and silicon-based nanoparticles improve biodiesel engine combustion quality. Nanoparticles have also been shown to increase enzyme activity in biofuel production; for example, in an experiment published by Srivastava et al., the addition of 1 mM nickel–cobaltite nanoparticles increased the activity of endoglucanase by 49%, xylanase by 19.8%, and β-glucosidase by 53%. There is currently a lack of research on the economic analysis and feasibility of the use of metal nanoparticles in large-scale biofuel production [22, 283].

Adding nanoparticles to biodiesel increases combustion efficiency, although delivering the fuel to the boiler requires much more power in the pump than diesel does. Furthermore, in an experiment concerning the total cost over a ten-year period published by Panbechi et al., diesel had an accumulative cost of $967332.5161, whereas biodiesel-nanoparticle fuel cost $124475.8381 more. Diesel is still the most cost-effective fuel, even when considering carbon taxes [285]. **Table 4** covers an approximate cost–benefit analysis, with the overall consensus being that cost efficiency can be reached only through high yield and a low amount of nanoparticles.

**Table 4.** Comparison of Efficiency, Costs, Emissions, and Other Essential Parameters in Nanomaterial-Enhanced Biofuels.

| **Biofuel** | **Nanomaterials Used or Added** | **Performances** | | | **Disadvantages** | **Ref** |
|---|---|---|---|---|---|---|
| | | **Improves Efficiency by:** | **Emissions** | **Cost** | | |
| Biodiesel | **Metal oxides:** CuO, $TiO_2$, $Al_2O_3$, $CeO_2$, $Fe_2O_3$, ZnO, $SiO_2$ | Improving break and thermal efficiency (BTE). Improved | Reduced smoke emission, CO, and unburned hydrocarbons. | Considered to be cost-effective with high economic | Particles tend to aggregate and cannot easily be dispersed throughout the liquid | [286] |
| | **Carbon-based:** CNTs and GO **Others:** $MoO_3$ and $Fe(C_5H_5)_2$ | atomization and combustion. Reducing fuel consumption in brakes (BSFC). | Reduced nitrogen oxides ($NO_x$) emissions. Reduced exhaust emissions. | viability. | fuel. | |
| Bioethanol | $TiO_2$ and $Al_2O_3$ | Increases BTE. Increases combustion performance. Improves fuel-air mixing. | Lower CO and hydrocarbon emissions. Lowers most tailpipe emissions and smoke levels. Higher $NO_x$ emissions. | High cost due to nanoparticles but the cost can be reduced. | High nanoparticle cost. Increase in $NO_x$ emissions. Agglomeration and sedimentation issues. Health and environmental risks. | [287] |
| Biogas | Fe, $Fe_2O_3$, $Fe_3O_4$, Ni, Co, WIP (waste iron powder), Nanocomposites. | Increased biogas production. Improved conversion of organic material. Faster production | Significantly reduced $H_2S$ levels. | Expensive due to nanoparticle cost but cost can be reduced. | Residual nanoparticles may be left over in soil and water, being left on the earth. | [280] |

## 6. Performance in terms of biofuel utilization

### 6.1. Nanocatalysts in Biofuel Cells

A fuel cell operates by transforming chemical energy to electrical energy, providing a cleaner alternative to harmful combustion and fossil fuel processes [288-290]. Biological fuel cells are supported by biocatalysts while operating similarly to conventional fuel cells, with a continuous flow of fuel to the anode and an oxidant to the cathode. Organic compounds, such as glucose, are used as fuel rather than traditional hydrocarbons. Biological fuel cells are generally categorized into two types on the basis of the biocatalyst involved: microbial fuel cells (MFCs), in which all microorganisms catalyze reactions, and enzyme-based fuel cells (EBFCs), in which enzymes are extracted from biological systems and purified to catalyze reactions [291]. BFCs are limited by challenges such as low power yield, long-term stability, biocompatibility and natural substrate availability [292-295].

#### 6.1.1. Nanomaterials in Microbial Biofuel Cells

MFCs function through an extracellular electron transport (EET) process, in which electrogenic microbes act as biocatalysts, transferring electrons generated from the oxidation of organic matter to the anode. The electrons then move through an external circuit to the cathode, while protons pass through a selective proton exchange membrane [293, 294]. As displayed in **Figure 10,** MFCs have several challenges that limit their industrial applications, including low power density and instability on the basis of temperature sensitivity (since microbes do not function under extreme conditions). Additionally, the production of MFCs is often associated with complexity, toxicity, and high expenses [22, 294, 296].

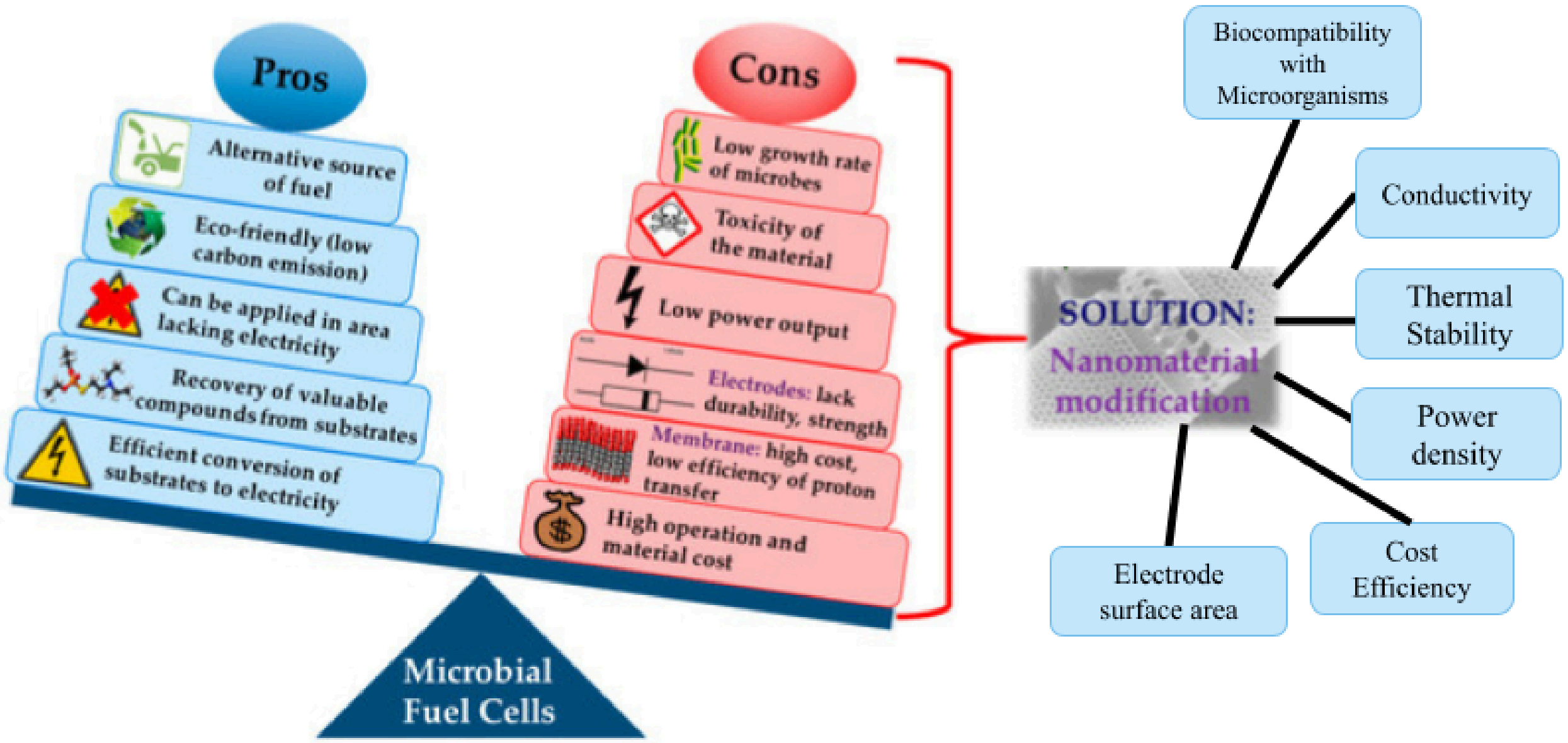

**Figure 10.** Benefits and drawbacks of Microbial Fuel Cells and advantages of Nanomaterials as a Solution. Comparison scale and solution adapted from [22].

Recent advances in nanotechnology have prioritized enhancing the performance and efficiency of MFCs. The integration of nanomaterials, particularly at electrodes, has improved characteristics such as microbial biocompatibility, cost efficiency, thermal stability, power density, electrode surface area, and conductivity [297]. Carbon-based nanomaterials and their composites are promising for increasing MFC performance because of their stability, conductivity, and large surface area, which are relevant factors for optimizing anode quality [298]. For example, a study enhanced a carbon felt electrode by incorporating a hybrid nanomaterial consisting of multiwalled carbon nanotubes (MWCNTs), gold, and platinum combined with an osmium redox polymer (OsRP) and *Gluconobacter oxydans* cells. This modified electrode acted as the bioanode in a two-chamber MFC, using an activated Nafion membrane as the proton exchange membrane. After the quantities of OsRP and the nanomaterials were optimized, the enhanced bioanode exhibited notable performance improvements, reaching a peak power density of 32.1 mW/m², a current density of 1032 mA/m², and a charge transfer efficiency of 22.3 [299].

In addition, it is necessary to consider the limitations of various nanomaterial structures. While two-dimensional nanomaterials provide good conductivity, their small pore size can lead to microbial clogging, limiting access to the interior of the anode and ultimately reducing efficiency. Three-dimensional anode structures have consequently been employed to increase the surface area. However, 3D materials may cause issues such as small pore sizes for microbes to penetrate and lower conductivity. To address these issues, various carbon-based nanoparticles have been applied as coatings on three-dimensional anode structures, improving their surface area and biocompatibility [301]. For example, graphene was used as an anode in MFCs powered by an *Escherichia coli* microorganism. Researchers compared three anodes: plain stainless steel mesh (SSM), PTFE-coated SSM (PMS), and graphene-coated SSM (GMS). The GMS significantly outperforms the other nonnanomaterials, achieving a maximum power density of 2668 mW/m², which is 18 times greater than that of the SSM and 17 times greater than that of the PMS. Scanning electron microscopy (SEM) revealed that this was a result of the larger surface area of the graphene anode and increased bacterial attachment [302].

### 6.1.2. Nanomaterials in Enzyme-based Biofuel Cells

EBFCs are particularly appealing because of their ease of operation and reduced functional size. These cells use enzymes as biological catalysts located within a membrane electrode assembly (MEA). Enzymes fixed in place drive the oxidation of organic compounds at the bioanode, releasing electrons that travel through an external circuit until they undergo a reduction reaction at the biocathode [298]. However, EBFCs face two primary challenges: enzymes produce low power densities due to poor electron transfer efficiency from the enzyme's active site to the electrode, and they often have short lifespans since their stability decreases outside of their native environment [303].

Enzymes are sensitive and can quickly lose activity when exposed to altered pH or temperature, making them difficult to reuse and potentially leading to system contamination. To address these limitations, enzyme immobilization is employed, in which enzymes are attached to solid supports to preserve their structure, increase their stability under varying conditions, and enable reuse. Owing to their large surface area, nanomaterials, such as carbon-based materials [304] and nanocomposites [298], can be integrated to improve overall efficiency and stability. Owing to their high porosity, conductivity, surface area, and consistency, carbon nanomaterials, which are commonly used as electrode components, have proven to be largely effective for immobilization to improve the stability and durability of enzyme attachment. [304].

For example, a three-dimensional graphene (3DG)/glucose oxidase (GOx) bionanocomposite was developed in a study to increase EBFC performance by improving enzyme immobilization, accelerating electron transfer, and increasing the enzyme lifespan. An increase in power output and enzymatic stability was possible because of the physical properties of 3DG, such as its high surface area, conductivity, and porosity. These nanomaterials offer a platform to facilitate electron transfer directly between the enzyme and electrode and help enzymes attach to the electrode. The study explained how the voltage and cell power output change with current, highlighting the enhancement of the EBFC function by graphene: the modified electrode can produce approximately 160 $\mu W/cm^2$ of power at 0.4 V, which is significantly greater than that reported in previous studies [305].

Furthermore, a case study incorporated DNA-wrapped single-wall carbon nanotubes (SWCNTs) in EBFCS that employ glucose oxidase (GOD) and laccase as biocatalysts. This highlighted the significant advantages of enhancing enzyme conditions and electrical performance under ambient conditions for the use of nanomaterials. The SWCNTs acted as mediators to improve electron transfer by increasing the conductivity between the enzymes and electrodes, which resulted in a significant increase in power generation to 442 $\mu W/cm^2$ at 0.46 V, which is 453% more power than an EBFC whose enzymes had not been immobilized by SWCNTs or DNA. The presence of nanomaterials addressed common EBFC challenges by increasing activity and contributing to greater stability for GOD and laccase enzymes [306].

Similarly, a study employed novel nanostructured multiwalled carbon nanotube (MWCNT) films fabricated via electrophoretic deposition (EPD) as electron supports in an EBFC to provide connections between the biocatalyst's active site and charge collectors. To produce bioelectrodes, flavin adenine dinucleotide-dependent glucose dehydrogenase and laccase enzymes were immobilized on the films. The EBFC achieved a high electrical performance of 0.43 mW/cm² at 0.59 V and demonstrated enhanced durability, maintaining 56% of its original power density after seven days [307].

### 6.2. Combustion Efficiency and Emission Reduction with the Use of Nanomaterials

The combustion efficiency refers to the proportion of fuel energy that is effectively converted into usable heat during combustion processes [308], whereas emission reduction refers to the process of minimizing the release of harmful gases and particulate matter into the atmosphere [309]. In biofuel applications, improving combustion efficiency is essential for maximizing energy output, which in turn lowers fuel usage and reduces expenses for operations [310]. Nanomaterials enhance biofuel production through several key properties, as outlined in section 2.2, such as their high surface area, which increases the number of reactive sites and enhances catalytic performance [311].

With respect to emission reduction, nanocatalysts contribute by storing oxygen, which enables more complete fuel combustion, thereby lowering greenhouse gas emissions [312]. Biofuels are increasingly relied upon as a cleaner replacement for fossil fuels, which is an indirect strategy for lowering emissions given that fossil fuels accounted for 91% of emissions in 2022 [313, 314]. Biofuels are often considered carbon neutral since the $CO_2$ released when they are burned is offset by the $CO_2$ taken in by plants during biomass growth, although some biofuels contain fossil-based carbon [315]. Additionally, the use of nanomaterials contributes to emission reduction in all energy sources, including biofuels and fuel cells, by enabling significant weight reduction in vehicles. Owing to their small size, nanoparticles reduce their overall mass, which, according to Newton's second law, decreases the force required for acceleration. This leads to lower fuel consumption and consequently reduced $CO_2$ emissions [310]. **Table 5** provides examples of numerous studies that have shown how incorporating nanomaterials into biofuel blends can greatly improve engine efficiency while simultaneously reducing harmful emissions, such as carbon monoxide (CO), hydrocarbons (HC), and nitrogen oxides ($NO_x$), in addition to certain outliers, thereby offering a promising approach to cleaner and more efficient fuel alternatives.

**Table 5.** Emission Reduction and Overall Efficiency of Nanomaterial-Enhanced Biofuels

| Biofuel Description | Nanomaterial | Emission Reduction | Efficiency | Ref |
|---|---|---|---|---|
| 40% biodiesel from waste cooking oil blended with diesel | Graphene Oxide at 150 parts per million (ppm) | Approximate 29% drop in nitrogen oxide ($NO_x$) emissions (from 1240 ppm to 884 pm) | Brake thermal efficiency (BTE) of 43.6% | [316] |
| Biogas used in dual fuel mode with diesel pilot fuel | Multiwalled carbon nanotubes (MWCNTs) at varying ppm (30, 60, 90 ppm) | CO and HC emissions similar with and without MWCNTs; slight increase in nitrogen oxide (NO) emissions | Improved BTE: 30.2% (30 ppm), 30.4% (60 ppm), 30.0% (90 ppm), all higher compared to without nanomaterial (27.9%) | [317] |
| 15% Bioethanol from | Aluminum Oxide | 75 ppm: 0.2% CO | 75 ppm: Maximum energy | [318] |
| waste rice straw blended with diesel (B15) | (Al2O3) at varying ppm (25, 50, 75 ppm) | emissions<br><br>50 ppm: minimum NOx emissions of 398 ppm | efficiency (34.3%) and exergy efficiency (65.27%) | |
| 20% Biodiesel from Mahua Oil blended with diesel. | $Al_2O_3$ and Cerium Oxide ($CeO_2$), at 100 ppm each | HC reduced by 30.73%, $NO_x$ reduced by 1.27%, CO reduced by 44.13% | Brake specific fuel consumption (BSFC) reduced by 3.25%<br><br>BTE increased by 1.39%. | [319] |
| 30% biodiesel from waste cooking oil blended with diesel | Titanium dioxide ($TiO_2$) at varying ppm (40, 80, 120) | 120 ppm:<br>CO emissions 25.61486 kg/kWh<br><br>HC emissions 0.05289 kg/kWh<br><br>$NO_x$ emissions increased slightly | 120 ppm:<br><br>Minimum BSFC of 0.33994 kg/kWh<br><br>Maximum BTE of 25.90% | [320] |
| Biogas production from cattle manure | Iron (Fe), Nickel (Ni), and Cobalt (Co) at varying quantities | 30 mg/L Fe, 2 mg/L, and 1 mg/L Co decreased Hydrogen sulfide ($H_2S$) emission by 35.10% | 30 mg/L Fe and 2 mg/L Ni increased biogas production by 14.61%<br><br>30 mg/L Fe, 2 mg/L, and 1 mg/L increased methane production by 19.30% | [280] |
| Biogas production from cattle manure | $Fe_2O_3$ at 20 mg/L (R1) and 100 mg/L (R2)<br><br>$TiO_2$ at 100 mg/L (R3) and 500 (R4) mg/L<br><br>A mixture ($Fe_2O_3$ and $TiO_2$) at 200 mg/L and 500 mg/L (R5), as well as 100 mg/L and 500 mg/L (R6) | Compared to control (R0), $H_2S$ production was decreased by<br>2.13-fold for R1,<br>2.38-fold for R2,<br>2.27-fold for R3,<br>2.51-fold for R4,<br>2.64-fold for R5,<br>2.17-fold for R6 | Biogas and methane production increased by<br>1.09 and 1.105-fold for R1,<br>1.15 and 1.191-fold for R2,<br>1.07 and 1.097-fold for R3,<br>1.17 and 1.213-fold for R4,<br>1.10 and 1.133-fold for R5,<br>1.13 and 1.15-fold for R6 | [321] |
| 30% biodiesel from hemp seed oil blended with diesel | Titanium dioxide ($TiO_2$) at varying ppm (25, 50, 75, 100 ppm) | Up to 75 ppm reduced CO by 40.15%, also reducing HC<br><br>$NO_x$ emissions increased | 75 ppm: BTE increased by 27.942%<br><br>100 ppm:<br>BTE increased by 29.65% and BSFC reduced by 5.16% | [322] |
| Biodiesel (B100) and 50% biodiesel (B50) blended with commercial diesel | $Al_2O_3$ at varying ppm (50, 100, 150 ppm) | B100 with 50 ppm reduced CO emissions by 42.8% (compared to solely diesel fuel (B0)) | B50 with 100 ppm: 26.5% fuel conversion efficiency, which is a 28.6% improvement | [323] |
| | | B50 with 100 ppm: reduced $NO_x$ by 30.3% | | |
| Diesel, Biodiesel, Ethanol blend at varying quantities<br><br>(D20B70E10Ti: 20% diesel, 70% biodiesel, 10% ethanol) | $TiO_2$ at 100 ppm | D20B70E10Ti:<br>CO emissions decreased by 7.44%<br><br>$CO_2$ emissions decreased by 7.12%<br><br>$NO_x$ emissions decreased by 66.29% | D20B70E10Ti<br><br>BTE was 1.23% greater than diesel<br><br>BSFC of blends were lower than biodiesel but higher than diesel | [324] |
| Biodiesel blends, tested at various revolutions per minute (rpm) (2000 rpm, 2250 rpm, 2500 rpm, 2750 rpm) | Zinc oxide (ZnO) at varying concentrations (0.025%, 0.05%, 0.1%) | 2500 rpm: reduced $NO_x$ emissions by 10.67% and $CO_2$ by 7.6% | 2500 rpm increased BTE by 11.7% and reduced specific fuel consumptions (SFC) by 1.67% | [325] |

Additionally, a study explored the use of castor seed-derived biofuels enhanced with 2% diethyl ether and cerium oxide ($CeO_2$) nanoparticles (NPs). Several biodiesel–diesel fuel blends (B0, B5, B10, B15, and B25), with and without $CeO_2$ nanoparticles, were tested under constant load conditions, which revealed that $CeO_2$ nanoparticles significantly enhance the combustion efficiency and emission characteristics. Among the blends, B10 showed the greatest increase in thermal efficiency, achieving a 22.2% improvement. Furthermore, the B5 blend with $CeO_2$ resulted in a 12% decrease in carbon monoxide (CO) emissions and a 37.7% reduction in unburned hydrocarbons (HCs). However, nitrogen oxide emissions slightly increased by 4.3% in the B10 blend due to more complete combustion. As reflected in **Figure 11,** the combustion duration tends to increase with increasing biodiesel content, which is a consequence of slower burning due to increased oxygen levels and reduced volatility; however, the addition of nanoparticles reduced the combustion duration for all the blends [326, 327].

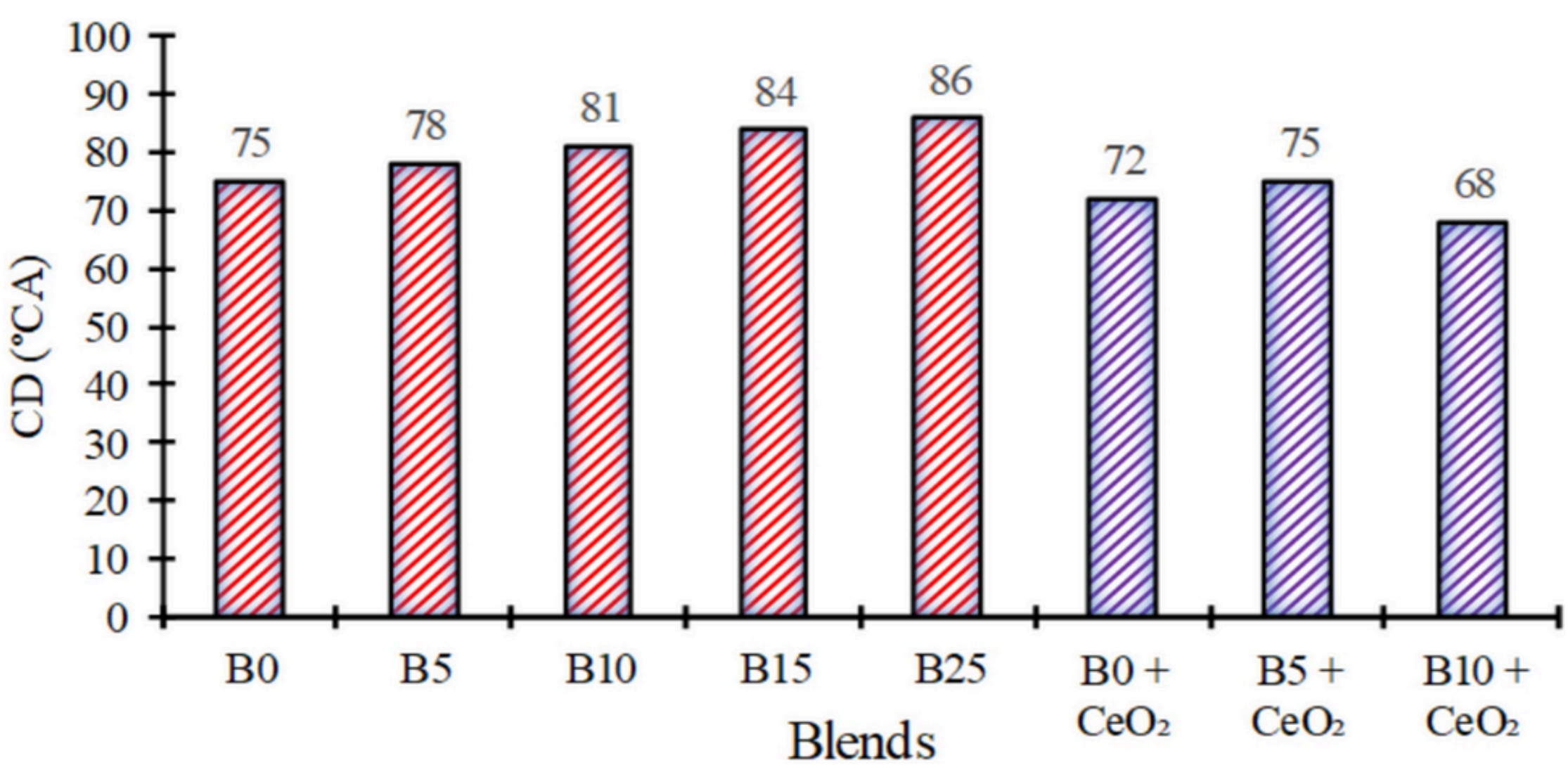

**Figure 11.** Combustion Duration for Various Biodiesel Blends. Adapted from [326].

### 6.3. Nanomaterials in the Storage and Transportation of Biofuels

Stable nanofuel blends and enhanced combustion characteristics are correlated with easier compliance with storage standards and fewer logistics chain issues. Carbon nanotubes and graphene-based structures, among other nanomaterials, possess excellent electrical and thermal conductivity, making them optimal for safe and efficient storage and transportation [28]. During long-term storage, the integration of nanoparticles as fuel additives in biofuels has been proven to reduce oxidation rates, microbial contamination, and sedimentation, contributing to overall increased stability during storage [22].

They increase thermal and oxidative stability (along with carbon-based nanostructures) under multiple operational and environmental conditions, allowing them to be suitable for transportation and storage over the long term [310]. Encouraging studies have set up solutions for portable, compact, and almost spill-proof transport using nanostructures (such as aerogels and MOFs) [327]. An additional benefit of nanostructure coatings that is being explored is their potential to resist corrosion (specifically fuel-induced degradation) and fuel–material interactions in storage tanks and pipelines, potentially increasing the lifetime of infrastructure involved in biofuel logistics [22, 28].

This mechanical reinforcement and thermal stability, which incorporates nanomaterials, improves the durability of energy storage units, which work best under differing ambient conditions [328]. When silver nanoparticles were added to pure diesel fuel in an experimental study [329], $NO_x$ emissions were reduced by 13%, and CO emissions were reduced by 20.5% [328]. In the same study, the addition of aluminum nanoparticles significantly reduced the concentration of smoke, which can be interpreted as the addition of nanoparticles resulting in a decrease in hazardous gas emissions. In a different study [330], the heat exchange efficiency of the engine was shown to increase with increasing $Al_2O_3$ nanoparticle concentration, which indicates an enhanced engine cooling effect.

Nanodiamond engine oil uses nanofluids to increase the performance of the engine by 1.15% while reducing the fuel consumption by 1.27%. Additionally, carbon nanotubes are examples of lightweight nanomaterials that can improve an automobile's overall durability and strength over the long term because of their relatively high thermal characteristics [328].

## 7. Challenges and Limitations

### 7.1. Environmental and toxicological concerns

Research into the health and environmental impacts of engineered nanoparticles (ENPs) has increased in recent years because of their increased potential in the field of biofuel as well as energy research. However, many widely used nanoadditives or other nanomaterial-based products have been less explored for their potential threats, such as toxicity, pollution or human hazards [331]. In the context of biofuels, dermal and inhalation exposure poses the greatest direct health risk from emissions. Airborne nanoparticles are typically produced by combustion sources such as biomass burning and engines [332]. For example, in Europe, nanoscale cerium dioxide has been approved and employed as a fuel additive in diesel engines because it helps reduce emissions from exhaust, such as carbon dioxide and carbon monoxide [333]. However, they present various toxicity risks, as these nanoparticles trigger the release of other emissions, particularly ultrafine particles. Higher concentrations of cerium in fuel lead to the emission of ultrafine particles with smaller diameters [333, 334]. Ultrafine particles are nanoparticles that pose human health risks because of their small size, since they can typically enter through inhalation into the lungs and have the ability to overcome typical defenses in the body, allowing them to be transferred to other organs and the bloodstream [31, 335]. In addition, dermal exposure serves as a pathway for entry into the body: nanoparticles can penetrate the skin barrier, particularly with prolonged or repeated exposure, potentially leading to systemic distribution or localized effects such as inflammation and irritation [336]. Evidence from ambient air pollution studies further supports this concern, as exposure to particulate matter ($PM_{2.5}$ and ultrafine particles) of similar size to ENPs has been associated with neurodegenerative disorders such as Alzheimer's disease and other systemic health effects. These particles may trigger various mechanisms, such as oxidative stress and adaptive immunity, that ultimately contribute to the onset of various respiratory and dermal conditions [337]. A study has shown that burning biomass, such as rice straw and pine stems, also releases fine particles, causing inflammation and reducing cell viability, suggesting that even biobased fuels can pose health risks [338].

Certain properties or stimuli-induced interactions of nanomaterials present potential risks in biofuels. For example, researchers have tested the durability of various nanoparticles and reported that gold and titanium dioxide nanoparticles dissolve more slowly, are more durable, and could have more long-term effects than silver [340]. This emphasizes the need for manufacturers to assess and ensure that the materials being used are fit for consumption owing to the durability and reusability of nanomaterials because there is a greater chance that certain nanomaterials may not breakdown and persist in the human body or environment, leading to detrimental effects [29]. Moreover, the large surface-to-volume area of nanomaterials increases their reactivity, allowing them to potentially hazardously engage with the environment. Their small size also provides them with multiple pathways to enter the environment and pose risks [340, 341], which was demonstrated by researchers using Leydig cells that enable testosterone synthesis in mice. This study exemplified the ability of nanomaterials to penetrate biological cells and cause damage. Diesel exhaust particles, titanium dioxide ($TiO_2$), and carbon black nanoparticles are absorbed by Leydig cells; among them, $TiO_2$ nanoparticles are the most toxic, impairing cell viability [342].

NPs can be absorbed by organisms through three main pathways: bioaccumulation, bioconcentration, and biomagnification. Even when immediate effects are not apparent, prolonged accumulation or magnification can lead to chronic problems such as organ damage and broader disruptions to ecosystems [343]. Bioaccumulation occurs when nanomaterials used in biofuels are unintentionally released into the environment and accumulate within organisms through exposure to contaminated water, soil, or air [343]. For example, if nanomaterials enter soil ecosystems, they can disrupt microbial communities, including their diversity and function. Such disruptions can impair nutrient availability, ultimately hindering the growth and productivity of plants used for biofuel feedstock [4]. Bioconcentration refers to the process by which the concentration of a substance becomes higher in an organism than in its surrounding environment [344]. A study revealed that $TiO_2$ nanoparticles threaten aquatic ecosystems by accumulating in organisms such as *Daphnia magna*. Prolonged exposure leads to bioconcentration because organisms have difficulty eliminating these particles. This causes growth delays, reproductive issues, and increased mortality, indicating lasting harm to aquatic life [345]. Biomagnification, the most significant and far-reaching path, occurs when the effects of a nanomaterial start at the base of the food chain and propagate upward through ecosystems [343, 346]. For example, ENPs such as cadmium selenide (CdSe) quantum dots have been found to biomagnify in food chains, reaching relatively high concentrations in predators that eat contaminated bacteria, such as protozoa [344].

### 7.2. Cost, and Industrial Feasibility

Large-scale biofuel production remains time-consuming and labor-intensive due to various limitations in existing technologies and increased operational costs [347]. Biofuel production is currently estimated to be approximately three times more costly than the production of traditional petroleum fuels. The higher cost is influenced by a range of factors including the cost of nanomaterial synthesis, pretreatment process, amount of land required for the cultivation of feedstock for biomass, labor expenses, transportation, distribution, and dealing with byproducts [348].

#### 7.2.1 Biomass Pretreatment Cost and Industrial Feasibility

A major barrier to scaling up biofuel production, even when abundant, local, low-cost feedstocks or nanocatalysts are used to speed up the process, is the high expenses associated with pretreatment. For example, pretreatment processes often require large amounts of water, which reduces efficiency and presents sustainability challenges [348]. Lignocellulose is highly abundant, representing approximately 57% of the biogenic carbon on Earth [349]. However, despite its prevalence, consistent research interest, and efforts to use lignocellulosic biomass for biofuel, it has not had a large impact on industrial applications [350]. Lignocellulosic biomass pretreatment accounts for approximately 40% of the total cost of biofuel production, making it a critical determinant of the economic feasibility of this form of biofuel [349]. Its limited success is attributed primarily to low market competitiveness due to the high costs associated with the equipment [351] and complex, multistage process to break down its structurally strong composition, purify the product, and often employ enzyme-based methods [351]. Enzyme costs also account for approximately 20% of the total expenses involved in producing fermentable sugars from lignocellulosic biomass [352].

#### 7.2.2 Nanomaterial Synthesis Cost and Industrial Feasibility

Transitioning nanoparticle synthesis to larger industrial scales presents various difficulties, despite being well established at the laboratory scale [12]. Several factors contribute to the elevated costs of producing nanomaterials that hinder their scale-up, including intricate and energy-intensive procedures (i.e., often require high temperatures) and expensive equipment. For example, metal–organic framework nanostructures have valuable properties, such as large surface areas and selective catalytic functions. However, their application is constrained by the expense of their synthesis, largely because research has prioritized enhancing performance over addressing factors such as cost-effectiveness, sustainability, and scalability. These methods involve expensive solvents, require specific equipment, and employ complex production techniques [353, 354].

Additionally, as discussed in section 4.2, homogeneous-based catalysts are widely employed for transesterification reactions in the biodiesel production process because they offer high catalytic efficiency and require less energy than acid catalysts do. Nevertheless, they present challenges such as complex purification processes, a reduced possibility for reuse, and large volumes of wastewater, all of which contribute to increased production costs. However, heterogeneous catalysts are increasingly favored because of their reusability, easy separation from reaction mixtures, and sustainable features [355].

While certain nanocatalysts may achieve excellent reaction efficiencies, their commercial use can be undermined by poor economic and environmental sustainability. Platinum-based catalysts are a notable example. Platinum is scarce, competes with critical technologies such as fuel cell electrocatalysts for limited global reserves, and its extraction is highly carbon-intensive (emitting approximately 14 t of $CO_2$ per kilogram extracted, prior to synthesis). The reliance on such high-cost, high-impact materials directly contradicts the goals of cost-effectiveness and carbon neutrality that nanomaterials are intended to help biofuels achieve [399].

### 7.3. Life Cycle Assessment of Nanomaterials in Biofuels

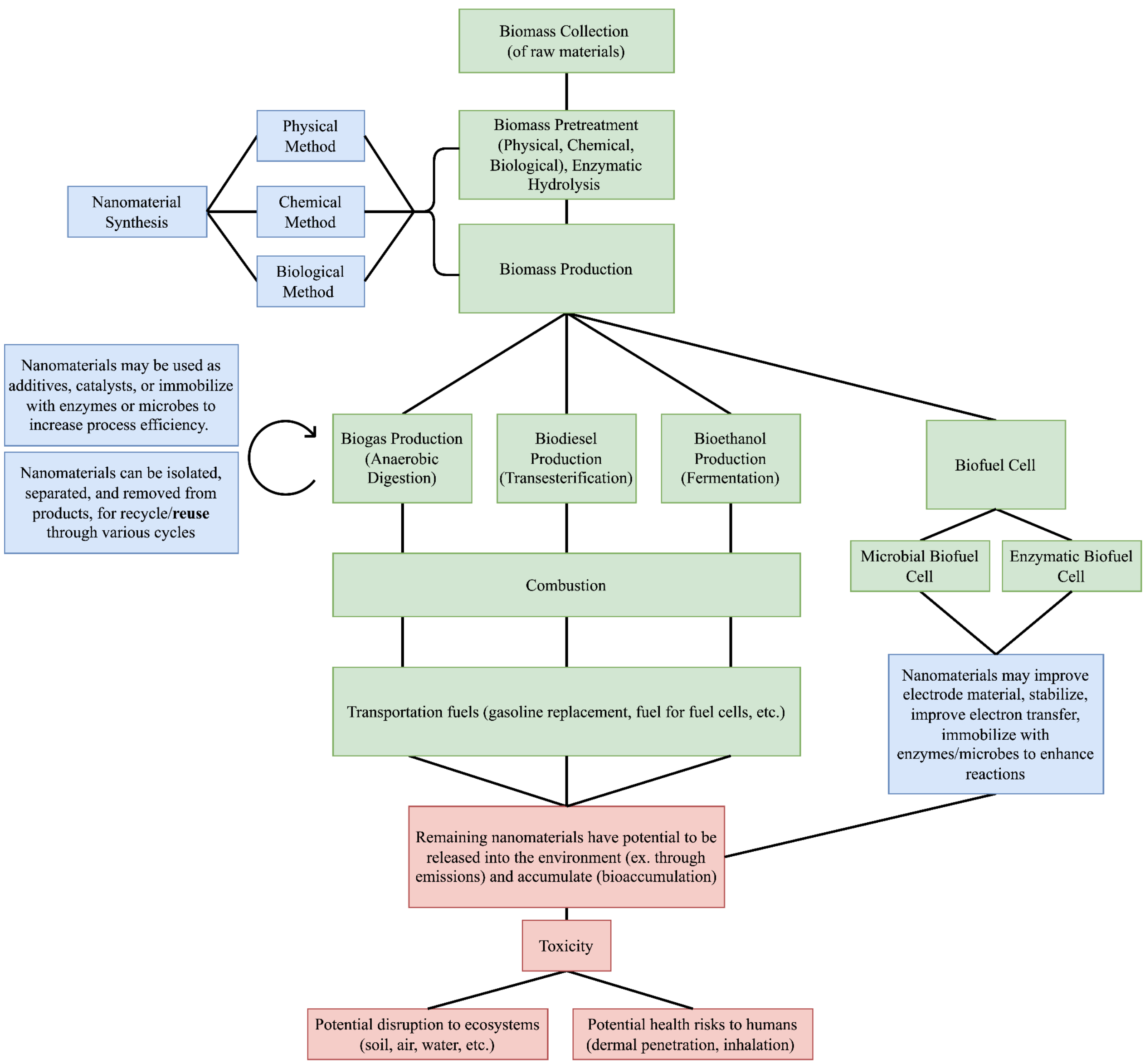

**Figure 12.** General Nanomaterials within the Biofuel Life Cycle

Life cycle assessment (LCA) plays a crucial role in evaluating the environmental impact of nanomaterials across their life cycle, from synthesis to disposal. Between 2001 and 2020, 71 LCA studies on nanomaterials were published, showing increased interest in this field. However, many studies overlook end-of-life impacts and rarely assess toxicity or exposure pathways. More comprehensive and detailed LCAs are needed to better understand and manage the environmental footprint of nanomaterials [356]. **Figure 12** shows a general example outlining the life cycle of nanomaterials used in biofuel applications without specifying the type of nanoparticle, synthesis approach, or biofuel application.

As described in section 3, nanomaterials can be produced via various synthesis processes depending on their desired properties and use. These materials can be integrated into the biofuel production process at various stages, depending on their purpose. As discussed in section 4.1, nanoparticles can support the breakdown of mass at the biomass pretreatment stage or may serve as enzyme immobilizers or catalysts during the production stage (transesterification, fermentation, anaerobic digestion). They can also be used as materials in components of biofuel cells, such as electrodes, as discussed in section 5.1.

Nanomaterials can improve various stages of biofuel production, such as pretreatment and catalysis/efficiency of production; however, the synthesis and disposal of nanomaterials themselves can result in environmental burdens. The nanomaterial synthesis process has raised environmental concerns because of its high resource demands. The production of nanomaterials can consume substantially more energy and raw materials than the production of other materials, such as fine chemicals or pharmaceuticals. These methods also often suffer from low efficiency and rely heavily on the use of toxic or hazardous chemicals and solvents [357].

As detailed in section 5.2, incorporating nanomaterials into biofuels can contribute to reduced emissions; however, the tradeoff is that the combustion of engineered nanoparticles is likely to release ultrafine particles, which present its own environmental and health-related problems, as described in section 7.1 [333]. Once the biofuel is produced, nanomaterials, along with their immobilizing supports, can often be isolated from reaction mixtures and recovered to be reused for various cycles of the production process, as detailed in section 2.2.2. However, once nanomaterials are utilized in fuels and are ready for consumption, such as in vehicles, there is limited information available on the best practices and standardized methods for identifying, monitoring, and managing nanowaste [358]. There is potential for nanomaterials that possess toxic qualities to be released into, and impact, the environment through various stages of the nanomaterial life cycle.

During production, improper handling or insufficient containment measures can lead to unintentional emissions through combustion, spills, or other indirect sources. Emissions may be intentional in certain cases, resulting from processes specifically designed to release nanoparticles under controlled conditions. At the end of their life cycle, nanomaterials are often disposed of in landfills, potentially degrading and being released into the surrounding environment [359]. This can contaminate surface, soil, and groundwater sources. For example, the prolonged buildup of metal oxide nanoparticles may inhibit photosynthesis and thus plant growth [81]. To reduce the environmental impact upon release, sustainable approaches such as the adoption of biodegradable alternatives, and the green synthesis of nanomaterials (as described in section 8.1) should be explored [360]. For instance, a comparative life cycle assessment showed that the green synthesis of iron oxide nanoparticles, using *Cymbopogon* *citratus* extract and sodium carbonate were shown to have a total environmental impact of $1.0 \times 10^{-9}$, substantially lower than $1.4 \times 10^{-8}$ for conventional coprecipitation; with large reductions in marine aquatic ecotoxicity ($7.6 \times 10^{-10}$ compared to $1.22 \times 10^{-8}$), during the nanomaterial production stage. These green-synthesized nanoparticles can then break down organic matter, thus enhancing biomass pretreatment, and have also shown to reduce hazardous wastes at the end of the life cycle [400].

Several nanomaterial disposal methods are currently being developed and utilized, although they do not address nanoparticles released during biofuel use, such as emissions from vehicles. For example, incineration is a potential strategy that involves burning waste at high temperatures. However, further research is needed to understand the behavior of nanoparticles at these temperatures, as nanoparticles may experience physical or chemical transformations that can influence their release or retention in ash and gases [362]. Modern incineration facilities are often equipped with specialized filtration systems to reduce ash emissions [359]. Additionally, nanowaste can be contained by embedding it within a solid matrix, making it easier to isolate and manage, or storing it in impenetrable containers to prevent soil contamination. Despite these approaches, there is still a lack of information regarding the safe detoxification and treatment of solid nanowaste [358].

Given the presently incomprehensible safety data regarding nanowaste management, a study has recommended certain guidelines, as follows [361, 362]:

- The emphasis is on assuming that all nanowaste is hazardous and using a broader nanoparticle definition, including particles up to 500 nm in size.
- Minimizing upstream nanowaste generation through methods such as machine learning.
- Minimizing airborne dispersion, as it is recommended that nanomaterials be stored in liquid forms or embedded in solid matrices.
- Whenever feasible, low-energy inactivation methods, such as hydrolysis or biodegradation, are encouraged; otherwise, more intensive treatments, such as pyrolysis at high temperatures (1200-1300°C), should be employed.

## 8. Future Perspective and Directions

### 8.1. Emerging Nanostructure and Green Synthesis Methods

Green nanotechnology refers to the environmentally sustainable synthesis of nanoparticles. Unlike conventional chemical-based approaches, green synthesis relies on natural agents such as DNA, fungal enzymes, bacteria, and plant extracts, resulting in nanoparticles with reduced toxicity [359]. Green synthesis of nanoscale metals typically involves preparing a plant extract, combining it with a metal salt solution under defined conditions to induce metal particle reduction, and performing filtration and processing steps to obtain the desired nanometal product [363]. Owing to its environmental advantages, green synthesis has proven to be a promising method [35]. For example, a study has already employed ZnO nanoparticles integrated into a polyindole-MWCNT electrode, achieving high electrocatalytic activity and stability for enzymatic biofuel cells [366]. Compared with traditional synthesis methods, green synthesis has a lower impact on climate change (value of $5.27 \times 10^{-8}$) (value of $7.69 \times 10^{-7}$) [367]. These methods have the advantages of utilizing renewable feedstocks, reducing toxicity (when organic pesticides are used), and reducing environmental risks related to managing hazardous waste [367]. Green synthesis for nanoparticle fabrication also does not include strict operational conditions, unlike conventional chemical synthesis methods [368].

Plants synthesize numerous bionanoparticles that are used in multiple research categories, mostly for observing catalyst performance during the process of biofuel production. Plants are also the most preferred entity (although algae bacteria and fungi are also used), as they are easier to scale and handle. Additionally, certain advanced bionanomaterials provide vigorous and adaptable solutions for challenging wastewater treatment problems while also protecting the environment and public health and improving the sustainability, effectiveness, and efficiency of wastewater treatment [35].

This area of study is continually being developed, particularly to develop bionanoparticles for sustainable solutions and environmentally conscious technology. They can be synthesized more sustainably than regular nanoparticles because they are derived from biodegradable and renewable resources, minimizing their ecological impact. Certain areas still require further development, such as the efficiency of synthesizing nanomaterials from a collection of natural resources in a sustainable manner. There is a predicted shift toward green and biological synthesis methods, as they have the potential to keep up with the demand for environmental consciousness, nontoxic synthesis, and cost-effectiveness. However, short- or long-term negative impacts are not yet adequately understood. This opens a new focus for research in risk management in this area, specifically with respect to the production, preservation, processing, and discharge cycles. Research must also be conducted on bionanoparticles to explore and expand their applications and optimize the synthesis process. Research is being conducted to improve the efficiency of biofuel production and production when nanotechnology is involved. Bionanotechnology is a combination of nanotechnology and biology, which is more eco-friendly when used in energy production. Research thus far has provided results with improved biofuel production efficiency, leading to various applications for nanotechnology in industrial settings and in the energy sector in the future [35].

### 8.2. Integration of Smart Sensing in Biofuels Using Nanomaterials & Aiding Artificial Intelligence

Nanomaterials, artificial intelligence (AI), and machine learning (ML) are increasingly being combined to increase the production, efficiency, and sustainability of biofuels by enabling improved properties and monitoring, optimization, and control of production processes [348]. Studies have shown that AI can improve nanomaterial-enhanced biofuel in various ways, including feedstock research for biomass, increasing the efficiency and yield of biofuel (typically employing feedstock properties, catalyst concentration, and reaction conditions as variables), and minimizing emissions to ensure cleaner biofuels [369].

AI models can be used to estimate biomass availability, yield potential, and growth monitoring throughout the bioenergy system, which includes steps such as biomass cultivation, production, harvest, and pretreatment [370]. For example, a geographic information system and AI model was created to assess how much crop waste could be used as energy in different areas over time, enabling accurate cost estimations and ideal biogas plant placement [371, 372]. A study also compared multiple linear regression (MLR) and random forest (RF) algorithms and reported that RF made stronger predictions, indicating that feedstock composition had a major influence on the quality (77%) and quantity (60%) of bio-oil produced from biomass [373].

AI models are capable of increasing the yield and efficiency of biofuels containing nanomaterials. Researchers have employed artificial neural networks (ANNs) to optimize a zinc oxide nanocatalyst that converts waste palm oil into biodiesel. ANNs can understand complex nonlinear relationships between variables to predict optimal outputs, typically involving calculations with numerical data inputs. In the context of this study, they studied certain parameters (including the metal ratio, reaction time, temperature of calcination, and temperature during the reaction) and were then able to predict characteristics to optimize the catalyst, such as a size of 56.4 nm produced at 700°C, reacting for 18 minutes, and more, which ultimately led to a biodiesel yield of 96.11% [37]. Furthermore, ANNs can improve the process of turning industrial waste (sludge) into biogas. Compared with the AI-based model and a math-based regression model, the AI algorithm was better at predicting methane production, indicating that chemical industry sludge had the greatest impact on methane production (28.59%), with a standard prediction error of 2.51% [374].

AI-driven analysis can also optimize processes to minimize harmful emissions generated during fuel combustion. In one study, ANNs evaluated a biofuel blend of waste fish oil biodiesel and diesel fuel, with graphene quantum dots as a nanoadditive. The model predicted various outcomes, such as emission quantities and engine performance. The optimized fuel blend (comprising 10% waste of fish oil biodiesel, 6% ethanol, and 45 ppm graphene quantum dots) resulted in decreases in the contents of unburned hydrocarbons, carbon monoxide, and carbon dioxide of 21.70%, 24.67%, and 26.38%, respectively. Brake power also increased by 13.84%, whereas brake specific fuel consumption decreased, indicating enhanced fuel efficiency [38].

Smart sensing technologies monitor key aspects of biofuel processes, such as temperature, feedstock quality, and pH, in real time. These data can be analyzed by AI, enabling full process control and allowing for immediate adjustments to be made if necessary to optimize enzymatic and microbial activity, as well as to predict yields and potential failures. Nanomaterials play a key role in smart sensing by improving the sensitivity, specificity, and quick response time of biosensors. A property that supports their responsiveness is their high surface area-to-volume ratio, which allows them to have increased contact with their surroundings. Smart nanomaterials can also self-assemble in response to environmental stimuli [375]. For example, a study developed biosensors that used enzymes (alcohol oxidase and alcohol dehydrogenase) combined with graphene oxide-based electrodes and palladium nanoparticles. The sensors could measure butanol levels accurately at fermentation levels even at low electrical voltages. This approach helps optimize the fermentation process by providing precise, fast feedback on biofuel quality [376].

### 8.3. Development of Conventional Processes and Modifications

As the demand for sustainable and renewable energy intensifies, nanomaterial-enhanced biofuels are undergoing various developments and optimizations to reach industrial scales. For example, although nanomaterials have significantly advanced biomass pretreatment, ongoing efforts have focused on biomass selection itself to further optimize the process in terms of cost and efficiency. Currently, several biofuels are derived from both edible (such as corn or palm oil) and nonedible (such as Jatropha or Mahua oils) feedstocks. However, inexpensive nonedible waste oils are gaining interest as a sustainable and economically viable option since edible crops often compete with food supplies and demand extensive water and land resources for cultivation [378]. Additionally, as introduced in section 5.3, future research is expected to focus on efficient methods to utilize algae-based biomass, which has a high oil content, a rapid growth rate, and low net carbon emissions, thus providing potential for biofuel production [87, 91, 368].

Furthermore, understanding the interaction mechanisms of nanoparticles during catalytic reactions, particularly the characteristics of intermediate compounds and final products, is essential [6]. For example, heterogeneous catalysis is a useful technique in the transesterification process for biodiesel production (as discussed in section 5.2), although there is potential for further improvements. Development may involve refining reactor designs and optimizing nanocatalyst compositions to regenerate, resist damage, and increase stability under industrial conditions. Zinc-based catalysts such as zinc acetate ($Zn(OAc)_2$) have increased in popularity because of their high selectivity and feedstock processing ability. Nevertheless, continued research has focused on enhancing their heat resistance, as they are currently limited by low thermal stability, which reduces their effectiveness under high temperatures and prolonged processing conditions [13, 232]. Additional modifications in biofuel processes will also be provoked from research aimed at screening nanoparticles across a wide range of concentrations to assess their effects on microbial activity and identify optimal conditions for bioprocessing. This research can likely be extended to explore nanoparticles of varying shapes and sizes to better understand how structural and surface properties influence the different stages of biofuel production [379].

### 8.4. Multifunctional Nanohybrids and Biomimetic Systems

Biomimicry at the nanoscale aims to design nanomaterials and nanosystems to imitate natural structures or functionalities in biological organisms and can be applied to various fields, including energy systems [380]. Multifunctional nanohybrids can enable a single nanomaterial to perform multiple functions simultaneously, thereby addressing multiple interrelated challenges. As discussed in section 2.1.3., nanohybrids can present a combination of unique properties. Carbon-metal nanohybrids are widely studied because of their multifunctionality and overall enhanced properties, which are distinct from those of the initial components, such as electronic qualities and particle morphology [40]. Both biomimicry and nanohybrids are emerging methods to increase biofuel production.

For example, nanoflowers are organic–inorganic hybrid nanomaterials with a three-dimensional, flower–like structure consisting of metal ions (such as copper, manganese, or cobalt) combined with biomolecules such as enzymes or nucleic acids. Their configuration results in a large surface area that enables increased catalytic efficiency [381]. A study produced a lipase–inorganic hybrid nanoflower as a biocatalyst for biodiesel production via biomimetic mineralization, a process inspired by natural biomineralization in which organic–inorganic materials are formed by combining bioactive molecules with inorganic components. In this case, the enzyme lipase was encapsulated by copper ions. The resulting hybrid demonstrated excellent catalytic efficiency, reusability, and stability during the transesterification of sunflower oil. The conversion yield reached 96.5% initially and remained at 72.5% after five reuse cycles, indicating strong potential as a cost-effective catalyst for biodiesel production [382, 383].

Additional studies have also highlighted the benefits of nanohybrids for biofuel performance. For example, researchers extracted lipase enzymes from hot spring bacteria and attached them to magnetic graphene oxide to produce a nanocomposite biocatalyst. The nanocatalyst was used to support biodiesel production from castor oil, increasing yield three times, functioning at a wider range of temperatures and maintaining approximately 75% of its activity after 30 days [385].

Moreover, a study employed a platinum, iridium, ruthenium nanoalloy (PtIrRu) for the production of biodiesel from waste lemon seed oil, achieving a 98.2% yield under optimal conditions, with a low voltage and without toxic solvents [386].

To further the concept of immobilizing nanoparticles with enzymes to increase catalytic performance, as discussed in section 5, researchers are working to develop synthetic catalysts that replicate enzymes entirely. Nanomaterials often operate as enzyme mimics and are referred to as "nanozymes"; they aim to reflect four fundamental characteristics found in natural enzymes to increase catalytic efficiency, as follows [387]:

- Stabilizing the transition state of reactions to lower the energy barrier, thereby increasing the reaction speed
- Bind with reactants to hold them in place for reactions.
- Containing key functional groups in the enzyme's active site to support required reaction tasks, such as transferring protons.
- A particular environment, with respect to pH and solvation properties, is created to facilitate the reaction

A study employed a gold-bismuth vanadate nanocomposite ($Au/BiVO_4$) as a catalyst to enhance the performance of glucose biofuel cells. The nanocatalyst mimicked the behavior of the glucose oxidase enzyme and provided additional photoelectric functionality. The gold nanoparticles allow the material to absorb more light and move electrons faster, increasing efficiency. As a result, a high power output of 575 $\mu W/cm^2$ was achieved, turning both sunlight and glucose into electricity [388]. Moreover, nanozymes are being increasingly employed as alternatives to natural enzymes in bioelectrofuel systems, as they address key limitations in microbial biofuel production, such as improving microbial attachment and electron transfer on electrode surfaces. The biological synthesis of nanozymes by microbes such as bacteria and archaea also serves as an eco-friendly, cost-effective way to produce these catalysts. For example, microbes can synthesize metal and metal–oxide nanoparticles such as gold, silver, and platinum [389].

### 8.5. Integration into Existing Energy Infrastructure

To reduce costs and streamline implementation while avoiding potential complications, it would be most practical for nanomaterials to simply integrate and function with existing infrastructure first. Nanomaterials have been employed in various petroleum refining processes, such as nanomembranes to purify oil and separate gas or nanocatalysts to improve reaction rates and yields [42]. For example, catalytic cracking, a key process that breaks down heavy hydrocarbons into useful chemicals and fuels in oil refineries, is currently unable to produce high-quality fuels because it lacks the ability to convert specific molecules. A study developed a nanocomposite catalyst from cobalt, hafnium, oxide, and mesoporous silica for catalytic cracking, achieving 94% heavy oil conversion and 67% gasoline selectivity and producing high-purity fuels that meet industry standards [390]. As a cleaner, more sustainable option, biofuels are anticipated to progressively replace petroleum-based fuels in the future [13]. Nonetheless, they still remain underdeveloped and require further advancement for widespread adoption. For example, investigations into postprocess management and recollection of nanomaterials are needed, as discussed in section 7.3. Substrates such as microalgae and lignocellulosic biomass should be explored to optimize biogas production. Using an integrated approach, nanomaterials can be paired with well-established enhancement techniques, such as thermal or chemical pretreatment of feedstock and the addition of biological or inorganic agents, to increase the overall yield [273].

Furthermore, nanomaterials are increasingly used in engine combustion systems, particularly as fuel additives, to improve performance and reduce emissions, as discussed in section 6.2. As additives, nanoparticles can enhance performance by decreasing the ignition lag and accelerating fuel evaporation. They can optimize the mixing of air and fuel and enhance heat transfer, contributing to more efficient fuel use and lower consumption [391]. Nevertheless, several challenges remain that must be resolved for future applications, related to the environmental and health risks,  high production expenses, and the potential for engine damage [392]. For example, the large surface area of nanoparticles results in agglomerating tendencies when added to fuel, which can block engine injection nozzles, causing difficulties such as failure to start the engine or reduced power output [393].

### 8.6. Regulatory, Environmental, and Safety Considerations

The industrial-scale application of nanomaterial-enhanced biofuels relies heavily on strong policy and legal frameworks. Countries adopt various strategies to support biofuel development. For example, in the United States, the government has provided support for biofuels through research funding and financial incentives, and the U.S. Senate Committee even passed a bill in 2012 to extend algal-based fuels in tax credits for cellulosic fuels [394]. Globally, biofuel policies have also significantly evolved over the past two decades, such as the 2018 National Biofuels policy, which categorized biofuel generations and promoted waste-to-energy strategies using feedstocks such as cooking oil and agricultural waste for fuel conversion [395].

To ensure safety in nanomaterial applications, regulatory efforts and protocols such as the safe-by-design (SbD) approach are being increasingly developed. In the context of nanomaterials, SbD aims to develop safer nanomaterials and products by integrating safety considerations early in the innovation process. It involves early screening before full development via simple, predictive, cost-effective tests to assess potential risks, such as toxicity, throughout the nanomaterial's life cycle. The European Union's (EU) Green Deal promotes the development of sustainable and safe chemicals, with the SbD approach emerging as a strategy to increase the EU's ability to anticipate and manage unforeseen risks associated with nanomaterials [396, 397].

From an environmental perspective, nanotechnology is predicted to play a significant role in achieving several United Nations Sustainable Development Goals (SDGs), including through its role in biofuel production, by advancing clean and sustainable energy systems. For example, SDG 2 (Zero Hunger) can be addressed by prioritizing nonedible feedstock as biomass, thereby preserving edible crops and increasing food availability. Nanotechnology-based biofuel production can help achieve SDG 1 (no poverty) by stimulating local economies and creating employment opportunities in areas such as biofuel cultivation and processing. SDG 7 (affordable and clean energy) and SDG 7 (climate action) can also be supported since the integration of nanomaterials into fuels helps reduce harmful emissions [398].

## 9. Conclusions

Compared with conventional fossil fuels, biofuels are an emerging renewable energy source that is more environmentally friendly. Nanomaterials are increasingly being used to enhance biofuel, as they have been proven to increase sustainability, efficiency, stability, and yield. They contribute to better overall performance, reduce harmful emissions, and improve the storage and transportation of biofuels by increasing their stability and minimizing their degradation. Nanomaterials can be applied across different stages of various biofuel processes, including the pretreatment of biomass, transesterification for biodiesel production, fermentation for bioethanol production, anaerobic digestion for biogas production, and the function of enzymatic and microbial biofuel cells.

The unique properties of nanomaterials, including their large surface-to-volume ratio, capacity for functionalization, thermal and chemical stability, reusability, and other microstructural variations, enable greater catalytic activity and cost reductions. Different types of nanomaterials, including carbon-based, metal-based, metal-oxide, nanocomposite, and nanohybrid materials, have diverse properties suitable for various applications. Various chemical, green, biological, and enzymatic methods have been employed to synthesize nanoparticles, with extensively studied methods including sol-gel, CVD, hydrolysis, spray pyrolysis, and coprecipitation. Each method offers certain advantages and disadvantages, enabling the production of nanoparticles with distinct characteristics suited for different applications. Characterization techniques such as TEM, SEM, FTIR, and BET are used in research to distinguish between nanoparticles and provide insight into physical traits, such as porosity and surface chemistry, to help optimize the role of nanomaterials in biofuels.

While nanomaterials offer measurable gains in efficiency, yield, and emission reduction across biofuel processes, there are remaining challenges and contradictions that must be addressed before nanomaterial-enhanced biofuels reach the industrial scale, including cost, energy-intensive synthesis processes, underdeveloped research, and environmental and health concerns regarding potential toxicity. Life cycle assessments must be conducted more extensively to truly understand the impact of nanomaterials from synthesis to disposal. It is important to select materials for nanoparticles that avoid shifting environmental and economic burdens from the combustion stage to the supply change. Ongoing advancements aim to enable the smooth integration of these biofuels into existing infrastructure, using artificial intelligence for process monitoring and optimization. Nanomaterial functionality can be strengthened by taking inspiration from biological systems through biomimicry and green synthesis methods using natural, renewable, or waste agents to help increase sustainability at a potentially lowered cost of biofuel production. The incorporation of nanomaterials already lowers hydrocarbon emissions with improved efficiency, helping to lower environmental impacts. Government support and adherence to regulatory, safety, and environmental protocols are critical for biofuel development to ultimately reach the industrial scale.

## Acknowledgements

This study and its abstract were presented as a poster at the Undergraduate Engineering Research Day (UnERD) conference at the University of Toronto, Toronto, Canada, on 20th August 2025.

The conception, framework, and core development of this work stems from the corresponding author’s independent research carried out during their Undergraduate program (late 2022). This publication is entirely independent of the author’s current doctoral research, supervisory oversight, or institutional funding. In full compliance with the University of Toronto's Policy on Ethical Conduct in Research and Framework for the Responsible Conduct of Research, the author retains the unrestricted right to disseminate original work produced independently and ethically.

All co-authors have contributed within their defined academic roles and bear no responsibility for institutional or authorship-related disputes.

## Funding

There are no funding bodies/agencies involved in this research work, and the authors solely contribute to it.

## Declaration of Competing Interest

The authors declare that they have no known competing financial interests or personal relationships that could have appeared to influence the work reported in this paper.

## Data availability

No data was used for the research described in the article.

## Author Contribution Statement

- **Lydia Lonardi:** conceptualization, methodology, investigation, writing-original draft preparation, writing-review and editing, visualization, resources, data curation.
- **Caitlyn Lew-Tong:** conceptualization, methodology, investigation, writing-original draft preparation, writing-review and editing, visualization, resources, data curation.
- **Brian Dunsten Miranda:** conceptualization, investigation, methodology, software, formal analysis, investigation, writing-original draft preparation.
- **Chandraleka V.S.:** conceptualization, methodology, software, formal analysis, investigation, writing-original draft preparation.
- **Evelyn (Wenxuan) Li:** conceptualization, methodology, formal analysis, investigation, writing-original draft preparation.
- **Bhaavyaa Sahukari:** methodology, investigation, writing-original draft preparation, writing-review and editing.
- **Harshit Poddar:** methodology, investigation, software, writing-original draft preparation, visualization, data curation.
- **Keerthana Satheesh:** methodology, investigation, writing-original draft preparation.
- **Utkarsh Chadha:** conceptualization, methodology, formal analysis, validation, supervision, project administration, writing-review and editing, visualization.